\documentclass[preprint]{elsarticle}
\makeatletter
\def\ps@pprintTitle{%
 \let\@oddhead\@empty
 \let\@evenhead\@empty
 \def\@oddfoot{\centerline{\thepage}}%
 \let\@evenfoot\@oddfoot}
\makeatother
\usepackage[utf8]{inputenc}
\usepackage[T1]{fontenc}
\usepackage[rmargin=2.5cm , bmargin=2cm, lmargin=2.5cm, tmargin=2cm]{geometry}
\usepackage{lineno,hyperref}
\modulolinenumbers[5]
\usepackage[colorinlistoftodos,prependcaption,textsize=normalsize]{todonotes}
\usepackage{units}
\usepackage{subcaption}
\usepackage{tabu}
\usepackage{booktabs}
\usepackage{makecell}
\usepackage{dsfont}
\usepackage{url}
\usepackage{amsmath,amssymb,amsfonts,amsthm}
\usepackage{mathtools}
\usepackage{bm, bbm}
\usepackage{listings}
\usepackage{xcolor}
\usepackage{appendix}
\usepackage[noabbrev]{cleveref}
\usepackage[normalem]{ulem}

\usepackage[symbol]{footmisc}
\usepackage{stmaryrd}
\usepackage{tikz}
\usepackage{tkz-euclide}
\usetikzlibrary{matrix,chains,positioning,decorations.pathreplacing,arrows, calc, fit, through, shapes.geometric}
\tikzstyle{arrow} = [thick,->,>=stealth]
\usepackage{standalone}
\usepackage{multirow}

\definecolor{blush}{rgb}{0.87, 0.36, 0.51} 
\definecolor{lgray2}{rgb}{0.9, 0.9, 0.9}
\definecolor{WtalBlue}{Hsb}{209,1,.47}
\definecolor{pastelorange}{rgb}{1.0, 0.7, 0.28}
\definecolor{darkpastelblue}{rgb}{0.47, 0.62, 0.8}
\definecolor{darkgray}{rgb}{0.3, 0.3, 0.3}
\definecolor{darkraspberry}{rgb}{0.53, 0.15, 0.34}

\DeclareMathOperator*{\argmin}{arg\,min}

\newcommand{\x}{\bm{x}}
\newcommand{\z}{\bm{z}}

\journal{Journal of \LaTeX\ Templates}

\begin{document}

\begin{frontmatter}

\title{Comparison of Generative Learning Methods as Turbulence Surrogates}

\author{Claudia Drygala$^1$\footnote[1]{equal contribution}, Edmund Ross$^{1\ast}$, Mohammad Sharifi Ghazijahani$^2$,\\ Christian Cierpka$^2$, Francesca di Mare$^3$ and Hanno Gottschalk$^1$}
\address{$^1$Technical University of Berlin, Institute of Mathematics\\
$^2$Technische Universität Ilmenau, Institute of Thermodynamics and Fluid Mechanics\\
$^3$Ruhr University Bochum, Department of Mechanical Engineering, Chair of Thermal Turbomachines and Aero Engines\\
$\{$drygala, ross, gottschalk$\}$@math.tu-berlin.de, $\{$mohammad.sharifi-ghazijahani, christian.cierpka$\}$@tu-ilmenau.de, francesca.dimare@ruhr-uni-bochum.de}

\begin{abstract}
Numerical simulations of turbulent flows present significant challenges in fluid dynamics due to their complexity and high computational cost. High resolution techniques such as Direct Numerical Simulation (DNS) and Large Eddy Simulation (LES) are generally not computationally affordable, particularly for technologically relevant problems. Recent advances in machine learning, specifically in generative probabilistic models, offer promising alternatives as surrogates for turbulence. This paper investigates the application of three generative models — Variational Autoencoders (VAE), Deep Convolutional Generative Adversarial Networks (DCGAN), and Denoising Diffusion Probabilistic Models (DDPM) - in simulating a von Kármán vortex street around a fixed cylinder projected into 2D, as well as a real-world experimental dataset of the wake flow of a cylinder array. Training data was obtained by means of LES in the simulated case and Particle Image Velocimetry (PIV) in the experimental case. We evaluate each model's ability to capture the statistical properties and spatial structures of the turbulent flow. Our results demonstrate that DDPM and DCGAN effectively replicate all flow distributions, highlighting their potential as efficient and accurate tools for turbulence surrogacy. We find a strong argument for DCGAN, as although they are more difficult to train (due to problems such as mode collapse), they show the fastest inference and training time, require less data to train compared to VAE and DDPM, and provide the results most closely aligned with the input stream. In contrast, VAE train quickly (and can generate samples quickly) but do not produce adequate results, and DDPM, whilst effective, are significantly slower at both, inference and training time.
\end{abstract}
\end{frontmatter}
\nolinenumbers
\section{Introduction}
\label{sec:intro}

Turbulent flows are notoriously difficult to model. The structures involved can be found across a 
wide range of temporal and spatial scales, and the high degree of non-linearity as well as sensitivity to the initial conditions make this an especially challenging problem \cite{frisch1995turbulence}.

Furthermore, these kind of flows play a central role in the field of fluid dynamics, and therefore in diverse fields such as aerospace \cite{Spalart_Venkatakrishnan_2016}, astrophysics \cite{stars_turbulence}, quantum mechanics \cite{Barenghi_2014}, and even in immune system biology \cite{cancer_turbulence}, since almost all real world flows exhibit turbulence of some kind. Despite the best efforts and decades of research from physicists and mathematicians, a general analytic solution to the governing (Navier-Stokes) equations remains elusive, requiring a variety of computational methods to obtain even numerical solutions. Scale-resolving techniques, such as DNS and LES, strive to resolve the entire spectrum of length and time scales present in the flow, or at least the energetically most significant part (as is the case for LES), whereas conventional averaging approaches (such as Reynolds-averaged Navier-Stokes, or RANS) instead attempt to fit a statistical model. Whilst the former approaches provide the highest possible modeling accuracy, the latter are affected by inevitable loss of information, with non-generalizable closures, strongly affected by flow typology and boundary conditions. Even in techniques such as RANS, which considerably reduce the computational cost of complex simulations, they remain prohibitively high if a large number of computations must be carried out in relatively short time, such as in rapid prototyping and design optimization loops \cite{gopalan2013unified}. This makes the use of machine learning, especially generative probabilistic AI extremely promising, particularly when only statistical distributions resulting from stochastic initial conditions are required \cite{lumley2007stochastic, olson2001turbulence}.

The availability of large amounts of data, helped by a recent increase in computing power and specialized machine learning hardware, as well as the development of new, bleeding edge model architectures, allow the issue of large computation cost in machine learning to be addressed conveniently, although these are still not remotely comparable to the costs induced by full numerical simulations.

Ling et al.~\cite{Ling:2016} and Jiang et al.~\cite{Jiang:2020} pioneered the use of deep neural networks (DNNs) to determine the model constants of nonlinear algebraic vortex viscosity models, significantly improving the prediction of anisotropic turbulence effects.

While early work used machine learning mainly to improve the prediction quality of existing models, also known as ML-augmented turbulence modeling~\cite{Cheung:2011, Edeling:2014a, Edeling:2014b, Weatheritt:2016, Weatheritt:2017, Zhang:2018a, Zhang:2018b, Zhao:2020}, recent publications propose that turbulent flow can be modeled directly by generative models.  

In addition to modeling entire flow fields, the applications of generative models range from improving the resolution of flow fields \cite{deng2019super}, to reconstructing incomplete (gappy) flow fields using the in-painting technique \cite{zheng2024high}, to uncertainty quantification to assess the variability in predictions due to uncertainties in initial or boundary conditions, or indeed in the flow model itself \cite{cheng2023uncertainty, goswami2023uncertainty}. Also Echo State Networks were recently used to spatially and/or temporally predict turbulent flow fields based on temporally resolved and spatially under-resolved experimental data~\cite{Sharifi_Ghazijahani_2025_eaai,Sharifi_Ghazijahani_2024_scirep,Ghazijahani_2024_MLST}.

New generative models are one of the driving forces behind the growing application of generative learning to turbulence modeling. Kingma et al.~\cite{kingma2013} introduced the Variational Autoencoder (VAE), a first generative model that provides stochastic variational inference and a learning algorithm that scales to large data sets. Shortly thereafter, Goodfellow et al.~\cite{vanilla_gan} proposed a generative adversarial network (GAN) that uses a minmax game to train an optimal generator, providing a paradigm shift in generative learning. Recently, the diffusion probabilistic models (DDPM) was introduced by Ho et al.~\cite{ho2020denoising} as another powerful generative model, which is based on the principle of diffusion models~\cite{sohldickstein2015deep} where the distribution of the data is learned by an iterative Markov chain process. Recently, Teutsch et al.~\cite{Teutsch_2025} showed that convolutional autoencoder-based reduced-order models can not only be used for modeling turbulence but provide interpretable features that further help to understand the flow physics. 

The advantage of these generative models is that the production of the data at inference time is very inexpensive once the models are trained. This allows for a large volume of data to be created, producing a representation of the target distribution that would be simply impractical to obtain with traditional methods. 

The question that remains is whether the results of the generative models are of reasonable quality and which generative model is best suited for modeling turbulence, and in particular, a flow field around a cylinder and in the wake of a cylinder array. We compare three generative models: VAE, Deep Convolutional GAN (DCGAN) \cite{dcgan}, and DDPM, and discuss their capabilities and limitations in terms of visual quality, physics-based metrics and computational cost.

Our previous work \cite{drygala2022generative} already showed that GAN-synthesized turbulence match LES-flow excellently. We show in this paper that the DDPM's results are competitive with GAN, but require significantly more training data, supporting the findings of \cite{wang2024patch}, and are computationally intensive in terms of both training and sample generation. The average inference speed per sample of our DDPM model is $36.3s$ compared to that of $0.001s$ by the GAN, a speedup factor of more than $1,000$. The VAE can generate at $0.0003$ samples per second, but suffers from physical and visual inaccuracies. 

With this work we place at least one of the hyperparameters of the machine learning approach for turbulence (namely, the model architecture) on scientific footing. We also note that we are one of the first to train generative machine learning model on a high-fidelity experimental dataset. In particular, we only check to see if the models can accurately reproduce the statistics of the projected flow. While the majority of the ML community are moving towards diffusion and other flow-matching based models due to the success of models like StableDiffusion \cite{latentdiffusionmodel2022} in other domains, this article finds in favor of GANs when applied to turbulence.

\paragraph{Outline} 
The remainder of the paper is structured as follows. 
\Cref{sec:relwork} gives a review on key studies in the field using VAE, GAN or DDPM for applications of turbulence modeling. In \cref{sec:methodology} we describe our specific implementation of the VAE, GAN, and DDPM along with a summary of the theoretical foundations of these models. \Cref{sec:dataset} offers a brief description of the experimental and numerical datasets used in this study. In~\cref{sec:experiments}, we will discuss each model's performance using physics-based evaluation metrics to assess their effectiveness and examine the associated computational costs. Finally, \cref{sec:conclusion} will summarize the main findings, discuss their broader implications, and suggest potential avenues for future research.

\section{Related work}
\label{sec:relwork}
Turbulence modeling has recently seen the rise of data-driven and machine-learning-based methods that complement classical approaches such as RANS and LES, with reviews of both the historical development and contemporary ML techniques provided in the recent literature~\cite{duraisamy2019turbulence, zhangreview2023, fang2025breakthroughs}.
In the field of classical simulation methods, RANS can provide solutions at a comparably lower cost, but also paying the penalty of a reduced accuracy. Machine learning approaches represent a promising solution to overcome the problem of high computational cost without losing the details of turbulent structures \cite{wang_2017, Maulik_2018, shankar2023differentiableturbulenceii}.  
In this section we give a broad overview of contributions using VAE, GAN and, DDPM.

\paragraph{Variational Autoencoders (VAE)}

Over the last few years, modified and extended versions of the Variational Autoencoder have been developed to solve problems in a variety of scientific fields \cite{higgins2017betavae, DBLP:journals/corr/abs-1903-05789, ghosh2020variationaldeterministicautoencoders, NIPS2016_ddeebdee, oord2018neuraldiscreterepresentationlearning}.
Advanced VAE frameworks have also been applied to turbulence generation. Gunderson et al.~\cite{gundersen2021scvae} presented a semi-conditional VAE (SCVAE) to reconstruct nonlinear flow from spatially sparse observations, which also allows, due to the probabilistic reconstruction, uncertainty quantification of the prediction.
In 2020, Agostini~\cite{agostini2020aelow}, a two-stage approach is proposed where low-dimensional dynamics reconstructed by an autoencoder are enhanced by another high-resolution neural network.
The work of Wang et al.~\cite{wang2021flow} and Cheng et al.~\cite{CHENG2020113375} goes even further and uses a hybrid approach to model turbulence, combining an autoencoder with a multilayer perceptron network or a GAN to predict steady flow fields around supercritical airfoils and nonlinear fluid flows in varying parameterized space. As often the availability of data is problematic, Teutsch et al.~\cite{Teutsch2025_aplml} presented a pipeline that determines useful data augmentation strategies for different turbulent flows.  

\paragraph{Generative Adversarial Networks (GAN)}

Among the new approaches to turbulence modeling, many make use of GAN technology. 
In our previous work, we introduced GAN as a mathematically well-founded approach for synthetic modeling of turbulent flows \cite{drygala2022generative}.
Pioneering this field of research, King et al.~\cite{King:2017, King:2018} showed that GANs are capable of generating syntheses of 2D flow fields after having been previously trained on direct numerical simulation (DNS) data. The reproductions even satisfied statistical constraints of turbulent flows, such as the Kolmogorov - 5/3 law and the small-scale intermittency of turbulence.
Using an unsupervised trained combination of a GAN and a recurrent neural network (RNN), Kim and Lee synthesized   boundary conditions for turbulent flow \cite{Kim:2020} or generated stationary DNS flow fields \cite{kim2021unsupervised}.

With the use of more advanced GAN frameworks such as conditional GAN (cGAN), turbulence can be predicted and controlled \cite{Kim_Kim_Lee_2024}, even only at specific local points \cite{yan2023local}.
\cite{atmos15010060} investigated the task of inferring a velocity component from the measurement of a second one for a rotating turbulent flow. To do this, a GAN with context encoders takes the simulated data as input and adds a second loss term that measures the point-to-point distance between the GAN output and the ground truth configuration.

Another application of GAN is the large field of super-resolution reconstruction of turbulent flows.
These methods can be used to synthetically scale up low-resolution or noisy flow fields \cite{deng2019super, Fukami:2019, Fukami:2020, Liu:2020, Xie:2017, Werhahn:2019, youssif2021high, youssif2022superres}. For these problems, the so-called super-resolution GAN (SRGAN) \cite{ledig2017photo} or enhanced SRGAN (ESRGAN) \cite{wang2018esrgan} are used, which can be conditioned by additional physical information \cite{subramaniam2020turbulence}. 
Also the high-fidelity reconstruction of 2D damaged turbulent fields was addressed with GAN \cite{zheng2024high, Li_Buzzicotti_Biferale_Bonaccorso_Chen_Wan_2023}.

Finally, the generalization capabilities of GAN are of interest and have been investigated. In our previous work, we demonstrated the generalization capabilities of conditional deep convolutional turbulence generators when geometric changes occur in the flow configuration \cite{drygala2023generalization}. By changing parameters in the numerical setup, such as the Reynolds number, Nista et al. investigated the generalization capabilities of their proposed SRGAN \cite{nista2023investigation}. 
Bode et al.~\cite{bode2021using} combined high and low resolution flows to improve the generalization capability for a physics-informed super-resolution GAN (SRGAN). 

\paragraph{Denoising Diffusion Probabilistic Models (DDPM)}

The most recent of the generative models considered in our work are the Denoising Diffusion Probabilistic Models \cite{ho2020denoising} based on the idea of Sohl et al.~\cite{sohldickstein2015deep}.
In the field of turbulence modeling, diffusion models are gaining popularity, especially for probabilistic spatiotemporal forecasting. For example, Gao et al.~\cite{gao2024bayesian} implemented a Bayesian conditional diffusion model for versatile spatiotemporal turbulence generation. 
Whereas autoregressive learning with DDPM is often used to solve the task of probabilistic spatiotemporal prediction, cf. \cite{shu2024zero, kohl2024benchmarking, yang2023diffusion}, Rühling et al.~\cite{NEURIPS2023_8df90a14} introduced a dynamically informed diffusion model, which adapts the model to the dynamic nature of the data in order to achieve long-range prediction at inference time.
Another application of DDPM in turbulence modeling is the reconstruction of high-resolution turbulent flow fields from low-resolution flow data. For example, Qi et al.~\cite{Qi2024Combined} combined a conditional DDPM with an enhanced residual network, and Sardar et al.~\cite{sardar2023spectrally} developed a preprocessing to decompose flow fields into high and low wavenumber components to learn a conditional DDPM on this data.
Furthermore, DDPM were also investigated for training uncertainty-aware surrogate models for simulating turbulence as flows around differently shaped airfoils \cite{liu2024uncertainty}.

\section{Methodology}
\label{sec:methodology}
\subsection{Variational Autoencoders (VAE)}
\label{subsec:vae}
\begin{figure}[!ht]
    \centering
    \includegraphics[trim=0mm 0mm 0mm 12mm, clip, width=1\textwidth]{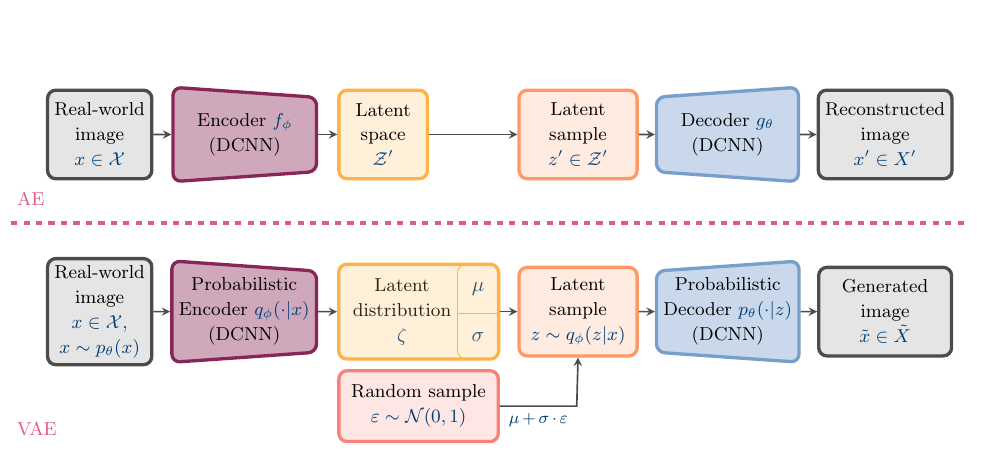}
    \caption{Comparison of the classical \cite{kramer1991nonlinear} (top) and the variational \cite{kingma2013auto} autoencoder (bottom) architectures. In the case of classical AE, the encoder $f_\theta$ receives as input a real-world image $x$ whose most important features are encoded in the latent space $\mathcal{Z}'$. Sending a point of this latent space back to the decoder $g_\phi$ should result in a reconstructed image $x'\approx x$. In contrast, in the case of the VAE, the real-world images have a chosen distribution, making the encoder and decoder probabilistic. The encoder learns the distribution parameters of the input images, and latent samples are drawn during training by the re-parameterization trick, making it possible to compute the gradients for $\mu$ and $\sigma$. Due to the stochasticity of the probabilistic decoder's input, the resulting images are generated rather than reconstructed, as they may differ from the original images.}
    \label{fig:vaeArch}
\end{figure}
An autoencoder consists of two components - an encoder $f:\mathcal{X} \rightarrow \mathcal{Z}'$ and a decoder $g:\mathcal{Z}'\rightarrow \mathcal{X}'$ with $\mathcal{X}$ the space of real-world images, $\mathcal{Z}'$ the latent space and $\mathcal{X}'$ the space of reconstructed images. In machine learning, both functions are represented by neural networks, and in our work specifically we use deep convolutional neural networks (DCNN), which are parameterized by weights $\phi$ and $\theta$. Due to the way an AE is implemented and the definition of its loss function, it can be trained end-to-end, meaning the encoder and decoder are optimized together in a single step, automatically learning useful internal representations without separate training.
The goal of training an AE is to learn a decoding function that is able to reconstruct an original image from a point in the encoded latent space  such that $x' = g_\theta(z') = g_\theta(f_\phi(x)) \approx x$ with $z' \in \mathcal{Z}', x\in \mathcal{X}$ and $ x' \in \mathcal{X}'$ (see \cref{fig:vaeArch}).
Once trained, the process of generating samples, which we also refer to as inference, is a simple matter of picking a point in the latent space and passing it through the decoding function.
Typically, the latent space in which important features are embedded is chosen to be small.
This property makes AE also a dimensionality reduction method and can be seen as a generalization of principal component analysis (PCA) \cite{ballard1987modular, kramer1991nonlinear}.

In recent years, more advanced variants of the AE have been developed \cite{pinaya2020autoencoders}, including the variational autoencoder (VAE) \cite{kingma2013auto}.
In contrast to a classical AE, the real-world input images $x \in \mathcal{X}$ are drawn from a chosen probability distribution $p_\theta(x)$ parameterized by the network weights $\theta$, making the encoder and decoder probabilistic and the latent space a latent distribution encoding the distribution parameters. In general, this underlying distribution is chosen to be the Gaussian $\mathcal{N}(x; \mu, \sigma)$.
Through the application of the powerful mathematical framework of Bayesian probability \cite{bayes1958essay, box2011bayesian}, the VAE can be interpreted as a model of the joint distribution $p_\theta(x,z) = p_\theta(x|z)p_\theta(z)$ of the real-world images $x \in \mathcal{X}$ and the samples of the latent distribution $z \in \zeta$, where the goal is to compute the posterior     
\begin{align*}
        p_\theta(z|x) = \frac{p_\theta(x|z)p_\theta(z)}{p_\theta(x)}~.
\end{align*}
Here, $p_\theta(x|z)$ is the likelihood of $x$ given $z$, $p_\theta(z)$ is the prior, and $p_\theta(x)=\int p_\theta(x|z)p_\theta(z)dz$ is the evidence, the total probability of observing $x$ under the model, averaged over all possible latent configurations $z$, which is generally expensive and intractable to compute.
The solution is to approximate the evidence with a family of distributions $q_\phi(z|x)$, where $\phi$ are the network weights for the distribution parameters for each data point $x$. 
Therefore, a computationally tractable approximation of the posterior can be derived by optimizing the evidence lower bound (ELBO), which leads to the following optimization problem for training the VAE:

\begin{align}
    \argmin_{\theta, \, \phi} \mathcal{L}_{\rm VAE}(x, z; \theta, \phi) =  \argmin_{\theta, \, \phi}{\left[
    -\mathfrak{d}_{KL}(q_\phi(z | x)\, || \, p_\theta(z)) + \mathds{E}_{z \sim q_\phi(z | x)}\log{p_\theta(x|z)}
    \right]}~,
    \label{eq:lossVAE}
\end{align}
where $p_\theta$ and $q_\phi$ are the probabilistic encoder and decoder respectively and $\mathfrak{d}_{KL}$ is the Kullback-Leibler (KL) divergence \cite{kullback1951information}, which quantifies the difference between two probability distributions by computing the expected logarithmic difference between them, i.e. how much information is lost when using $q_\phi(z | x)$ to approximate $p_\theta(z)$. Although it behaves like a distance between two distributions, it's not a true norm, being asymmetric and not satisfying the triangle inequality.
Note that the right-hand side of \eqref{eq:lossVAE} shows the reconstruction loss and the left-hand term forces the latent distribution to be close to the Gaussian prior.
Since we adopt the ubiquitous assumption of a standard normal prior $p_\theta(z)$ and Gaussian posterior $q_\phi(z | x)$, the KL-divergence can be calculated analytically and only the reconstruction loss needs to be estimated by sampling. Thus, we can rewrite the optimization problem in terms of
\begin{align}
    \argmin_{\theta, \, \phi}{\left[
    -\mathfrak{d}_{KL}(q_\phi(z | x)\, || \, p_\theta(z)) + \frac{1}{L}\sum^L_{l=1}{\log{p_\theta(x | z^{(l)})}}
    \right]}~.
\end{align}
With our prior and posterior assumptions, we can simply sample a point $z$ from the standard normal distribution and run it through the trained probabilistic decoder at inference time.
However, for sampling at training time, we need to use the so-called reparameterization trick (see \cref{fig:vaeArch}).
We want to optimize the distribution parameters $\mu$ and $\sigma$ of the latent distribution $\zeta$ from which we want to sample the inputs for the decoder.
The problem is that we can't backpropagate \cite{backprop} through a stochastic sample, i.e., we cannot compute gradients of the loss with respect to network parameters if $z$ is sampled directly.
Hence, the idea of reparameterization is to treat $\mu$ and $\sigma$ as deterministic variables, sample a point $\varepsilon$ from the standard normal distribution, and define a new latent sample $z=\mu+\sigma\cdot\varepsilon$.
This disentangles the stochasticity of the latent variable from the parameters, allowing gradients of $\mu$ and $\sigma$ to be computed while preserving randomness through $\varepsilon$.
Importantly, since the decoder's inputs are stochastic, it does not merely reconstruct the original images pixel-by-pixel, but can generate new samples within the same domain (e.g., turbulent flow data), capturing the variability learned during training.

\subsection{Deep Convolutional Generative Adversarial Networks (DCGAN)}
\label{ssec:DCGAN}
\begin{figure}
    \centering
    \includegraphics[width=0.96\linewidth]{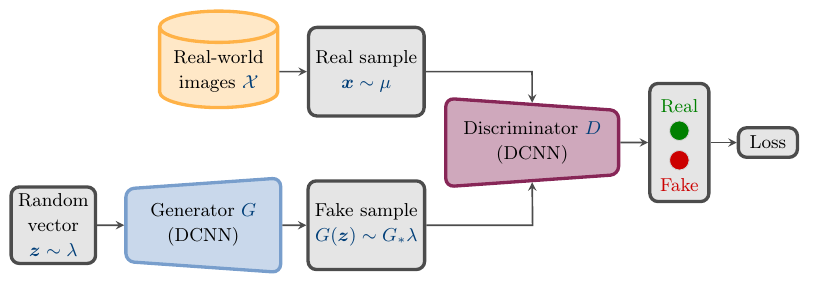}
    \caption{Architecture of a deep convolutional GAN \cite{vanilla_gan, dcgan}.
    While the training data is given by the real-world data $\mathcal{X}$, the fake samples $G(\z)\sim G_*\lambda$ are produced by the generator from a noise vector $\z$.
    The discriminator's inputs are GAN-synthesized and real-world samples, and it's task is to estimate the probability in a range of $[0,1]$ that an input sample comes from $\mathcal{X}$ rather than being generated by $G$. During training, the feedback from the discriminator reaches the generator by backpropagation. The entire GAN framework can be backpropagated at once, since $G$ and $D$ are both fully differentiable and trained end-to-end. If the unknown distribution of the real-world data is approximated by $G$ (i.e. $G_*\lambda\approx\mu$) and $D$ is only able to guess the real-world from the fake samples (i.e. $D(\cdot)\approx\frac{1}{2}$), the problem \eqref{eq:vanilla_gan_optimization_problem} reached its optimum.}
    \label{fig:ganArch}
\end{figure}

Similar to AE, generative adversarial networks (GAN) are basically composed of two mappings - a generator $G:\Lambda \rightarrow \Omega$ and a discriminator $D:\Omega \rightarrow [0,1]$, where $\Lambda$ is the space of latent variables with an easily simulated probability measure $\lambda$ as Gaussian noise, $\Omega=\{\mathcal{X}, \{G(z)\}\}$ is the space of real-world and generated images, and the interval $[0,1]$ gives the probability whether a sample is from the real-world images or a generated one. 
The generator $G$ transforms the noise measure $\lambda$ to the image measure $G_*\lambda$.
The objective of adversarial learning is to train a mapping $G$ using feedback from the discriminator $D$, such that $D$ cannot differentiate between synthetic samples generated by $G_* \lambda$ and real samples drawn from the unknown target distribution $\mu$. The discriminator $D$, in turn, is trained as a classifier to assign high probabilities to real-world data and low probabilities to synthesized data. If the generator $G$ has been trained so effectively that even the most optimal discriminator $D$ is unable to distinguish between samples from $\mu$ and $G_* \lambda$, then the generative learning process is considered successful, see \cref{fig:ganArch}.

In practice, both the generator $G$ and the discriminator $D$ are neural networks. The feedback from $D$ to $G$ is propagated backward \cite{backprop} through the composite mapping $D \circ G$, allowing the neural network weights of $G$ to be updated. Moreover, the universal approximation theorem for (deep) neural networks ensures that any mappings $G$ and $D$ can be approximated to a desired level of precision if the neural networks have a sufficiently wide and deep architecture. For qualitative and quantitative results, see \cite{vanilla_gan, asatryan2020convenient, pix2pixHD, CycleGAN2017, wang2018esrgan, karras2019style}. 
This work investigates the advanced deep convolutional GAN (DCGAN) \cite{dcgan} framework. As the name suggests, the generator $G$ and the discriminator $D$ are deep convolutional neural networks (DCNN), which have been successfully applied to image processing \cite{cnn_impact, cnn_explanation}. For guidelines on how to properly integrate DCNN into GAN to ensure stable training at high resolution with deeper architecture, see \cite{dcgan}. 

The training of a GAN is structured as a minmax game between the discriminator $D$ and the generator $G$, which is mathematically represented by the min-max optimization problem
\begin{align}
    \min_G \max_D \mathcal{L}(D, G) = \min_G \max_D \big( \mathbb{E}_{\x\sim \mu}[\log(D(\x))]+\mathbb{E}_{\z\sim \lambda}[\log(1-D(G(\z)))]\big)~.
    \label{eq:vanilla_gan_optimization_problem}
\end{align}
Note that the loss function $\mathcal{L}(D, G)$ is commonly known as binary cross-entropy \cite{bce}. 
As observed in \cite{vanilla_gan}, taking the maximum over a sufficiently large hypothesis space $\mathcal{H}_D$ of discriminators yields to
\begin{align}
    \label{eq:loss_JS}
    \max_{D\in\mathcal{H}_D}\mathcal{L}(D, G)=\mathfrak{d}_{\text{JS}}(\mu\|G_*\lambda)+\log(4)~,
\end{align}
where, $\mathfrak{d}_{\text{JS}}(\mu\|G_*\lambda)$ is the Jensen-Shannon (JS) divergence \cite{menendez1997jensen}, which is an information-theoretic pseudo-distance between the invariant measure $\mu$ and the generated measure $G_*\lambda$ and is defined as
\begin{align}
    \label{eq:Jenson-Shannon}
    \mathfrak{d}_{\text{JS}}(\mu\|G_*\lambda)=\mathfrak{d}_{\text{KL}}\left(\mu\left\|\frac{G_*\lambda+\mu}{2}\right.\right)+\mathfrak{d}_{\text{KL}}\left(G_*\lambda\left\|\frac{G_*\lambda+\mu}{2}\right.\right)~.
\end{align}
Here, the Kulback-Leibler distance between the measures $\nu$ and $\mu$ is given by $\mathfrak{d}_{\text{KL}}(\mu\|\nu)=-\mathbb{E}_{\x\sim \mu}\left[\log\left(\frac{f_\nu}{f_\mu}(\x)\right)\right]$ with continuous probability densities $f_\mu$ and $f_\nu$, respectively. Note that $\mathfrak{d}_{\text{KL}}(\mu\|\nu)=0$  is only true if and only if $f_\mu(\x)=f_\nu(\x)$ with $\mu$-probability one and hence $\mu=\nu$. 
Consequently, $\mathfrak{d}_{\text{JS}}(\mu\|G_*\lambda)$ also measures the distance between $\mu$ and $G_*\lambda$, and, being symmetric and bounded, it provides a more stable measure of distributional difference than the KL divergence alone - a property exploited in GAN training.

Once trained, the generator is able to synthesize samples from the distribution $G_*\lambda\approx\mu$ by simply sampling a noise vector $\z$ and passing it through $G$.

Note also that GAN have been shown to be a sound mathematical approach to turbulence modeling, as one can theoretically prove that they converge for ergodic systems\cite{neumann1932proof, birkhoff1931proof}, which are systems in which a single deterministic trajectory explores the entire phase space and thus allows for a statistical description of the system’s behavior \cite{Buzzi2009, eisner2015operator}. For more details, see our previous work \cite{drygala2022generative}.

\subsection{Denoising Diffusion Probabilistic Models (DDPM)}
\label{subsec:diffusion}
\begin{figure}
    \centering
    \includegraphics[width=0.96\textwidth]{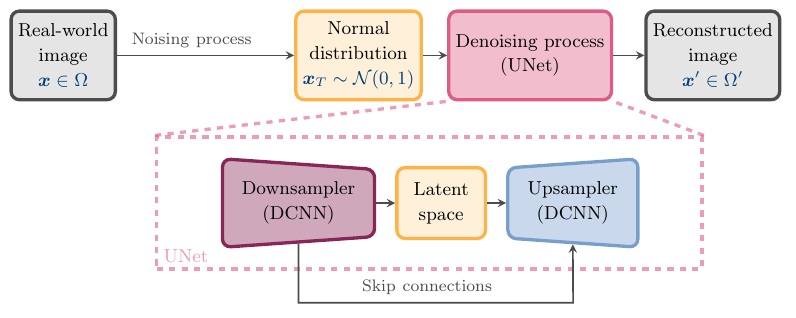}
    \caption{Denoising diffusion probabilistic (DDPM) model architecture \cite{ho2020denoising}. At training time, a sample from the dataset $x \sim \Omega$ is noised to a random step $t$ in the noising process, producing a partially noised sample $x_t \sim \mathcal{N}(x_{t-1}\sqrt{1 - \beta_t}, \beta_t I)$. The UNet is trained to produce $x_{t-1}$ from $x_t$ with the parameter $t$ given as a positional embedding. At inference time, a normal sample $x_T = z \sim \mathcal{N}(0, 1)$ is produced and fed to the UNet, which reverses the noising steps one at a time and produces a new sample $x' \sim \Omega'$.}
    \label{fig:ddpm_arch}
\end{figure}

Diffusion models convolve the distribution we are interested in into noise via an iterative Markov chain process \cite{sohldickstein2015deep, ho2020denoising}, by adding noise at each timestep. The model then learns the backwards map, that is, the function that returns the noise to the original distribution. In other words, the forwards process progressively corrupts the data with noise, while the backwards process denoises it to reconstruct the original distribution. The forwards process contains no learnable parameters, and the backwards process is modeled with a neural network, parameterized by weights $\theta$. Figure \ref{fig:ddpm_arch} gives an outline of the model architecture.

\subsubsection{VAE Interpretation}
Broadly speaking, there are two schools of thought by which to characterize the theory of diffusion models. We will begin with the VAE interpretation, which DDPM also subscribes to. Adapting the minimization problem \eqref{eq:lossVAE} from the VAE section above gives
\begin{align}
    \argmin_{\theta}{\left( 
    \mathfrak{d}_{KL}(q(x_T | x_0)\, || \, p(x_T)) + \sum^{T}_{t=1}{\mathfrak{d}_{KL}(q(x_{t-1} | x_t, x_0)\, || \, p_\theta(x_{t-1} | x_t)) - \ln{p_\theta(x_0 | x_1)}}
    \right)}
    \label{eq:ddpm_loss}
\end{align}
Here $x_t$ represents the state of the datapoint $x_0$ at timestep $t$ in the Markov chain, which has $T$ total steps. $x_T$ is equivalent to the latent variable $z$, the fully noised datapoint, assumed to be drawn from a standard unit normal with dimension equal to that of the datapoint.
The particular implementation used in this paper, denoising diffusion probabilistic models (DDPM) \cite{ho2020denoising}, parametrizes the forward noising process with a normal distribution
\begin{align}
    q(x_t | x_{t-1}) = \mathcal{N}(x_t; \, x_{t-1} \sqrt{1 - \beta_t}, \beta_t I).
\end{align}
The $\{\beta_t\}$ represent a fixed noise schedule, often linear in $t$, which controls the amount of noise added at each timestep and is chosen so that $q(x_T | x_0) \approx \mathcal{N}(0, I)$ \cite{song2021scorebased}.
This gives rise to the backwards model, notably also a normal distribution, with analytical form
\begin{align}
    q(x_{t-1} | x_t, x_0) = \mathcal{N}(x_{t-1} ; \, \tilde{\mu}_t(x_t, x_0), \tilde{\beta}_t I)
\end{align}
We can do one better and give the forms of $\tilde{\mu}_t(x_t, x_0)$ and $\tilde{\beta}_t$. The full derivation can be found in \cite{kong2021diffwaveversatile},
\begin{align}
    \tilde{\mu}_t\left(x_t, x_0\right)=\frac{\sqrt{\bar{\alpha}_{t-1}} \beta_t}{1-\bar{\alpha}_t} x_0+\frac{\sqrt{\alpha_t}\left(1-\bar{\alpha}_{t-1}\right)}{1-\bar{\alpha}_t} x_t
\end{align}
\begin{align}
    \tilde{\beta}_t=\frac{1-\bar{\alpha}_{t-1}}{1-\bar{\alpha}_t} \beta_t
\end{align}
with the following definitions
\begin{align}
     \alpha_t := 1 - \beta_t, \, \bar{\alpha}_t := \prod^t_{s=1} \alpha_s
\end{align}
We can model this with the generative forward process
\begin{align}
    p_\theta(x_{t-1} | x_t) = \mathcal{N}(x_{t-1}; \, \mu_\theta(x_t, t), \Sigma_\theta(x_t, t))
\end{align}
and make the choice that $\Sigma_\theta(x_t, t) = \sigma_t^2 I$.
We are finally ready to derive the ELBO, as we did with the VAE. Inserting into the minimization problem \ref{eq:ddpm_loss}, and noting that the form of the term in inside the sum can be written \cite{kong2021diffwaveversatile, sohldickstein2015deep, ho2020denoising}
\begin{align}
    \frac{1}{2 \sigma_t^2}\left\|\tilde{\mu}_t\left(x_t, x_0\right)-\mu_\theta\left(x_t, t\right)\right\|^2+C
\end{align}
for some $C$ independent of the parameters $\theta$. By itself, this is suitable for optimization, but we can go one step further, using the reparameterization trick mentioned in \ref{subsec:vae} \cite{kingma2013}. We separate the noise from the data to get
\begin{align}
    x_t(x_0, \epsilon) = \sqrt{\bar{\alpha}_t} x_0 + \epsilon \sqrt{1 - \bar{\alpha}}_t
\end{align}
where $\epsilon$ is distributed with a standard unit (multi-dimensional) normal. It turns out that predicting $\epsilon$ leads to better results. Finally, we get
\begin{align}
    \frac{\beta_t^2}{2 \sigma_t^2 \alpha_t\left(1-\bar{\alpha}_t\right)}\left\|\epsilon-\epsilon_\theta\left(x_t\left(x_0, \epsilon\right), t\right)\right\|^2
    \label{eq:ddpm_final_obj}
\end{align}
for our minimization objective. In practice, we also drop the prefactor since it only scales the loss. In \cite{ho2020denoising} it is noted that, aside from being simpler, this empirically leads to better results.\\

\subsubsection{Continuous Limit}
So far we have given the model in terms of discrete Markovian processes, in the sense that
\begin{align}
    x_t = \sqrt{1 - \beta_t} x_{t-1} + \sqrt{\beta_t} \epsilon_t, \,\,\,\, 1 \leq t \leq T
\end{align}
for some ${\epsilon_t}$, each independently identically distributed by the unit normal. It is natural to ask what happens in the continuous limit, namely as $T$ becomes very large. \cite{song2021scorebased, 2023diff_survey, mcallester2023maths_diff} show that we arrive at the stochastic differential equation (SDE)
\begin{align}
 dX_t = -\frac{1}{2}\beta(t)X_t dt + \sqrt{\beta(t)} dW_t, 
\end{align}
where $\beta(t)$ is an appropriate continuation of the discrete noise schedule, and $W_t$ is the standard Wiener process, i.e., a continuous-time stochastic process with independent Gaussian increments, modeling the accumulation of noise over time.
The reverse is also true, as discretizing this equation again returns us to the case above \cite{mcallester2023maths_diff}. In this sense SDEs can be seen as an overarching framework for diffusion models. Given that the noise schedule increases sufficiently quickly so that $x_T \approx \mathcal{N}(0, I)$, the time reversed (see the main theorem in \cite{anderson1982reverse_sde}) SDE is then
\begin{align}
    dY_t = \left(\frac{1}{2}\beta(T - t)Y_t + \beta(T - t)\nabla\log{p(x_{T - t}, Y_t)} \right)dt + \sqrt{\beta(T - t)} dW_t
    \label{eq:sde_reverse}
\end{align}
this is known as a `variance preserving' (VP) SDE. The probability distribution $p(\cdot; x_t)$ (here $t$ is a continuous parameter) satisfies the Fokker-Planck equation, also known as Kolmogorov's forwards equation, which is an ODE describing how the probability density decays with time \cite{song2021scorebased}. This equation, like equation \ref{eq:sde_reverse}, can be used to sample from the original data distribution \cite{2023diff_survey}.

\subsubsection{Score-Matching}
Another possible interpretation of diffusion models is to understand them by analogy with score-matching methods, which learn the `score' (the gradient of log probability) with methods such as Langevin dynamics \cite{song2020generativemodelingestimatinggradients}. Note that the score is known for a Gaussian. Indeed, in the formalism we have used so far,
\begin{align}
    \nabla \ln{\mathcal{N}(x; \, \mu, \sigma^2)} = \frac{\mu - x}{\sigma^2}
\end{align}
Apply the reparameterization trick from \ref{subsec:vae} to the distribution of $x_t$ and obtain
\begin{align}
  \nabla \ln{\mathcal{N}(x_t; \, \sqrt{\bar{\alpha}_t}x_0, (1-\bar{\alpha}_t)I)} = \frac{\sqrt{\bar{\alpha}_t}x_0 - x_t}{(1-\bar{\alpha}_t)} = -\frac{\epsilon}{\sqrt{1 - \bar{\alpha}_t}}  
\end{align}
Learning this score function with mean-squared error leads to the same optimization problem as \ref{eq:ddpm_final_obj}. Philosophically, this is much the same as models formulated under the VAE interpretation. Where the schools of thought differ is in the noising process (which in turns leads to a different sampling process). SBM (score-based matching) models utilize
\begin{align}
    x_t = x_{t-1} + \sqrt{\sigma^2_t - \sigma^2_{t-1}}\epsilon_t \,\,\,\, 1 \leq t \leq T
\end{align}
for some noise scales $\{\sigma_t\}$ typically set as a geometric sequence, which again ensure in the limit that the fully noised distribution is equivalent to $\mathcal{N}(0, I).$
Performing the limit procedure above on this process also leads to a SDE, known as a `variance exploding' (VE) SDE, a careful treatment of which can be found in \cite{song2021scorebased}.

\section{Dataset}
\label{sec:dataset}

\subsection{Setup of Simulations and Measurements}

We investigate different test cases, comprising a high-resolution dataset obtained from Large Eddy Simulation (LES) representing the flow around a cylinder, and lower-resolution datasets from experimental Particle Image Velocimetry (PIV) capturing the flow behind arrays of seven cylinders, which are described in the following.

\subsubsection{Flow around a cylinder}

\begin{figure}[!t]
    \centering

    \begin{subfigure}[b]{0.45\textwidth}
        \centering
        \includegraphics[width=\textwidth]{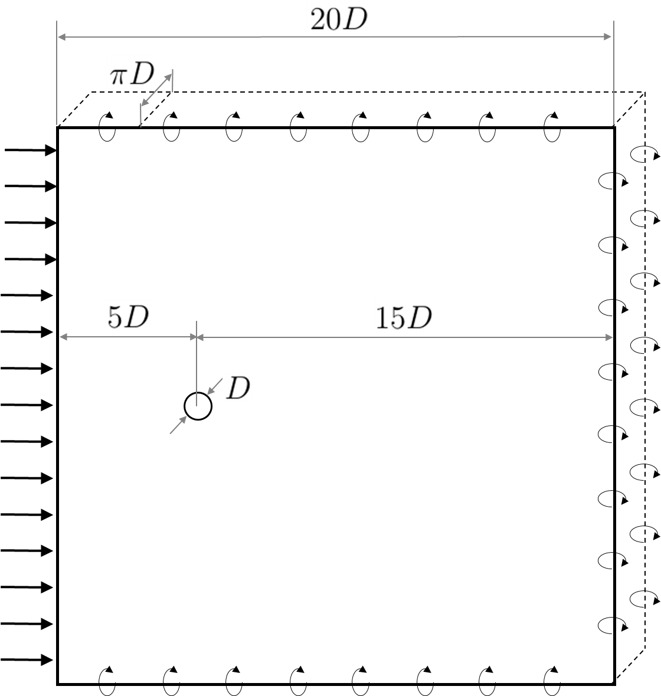}
        \caption{Flow around a cylinder}
        \label{fig:numericalsetupA}
    \end{subfigure}
    \hfill
    \begin{subfigure}[b]{0.45\textwidth}
        \centering
        \includegraphics[width=\textwidth, trim=0.2cm 0cm 0.2cm 0cm, clip]{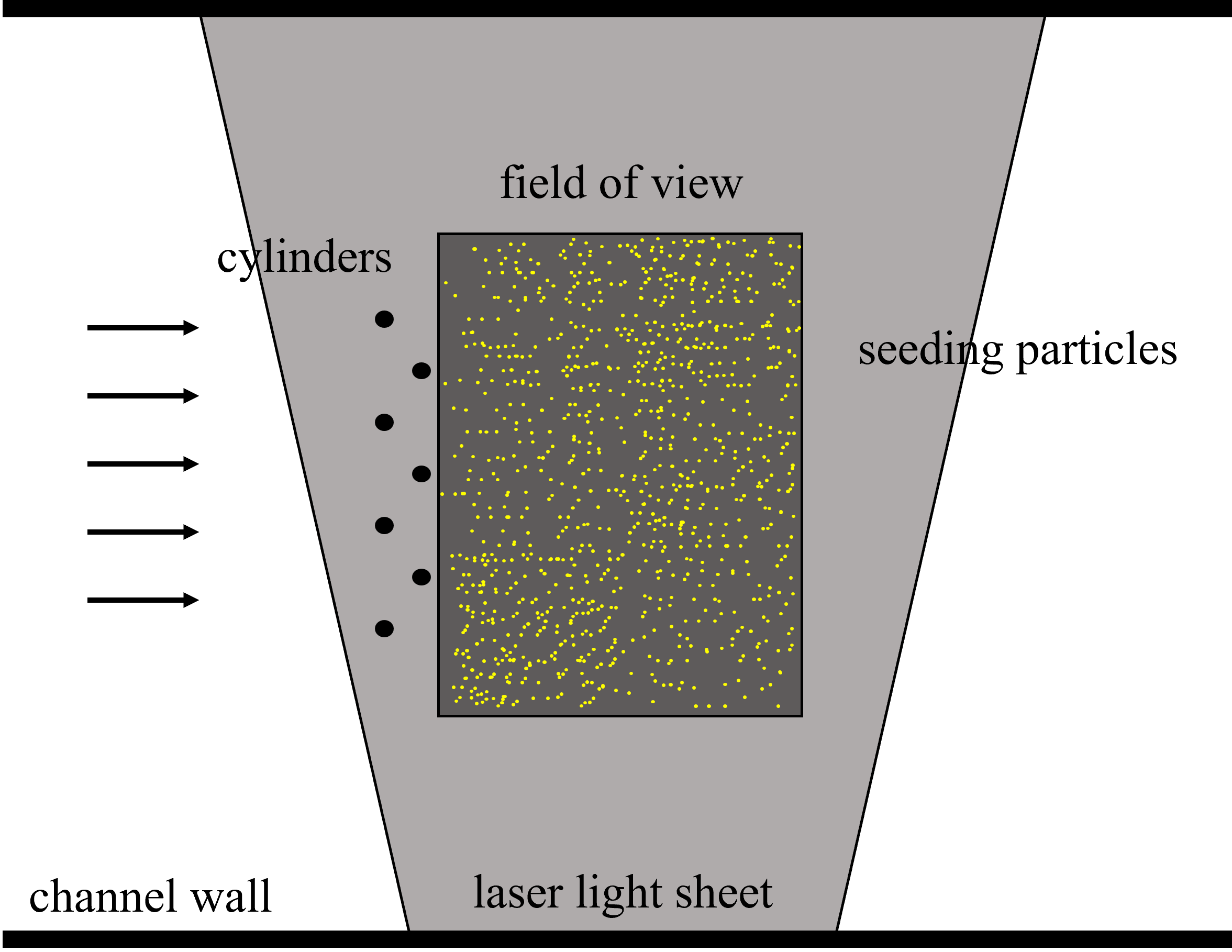} 
        \caption{A schematic sketch of the experiments for flow behind cylinder arrays.} 
        \label{ExpSetup}
    \end{subfigure}

    \vspace{0.5cm} 

    \begin{subfigure}[b]{\textwidth}
        \centering
        \includegraphics[width=0.85\textwidth, trim=1cm 2cm 1cm 1.5cm, clip]{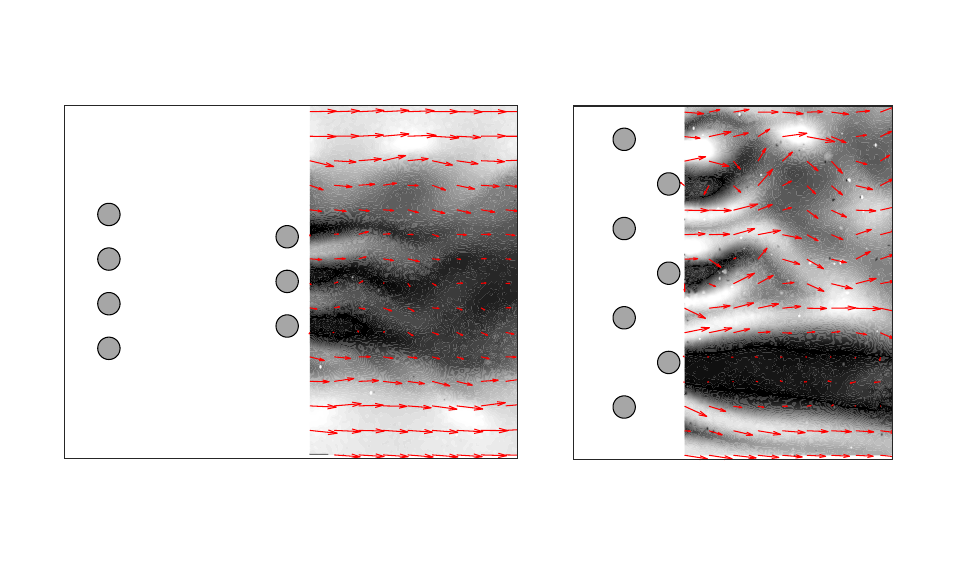}
        \caption{Flow behind array of seven cylinders in two different configuration of 2V8H, and 4V2H from left to right, respectively. For clarity, only one out of every 256 vectors is displayed.}
        \label{AbsV}
    \end{subfigure}

    \caption{Numerical domain (a) and field of view for the investigated test cases numerical and experimental test case, respectively. Flow fields for the cylinder arrays (c).}
\end{figure}

\begin{figure}[t]
    \centering
    \includegraphics[width=\textwidth]{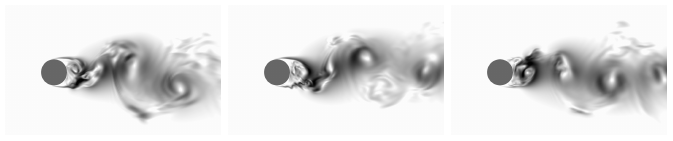}
    \caption{Examples from the dataset of LES simulated flow around a cylinder.}
    \label{fig:dataExample}
\end{figure}
First, we investigate the flow around a circular cylinder at a Reynolds number of 3900, a classical test case that exhibits complex fluid dynamic behavior due to the coexistence of laminar and turbulent regimes. This flow is characterized by vortex shedding, forming a Kármán vortex street in the cylinder’s wake \cite{von1911mechanismus,Parnaudeau:2008,Norberg:1994,Ong:1996,Beaudan:1995,Kravchenko:2000}. The vortex street consists of a coherent system of vortices whose rotational axes are aligned with the cylinder axis. This case has been extensively studied both experimentally and numerically, making it suitable for validation and data-driven modeling.

The viscous fluid is governed by the incompressible Navier-Stokes equations (NSE), which originate from the conservation of mass and momentum. Under incompressible and isothermal conditions, the mass conservation reduces to the divergence-free condition of the velocity field, and the energy equation can be neglected. Let $\mathbf{u} = (u, v, w)$ denote the velocity vector. The NSE in Cartesian coordinates are then given by
\begin{equation}
\nabla \cdot \mathbf{u} = 0, \quad 
\frac{\partial \mathbf{u}}{\partial t} + (\mathbf{u} \cdot \nabla)\mathbf{u} = -\frac{1}{\rho} \nabla p + \nu \nabla^2 \mathbf{u},
\end{equation}
where $\rho$ is the fluid density, $\nu$ the kinematic viscosity, and $p$ the pressure field. While ergodicity in turbulent flows is not guaranteed in general, it can reasonably be assumed for this setup \cite{frisch1995turbulence}.

Direct numerical simulation of the NSE for this test case is computationally infeasible due to the wide range of scales in turbulence. Consequently, Large-Eddy Simulation (LES) is employed using the commercial solver ANSYS Fluent \cite{winhart2021large}. LES resolves the large, energy-containing eddies explicitly, while the effects of smaller subgrid-scale motions are modeled \cite{sagaut2005large,Hirsch:2007}. This is achieved through spatial filtering, which decomposes a flow variable $\varphi_t(x)$ into a filtered (resolved) component $\overline{\varphi}_t(x)$ and a subgrid-scale component $\dot{\varphi}_t(x)$,
\begin{equation}
\varphi_t(x) = \overline{\varphi}_t(x) + \dot{\varphi}_t(x),
\end{equation}
with the filter defined implicitly by the computational grid. The filtered NSE then take the form
\begin{equation}
\nabla \cdot \overline{\mathbf{u}} = 0, \quad
\frac{\partial \overline{\mathbf{u}}}{\partial t} + (\overline{\mathbf{u}} \cdot \nabla) \overline{\mathbf{u}} = -\frac{1}{\overline \rho} \nabla \overline{p} + \frac{1}{\mathrm{Re}} \nabla^2 \overline{\mathbf{u}},
\end{equation}
where $\mathrm{Re} = U_\infty D / \nu$ is the Reynolds number, $U_\infty$ the freestream velocity, and $D$ the cylinder diameter. The LES approach is particularly suitable because the large vortex structures carry most of the turbulent kinetic energy (TKE), while smaller structures can be assumed to be isotropic and homogeneous in the bulk flow according to Kolmogorov’s local isotropy assumption \cite{Ferziger:2008,Kolmogorov:1991}.

The computational setup consists of a grid with approximately 15 million cells and a time step chosen to maintain a CFL number on the order of unity. A schematic representation of the numerical domain is shown in Fig. \ref{fig:numericalsetupA}. The simulation was performed for 25,000 time steps after the initial transients, corresponding to a physical time of roughly 1.45 seconds. Time integration is carried out using a non-iterative advancement scheme combined with a fractional-step method for pressure-velocity coupling. Advective fluxes are treated with a bounded central scheme to minimize numerical dissipation and prevent unphysical damping of small-scale turbulence \cite{winhart2021large}. Snapshots of the flow field are sampled at a frequency of 68.9 Hz, with sufficient temporal separation to reduce correlation between frames.

For data-driven modeling, the transient LES velocity fields are post-processed to generate two-dimensional grayscale images representing the fluctuating velocity magnitude (see \cref{fig:dataExample}). Let $V(\xi,t) = (V_x, V_y, V_z)$ denote the local velocity vector. The grayscale intensity at a spatial location $\xi$ corresponds to the absolute deviation of the local velocity magnitude
\( c(\xi,t) = \sqrt{V_x^2 + V_y^2 + V_z^2} \)
from its time average
\(\overline{c}(\xi) = \frac{1}{T}\int_0^T c(\xi,t) \, dt,\)
i.e.,
\(c'(\xi,t) = |c(\xi,t) - \overline{c}(\xi)|.\)

The fluctuation magnitude is then linearly scaled to an 8-bit intensity range $[0,255]$, such that darker regions correspond to nearly steady flow with weak fluctuations, whereas brighter regions mark locations of strong unsteady motion in the wake, providing a physically interpretable field of velocity fluctuation strength.
This projection mapping preserves ergodicity in the reduced state space. The dataset contains $100,000$ images at $1,000 \times 600$ resolution and is available at \cite{winhart2024data}. 
The transient velocity fields were mapped onto structured evaluation grids, which define the sampling points for the images. Given the numerical domain length of $20D$ (see \cref{fig:numericalsetupA}) and an image resolution of $1,000$ columns in the streamwise direction, one image column corresponds to a physical distance of $0.02D$. Because the evaluation grids were adapted to the required spatial resolution, the image sampling does not introduce scales smaller than those resolved by LES, ensuring that the image resolution is consistent with the LES filter scale and reflects the number of physically resolved degrees of freedom \cite{winhart2021large}. 

\subsubsection{Flow behind arrays of seven cylinders}

The flow behind two arrays of seven cylinders each with different configurations was chosen, as although the Reynolds number based on a single cylinder is only $Re \sim 100$, it is already unsteady. The arrangement of multiple cylinders in the flow results in multiple individual wake flows that interact with each other. The vortex shedding in the individual wake flows does not necessarily show the same frequency, resulting in very broad and rich dynamics depending on the vertical and horizontal distances between the cylinders. If they are very close, they act as a single bluff body (where one may redefine the Reynolds number). On the other end, if they are far apart, the resulting flow can be characterized as co-shedding with specific interactions between the individual wakes that might differ in frequency and phase depending further on the distances. 

The most complex case is the transitional state where the distances between the cylinders are large enough to prevent the array from acting as a single bluff body and still not large enough to let each cylinder develop its individual wake. As a result, irregular and asymmetric flow patterns arise behind the array, with varying vortex shedding frequencies over space and time. For further details on the flow physics, the interested reader is referred to~\cite{SharifiGhazijahani2023}. 
For this reason, the datasets 2V8H and 4V2H were chosen for further processing. The cylinders were arranged in two columns: four in the front and three in the rear, where the horizontal distance between these two columns (H) and the vertical distance between cylinders inside each column (V) are given in cylinder diameters.    

A schematic of the experimental setup is presented in Fig.\ref{ExpSetup}. All rigid cylinders had a diameter of 1 mm. The cross-section of the water channel was $50 \times 50 $mm$^2$. For the particle image velocimetry, the flow was seeded with polyamide particles of $5 \mu$m in diameter and a density of 1.03 g/cm$^3$ that followed the flow faithfully. A vertical laser light sheet generated by a cw-laser (Laserworld Green-200 532) illuminates the 
mid-plane of the channel. Images of the distributed particles were observed with a high-speed camera (HS 4M by LaVision GmbH) at a frame rate of 200 Hz perpendicular to the laser sheet from outside of the channel. The displacement between particle image distributions in successive snapshots was determined via cross-correlation. To convert image coordinates to physical space, a calibration plate with known marker spacing was applied. The flow velocity was set to $u_{\infty} = 133$ mm/s, which gives a Reynolds number of $ Re = D u_{\infty} / \nu = 100$, based on the diameter of the single
cylinder ($D$) and the kinematic viscosity of water $\nu$. Within a measurement duration of 15 s, a total of 90 to 270 vortex shedding events were recorded for the lowest and highest frequencies observed for the two different cases. This represents a certain statistical confidence, which is necessary for a successful training of AI methods~\cite{Teutsch2025_aplml}.
Figure \ref{AbsV} shows the absolute velocity field along the velocity vectors obtained from particle image velocimetry measurements.
In total, the dataset contains $3,000$ images at $138 \times 231$ resolution and is available at \cite{sharifi_ghazijahani_2025_16794036}. 

\subsection{Computational Cost}

At this stage, it is worth briefly addressing the computational effort involved in the datasets presented in this paper, as this is a key factor determining the feasibility of high-resolution LES and experimental PIV measurements.

\subsubsection{Flow around a cylinder}
The LES was run on a partition of the High-Performance Computing (HPC) cluster of the Chair of Thermal Turbomachines and Aero Engines with Intel Xeon ``Skylake'' gold 6132 CPUs of 2.6 GHz and 96 GB RAM.
For the simulation, $20$ nodes with $28$ cores each had to be allocated and the computation time was about $20$ days, which corresponds to $1,440$ core weeks.

\subsubsection{Flow behind array of seven cylinders}

With the above-mentioned experiment, the data were collected at a 200 Hz frame rate, which converts to a measurement time of 15 s. However, the subsequent image processing took several hours on a desktop PC equipped with an Intel(R) Xeon(R) W-3223 CPU @ 3.50 GHz. In general, the advantage of experiments is that the processing time is not dependent on the Reynolds number, and significantly larger Reynolds numbers (i.e., higher levels of turbulence) can be characterized in a shorter time as compared to simulations. Major difficulties are the precise control of the boundary conditions and the fact that any measurement comes with an uncertainty that the AI methods need to deal with. 

\section{Experiments}
\label{sec:experiments}
\subsection{Implementation Details}

\begin{table}[t]
\begin{center}
\resizebox{0.8\textwidth}{!}{%
    {\tabulinesep=1.2mm
        \begin{tabu}{|l|l|l|l|}
            \hline
            Model & VAE & DCGAN & DDPM (with attention)\\
            \hline
            Total Parameters & $3,939,085$ & $212,263,362$ & $135,764,353$\\
            \hline
            Layer Sizes & $\{2^n \, | \,\,  3 \leq n \leq 9, n \in \mathbb{N}\}$
            & $\{2^n \, | \,\,  4 \leq n \leq 21, n \in \mathbb{N}\}$
            & $\{2^n \, | \,\,  6 \leq n \leq 10, n \in \mathbb{N}\}$\\ 
            \hline
            Latent Dim Size & $128$ & $100$ & $512 \times 512$ \\
            \hline
            Sampling Timesteps & N/A & N/A & $1,000$\\
            \hline
            Effective Batch Size & $512 \times 1$ & $20 \times 1$ & $11 \times 3$\\ 
            \hline
            Initial Learning Rate  & $1 \times 10^{-3}$ 
            & $2 \times 10^{-4}$ 
            & $1 \times 10^{-5}$ \\ 
            \hline
            Epochs & $150$ & $2,000$ & $200$\\
            \hline
            Optimizer & Adam & Adam & Adam \\ 
            \hline
            Use EMA & No & No & Yes\\ 
            \hline
            Training Images 
            & \begin{tabular}{@{}c@{}}100,000 (A) \\ ~~~3,000 (B)\end{tabular} 
            & \begin{tabular}{@{}c@{}}5,000 (A) \\ 3,000 (B)\end{tabular} 
            & \begin{tabular}{@{}c@{}}100,000 (A) \\ ~~~3,000 (B)\end{tabular} \\
        \cline{2-4}
            \hline
            Image Resolution 
           & \begin{tabular}{@{}c@{}}512×512 (A) \\ 256×256 (B)\end{tabular}
           & \begin{tabular}{@{}c@{}}512×512 (A) \\ 128×128 (B)\end{tabular}
           & \begin{tabular}{@{}c@{}}512×512 (A) \\ 256×256 (B)\end{tabular} \\
            \hline
        \end{tabu}}}
    \end{center}
    \caption{Summary of the parameter settings of the generative models VAE, DCGAN, and DDPM, investigated using two datasets: (A) flow around a cylinder and (B) flow behind array of seven cylinders. The effective batch size is the actual batch size multiplied by the gradient accumulation.}
    \label{tab:implementdet}
\end{table}
In this work, we investigate VAE and DDPM as alternative generative learning approaches for turbulence modeling and compare them to the DCGAN turbulent flow generator developed in our previous work \cite{drygala2022generative}, where the implementation details can also be found. 
For the training of the VAE and the DDPM, the architecture proposed by \cite{kingma2013} and \cite{ho2020denoising}, respectively, was adopted. In particular, the VAE consists of an encoder and decoder of similar structure with seven hidden layers containing $\Sigma_{n=3}^9{2^n}$ neurons.
We used a more advanced DDPM model with attention, which means it is combined with a transformer \cite{vaswani2023attentionneed} and shadowed by an Exponential Moving Average (EMA) model \cite{Karras2023AnalyzingAI}.
The generative models require square images as input for training. The images were resized to a resolution of $m \times m, m\in \mathbb{N}^+$ pixels. For the flow-around-a-cylinder dataset, an additional pre-processing step removed 150 pixels in front of the static wake where the flow is laminar and no turbulence is present. This resizing did not alter the LES- or PIV-resolved scales and represented a purely computational adjustment to meet the input requirements of the neural networks. Only the first $5,000$ images of the cylinder dataset were considered for GAN training.
Finally, all generative networks were built and trained with PyTorch \cite{pytorch2019}.
A summary and comparison of the relevant training parameters, such as learning rate or effective batch size, of the three generative models investigated can be found in \cref{tab:implementdet}.
At inference time, the input for all three generative models is a latent vector drawn from a simple prior distribution - here, the standard Gaussian - with the same dimensionality as during training (see \cref{tab:implementdet}). This latent vector is mapped by the trained generator to a turbulent flow field sample. In GANs and VAEs through a single forward pass, and in DDPMs through an iterative denoising process. Because training was performed without a temporal component, the generative models do not capture time dependencies but instead produce temporally independent state snapshots of turbulent flows.

\subsection{Physics-based Evaluation Metrics}
\begin{figure}
\begin{subfigure}[b]{.49\textwidth}
  \centering
  \includegraphics[width=1.2\textwidth]{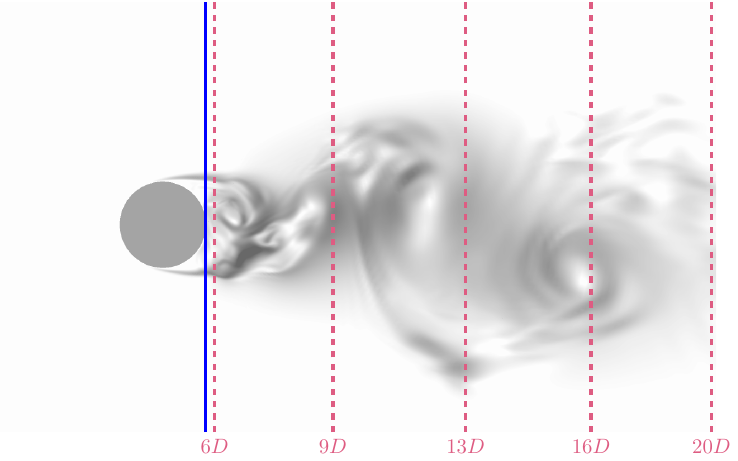}
  \caption{Flow around a cylinder}
  \label{fig:300px_and_500px_lines}
\end{subfigure}%
\hspace{1cm}
\begin{subfigure}[b]{.49\textwidth}
  \centering
  \includegraphics[width=0.5\textwidth]{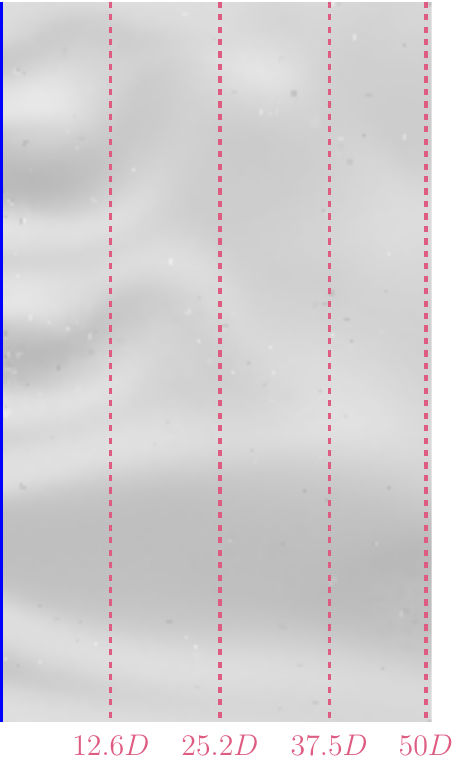}
  \caption{Flow behind array of seven cylinders.}
  \label{fig:35px_and_70px_lines}
\end{subfigure}
\caption{Evaluation regions for both datasets, where the vertical blue line marks the start of the investigated wake behind the cylinders. 
For the flow around a single cylinder (a), the red lines indicate regions located at $6D$ (small near-wake region), $9D$, $13D$, $16D$, and $20D$, corresponding to approximately 25\%, 50\%, 75\%, and 100\% of the turbulent wake. 
For the flow behind an array of seven cylinders (b), these regions are located at $12.6D$, $25.6D$, $37.5D$, and $50D$. 
The specific locations within the numerical domains are computed separately for each dataset.
}
\label{fig:eval_regions}
\end{figure}
To assess the performance of the generative models, we compute two physically interpretable metrics - the mean local velocity fluctuation magnitude and its variance - at each spatial location $\xi$, across the turbulent wake. 

For both datasets, the LES flow around a single cylinder and the PIV flow behind an array of seven cylinders, the evaluation is performed by progressively increasing the portion of the downstream wake considered. Specifically, we analyze 25\%, 50\%, 75\%, and 100\% of the turbulent wake, allowing us to assess model performance across regions of varying turbulence intensity (see \cref{fig:eval_regions}. In addition, for the LES dataset, we include a very small near-wake region located at $6D$ within the numerical domain, directly behind the cylinder, where turbulence intensity is highest and coherent vortex shedding dominates the flow dynamics. 

For the PIV dataset, the velocity components in the streamwise and transverse directions ($u$ and $v$) are combined into the local velocity magnitude, $\sqrt{u^2 + v^2}$, to provide a scalar measure of instantaneous velocity fluctuations, analogous to the quantity computed for the LES data. 

The evaluation is based on 5{,}000 LES snapshots and 5{,}000 images generated by each of the VAE, DCGAN, and DDPM models for the single-cylinder case, and on 3{,}000 PIV snapshots and 3{,}000 synthesized images for each test case of the cylinder-array flow. These selections ensure that both near- and far-wake turbulence are captured, enabling a consistent and physically meaningful comparison between the model-generated and reference flows across increasing wake regions.

A detailed discussion of the physical relevance and justification for these metrics in cylinder wake flows is provided in \cite{drygala2022generative}.

It is worth noting that no known method is capable of reproducing exactly the turbulent flow for a given set of initial conditions, if such a thing even exists, and instead, produce stochastic realizations of what the flow may look like. All of these realizations contain some sort of error, and as such, the models trained on this flawed data will never be able to improve on the error that the data contains. Thus, we do not claim to be able to approximate the flow itself, but rather, we are attempting to approximate the LES and PIV data.

\subsection{Results Discussion}
Below, we discuss the results of the conducted experiments. For improved readability, only the evaluation results of the near-wake regions are included in the main part of the discussion. Additional evaluation plots are provided in \cref{sec:appendix_eval}.

\subsubsection{Flow around a cylinder}
\begin{figure}[!b]
    \centering
    \includegraphics[width=\textwidth]{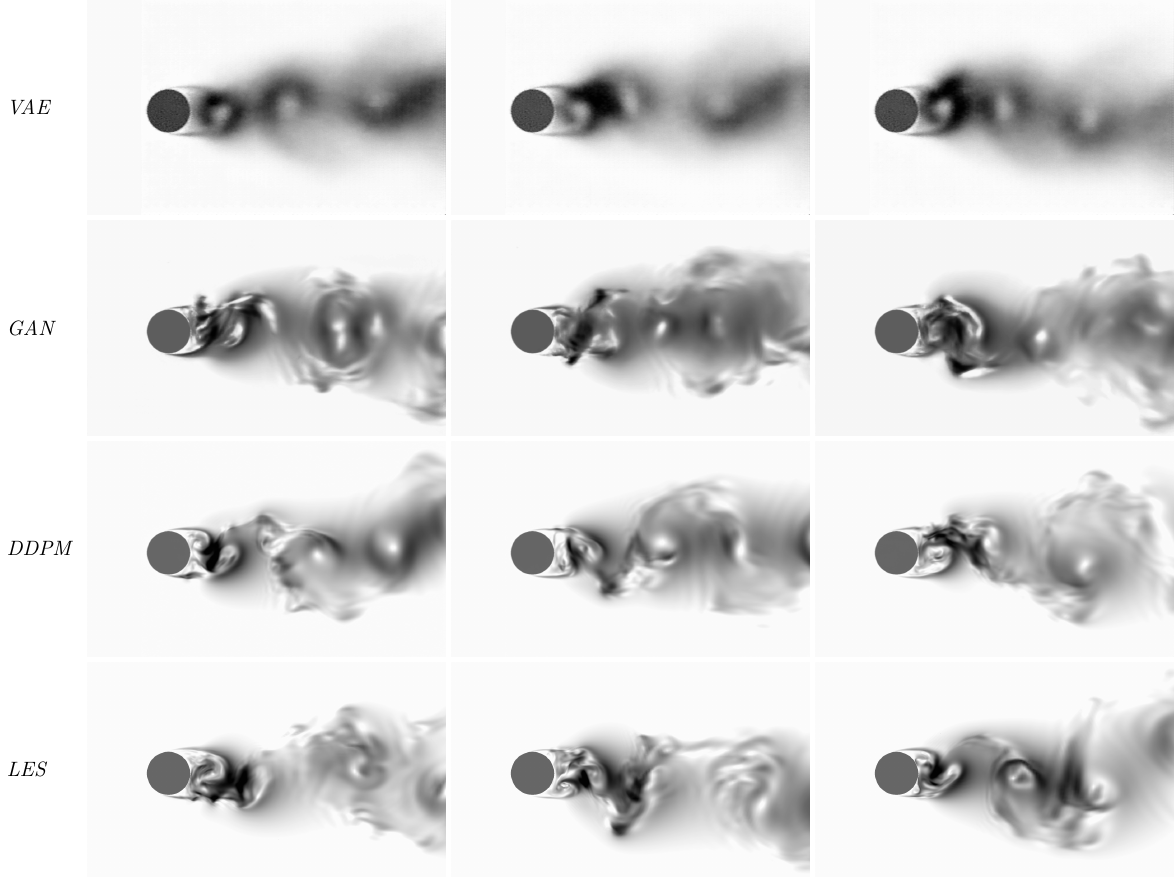}
    \caption{Examples of, from top to bottom, the fully trained VAE, DCGAN, and DDPM, with the LES dataset at the bottom for comparison.}
    \label{fig:examples_grid}
\end{figure}

\Cref{fig:examples_grid} shows example outputs generated by the three models, highlighting differences in their ability to reproduce turbulent flow structures. The VAE model is the weakest performer, as it reproduces the large-scale wake location but underestimates the amplitude of local velocity fluctuations and underrepresents small-scale vortical structures.

\begin{figure}[!t]
\centering
\begin{subfigure}[b]{.5\textwidth}
  \centering
  \includegraphics[width=\linewidth]{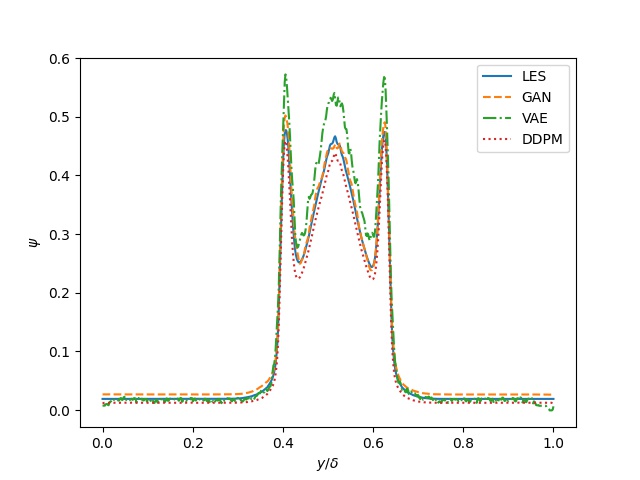}
  \caption{Mean local velocity fluctuations}
  \label{fig:mean_v_300}
\end{subfigure}%
\begin{subfigure}[b]{.5\textwidth}
  \centering
  \includegraphics[width=\linewidth]{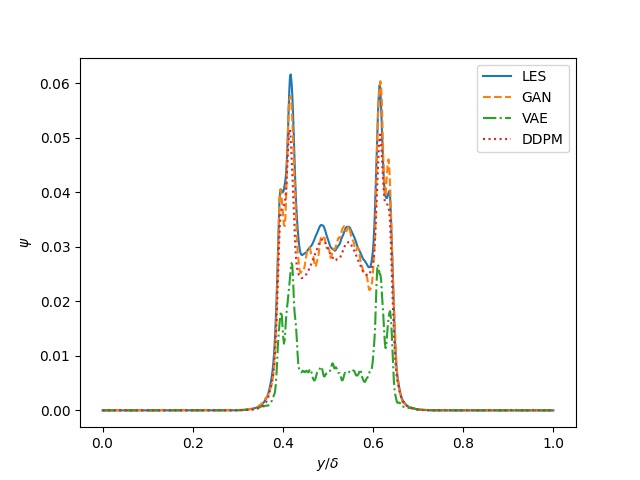}
  \caption{Variance of local velocity fluctuations}
  \label{fig:var_v_300}
\end{subfigure}
\caption{Comparison of (a) the mean local velocity fluctuation magnitude and (b) the variance of the local velocity fluctuations, evaluated over the near-wake region extending to $6D$ within the numerical domain behind the cylinder (see \cref{fig:300px_and_500px_lines}). Each value represents the deviation of the local flow velocity from the background or mean flow. All datasets were normalized prior to evaluation.}
\label{fig:v_300}
\end{figure}

\begin{figure}[!t]
\centering
\begin{subfigure}[b]{.5\textwidth}
  \centering
  \includegraphics[width=\linewidth]{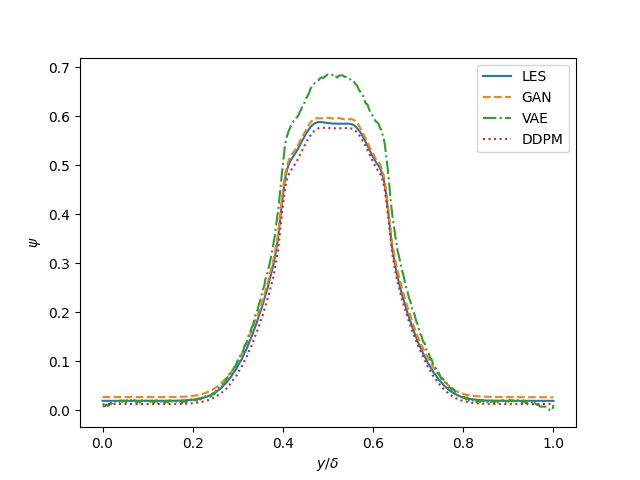}
  \caption{Mean local velocity fluctuations}
  \label{fig:mean_v_500}
\end{subfigure}%
\begin{subfigure}[b]{.5\textwidth}
  \centering
  \includegraphics[width=\linewidth]{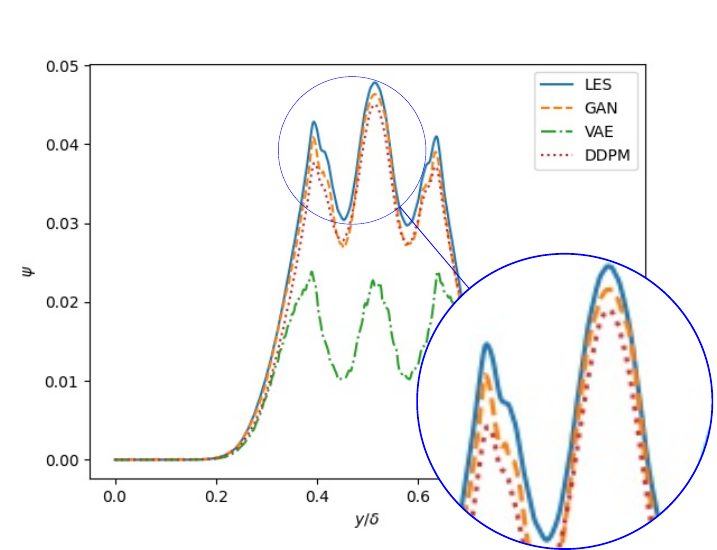}
  \caption{Variance of local velocity fluctuations}
  \label{fig:var_v_500}
\end{subfigure}
\caption{Comparison of (a) the mean local velocity fluctuation magnitude and (b) the variance of the local velocity fluctuations, with a zoomed view, evaluated over the wake region comprising 25\% of the wake flow directly behind the cylinder, extending to $9D$ within the numerical domain (see \cref{fig:300px_and_500px_lines}). Each value represents the deviation of the local flow velocity from the background or mean flow. All datasets were normalized prior to evaluation.}
\label{fig:v_500}
\end{figure}

Differences between the LES, DCGAN, and DDPM outputs are less apparent by visual inspection alone. Therefore, we turn to the physics-based evaluation metrics shown in \cref{fig:v_300} and \cref{fig:v_500}, which display the mean local velocity fluctuation magnitude and its variance along the $y$-axis across the downstream wake, evaluated at a small region directly behind the cylinder ending at position $6D$ in the numerical domain, and a near-wake region capturing up to 25\% of the downstream wake ending at position $9D$. These figures confirm the earlier assessment: the VAE systematically overestimates the mean local fluctuation magnitude while underpredicting the variance, indicating that it fails to capture both the average turbulent intensity and the statistical fluctuations associated with coherent and small-scale turbulence.


A noticeable issue with the DCGAN outputs is the presence of artificially elevated velocity fluctuations in regions of the flow that should remain laminar, corresponding to areas outside the primary wake structures. These deviations are manifested as slight offsets in the tails of the mean fluctuation distributions. The background flow in these regions is physically nearly steady, so any nonzero fluctuations reflect artifacts introduced by the generative model rather than actual turbulence. This behavior likely arises from the DCGAN’s emphasis on generating prominent vortical structures, sometimes at the expense of accurately representing uniform, laminar regions, since the discriminator may prioritize regions with more variability as indicators of realism. Additionally, the generative process inherently introduces small-scale noise, which becomes more noticeable in these near-laminar areas. The DDPM outputs show a similar effect, but the magnitude of the artificial fluctuations in the laminar background is smaller.

Both DCGAN and DDPM better reproduce the mean and variance of the LES flow, though neither is perfect. The DCGAN captures the chaotic nature of the turbulence more effectively, particularly in the near wake, as seen in its ability to reproduce peaks and inflection points in the variance distributions (\cref{fig:var_v_500}). The DDPM underpredicts variance for high-intensity turbulent regions, indicating a reduced representation of extreme turbulent events.

Considering the near-wake regions, the DCGAN slightly outperforms the DDPM in reproducing areas of high-intensity velocity fluctuations. Examining the downstream evolution of the variance, particularly in the far wake beyond $16D$ (see \cref{fig:eval_kvs_large_mean} and \cref{fig:eval_kvs_large_var}), both models tend to underestimate the highest-intensity fluctuations, with this underestimation becoming most pronounced in the final 25\% of the far-wake region. Interestingly, while the DCGAN captures the overall shape of the LES variance curve, it underestimates high-intensity turbulence in the far wake more than the DDPM. This suggests that the DCGAN better reproduces near-wake dynamics, whereas the DDPM retains somewhat higher fidelity in representing intense turbulent fluctuations further downstream.

\subsubsection{Flow behind array of seven cylinders}

\begin{figure}[!b]
    \centering
    \includegraphics[width=0.55\textwidth]{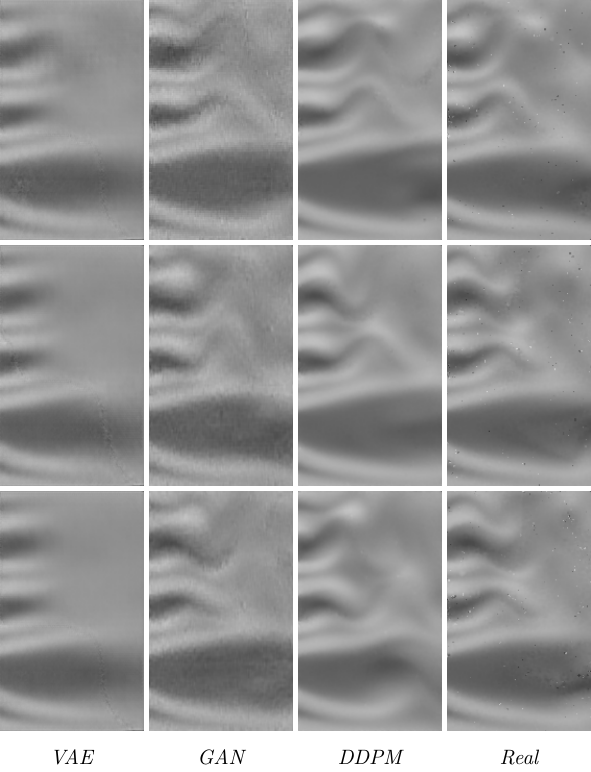}
    \caption{Examples from the 4V2H test case showing, from left to right, the fully trained VAE, DCGAN, and DDPM, with the PIV dataset on the right for comparison.}
    \label{fig:examples_grid_4V2H}
\end{figure}
\begin{figure}[!ht]
\centering
\begin{subfigure}{.5\textwidth}
  \centering
  \includegraphics[width=0.9\linewidth]{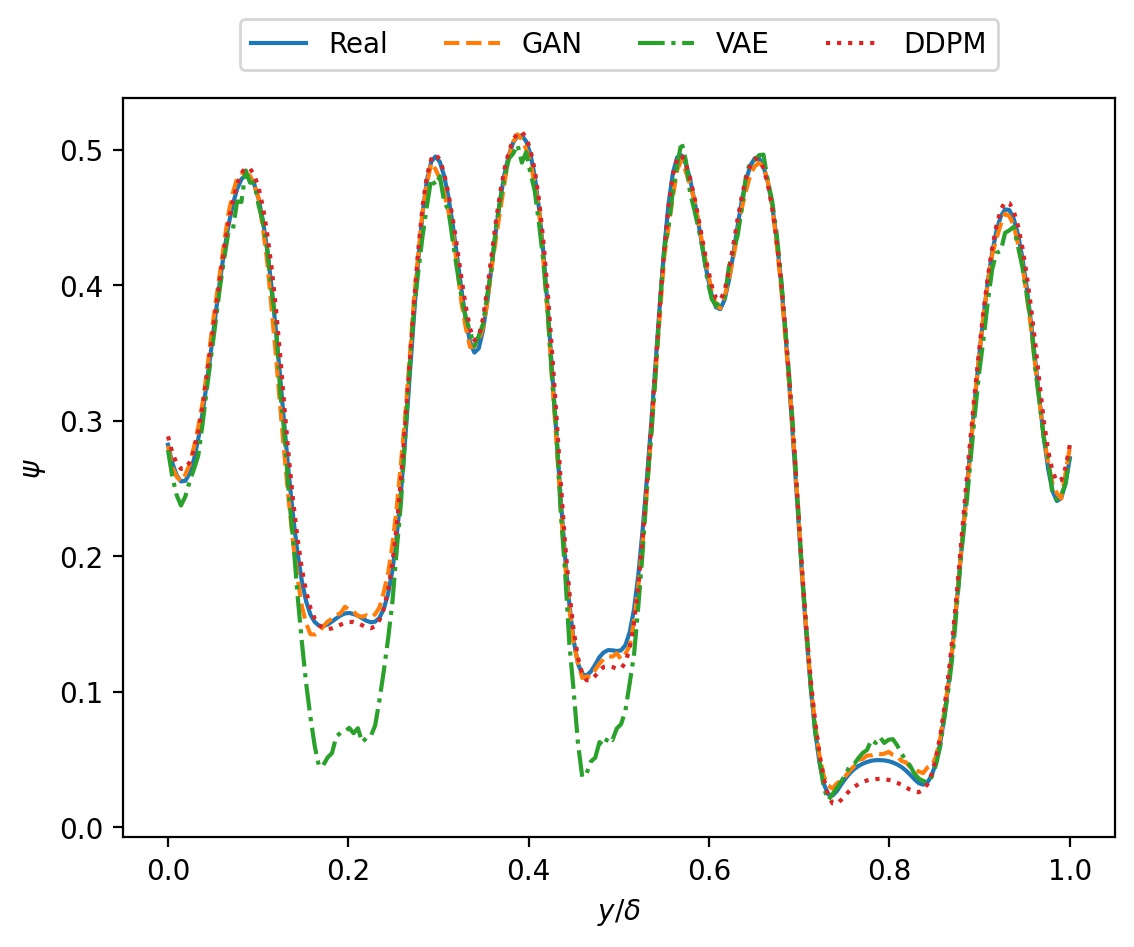}
  \caption{Mean local velocity fluctuations}
  \label{fig:mean_4V2H_small}
\end{subfigure}%
\begin{subfigure}{.5\textwidth}
  \centering
  \includegraphics[width=0.9\linewidth]{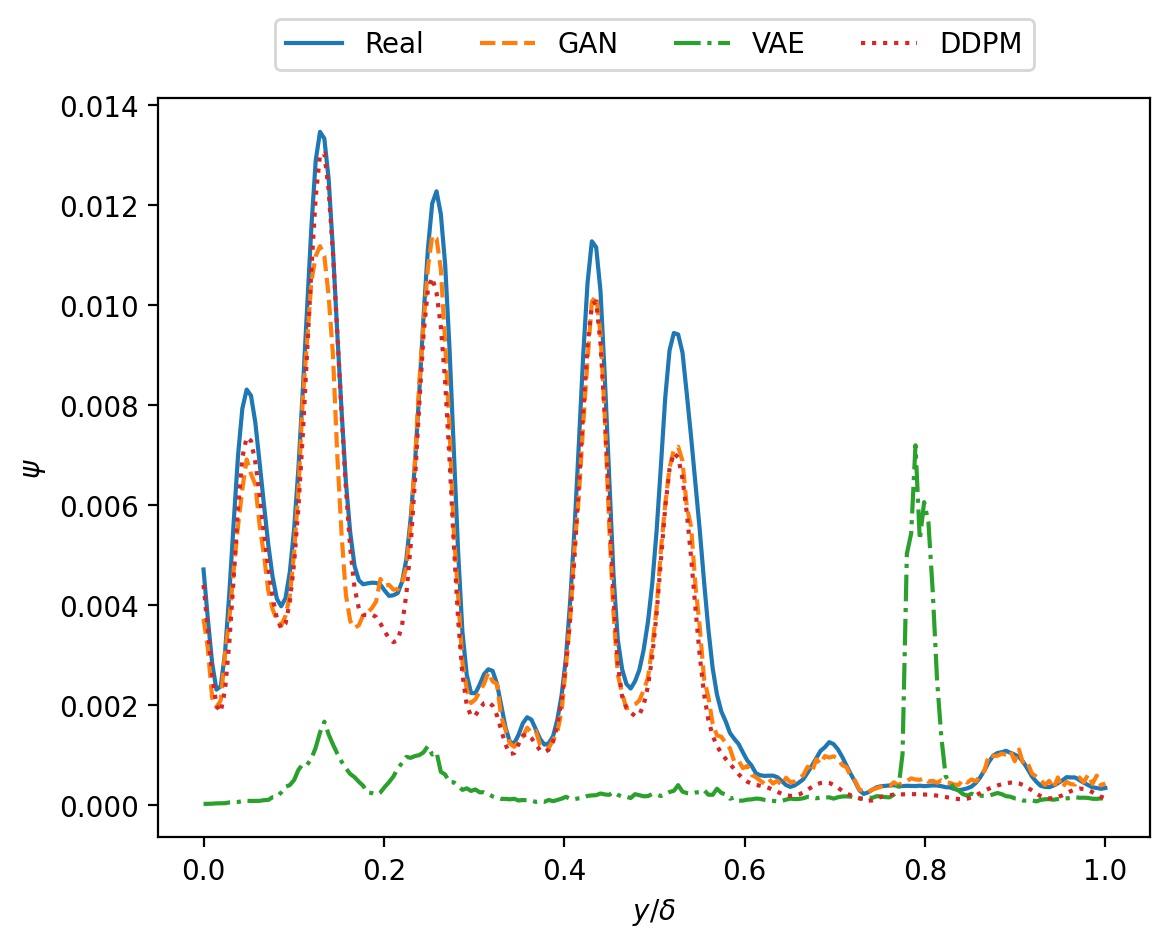}
  \caption{Variance of local velocity fluctuations}
  \label{fig:var4V2H_small}
\end{subfigure}
\caption{Comparison of (a) the mean local velocity fluctuation magnitude and (b) the variance of the local velocity fluctuations, evaluated over the wake region comprising 25\% of the wake flow behind the array of seven cylinders in the 4V2H test case, extending to \(12.6D\) within the numerical domain (see \cref{fig:35px_and_70px_lines}). Each value represents the deviation of the local flow velocity from the background (mean) flow. All datasets were normalized prior to evaluation.}
\label{fig:eval_4V2H_small}
\end{figure}

For the flow behind the array of seven cylinders, two test cases, 4V2H and 2V8H, were investigated. Starting with the 4V2H configuration, the results reveal behavior consistent with observations from LES data. In \cref{fig:examples_grid_4V2H}, a visual inspection of the streamwise velocity component \(u\) shows that the VAE captures the large-scale structures of the downstream wake but struggles to resolve finer vortex details. In contrast, both the DCGAN and the DDPM are able to reproduce small-scale vortex features more accurately. It can be observed, however, that the DCGAN-generated wakes contain noticeable noise, which can be traced back to the inherent measurement noise in the experimental PIV data. During training, the DCGAN appears to have learned that these small-scale fluctuations are characteristic of realistic samples and therefore reproduces them in its generated outputs.

The evaluation metrics support these qualitative impressions. Considering the near-wake region (see \cref{fig:eval_4V2H_small}), comprising the first 25\% of the total wake length, the VAE tends to locally underestimate the mean local velocity fluctuation magnitude and fails to fully capture the statistical variability associated with coherent and small-scale turbulence. For the 4V2H case, both DCGAN and DDPM closely match the statistical properties of the measured PIV data. The DCGAN performs particularly well in the lower wake region, where the flow exhibits a large, steady vortex with weak fluctuations. Nevertheless, in the second high-intensity turbulence region, the DDPM slightly outperforms the DCGAN. Interestingly, in the far-wake region, comprising the final 25\% of the wake, all generative surrogates tend to underestimate again the highest-intensity fluctuations, with this underestimation becoming most pronounced toward the end of the wake (see \cref{fig:eval_4V2H_large_var}).

\begin{figure}[!h]
    \centering
    \includegraphics[width=0.53\textwidth]{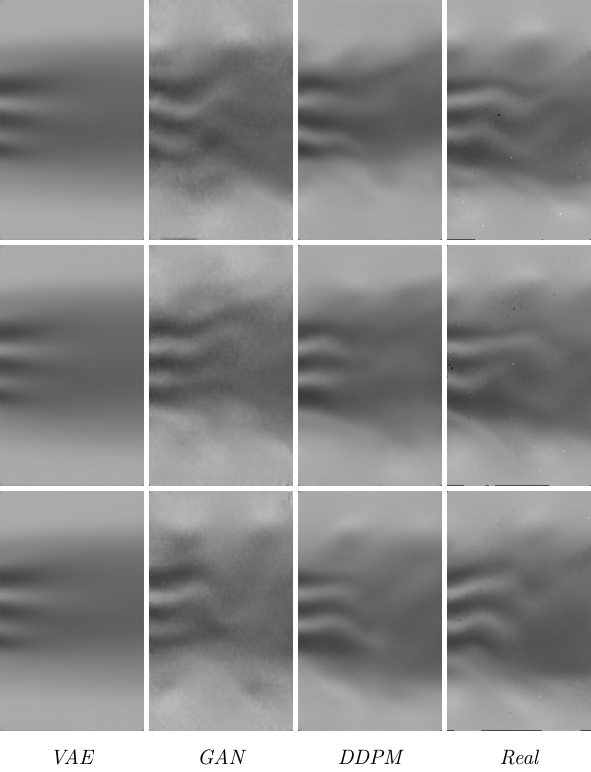}
    \caption{Examples from the 2V8H test case showing, from left to right, the fully trained VAE, DCGAN, and DDPM, with the PIV dataset on the right for comparison.}
    \label{fig:examples_grid_2V8H}
\end{figure}

\begin{figure}[!h]
\centering
\begin{subfigure}{.5\textwidth}
  \centering
  \includegraphics[width=0.9\linewidth]{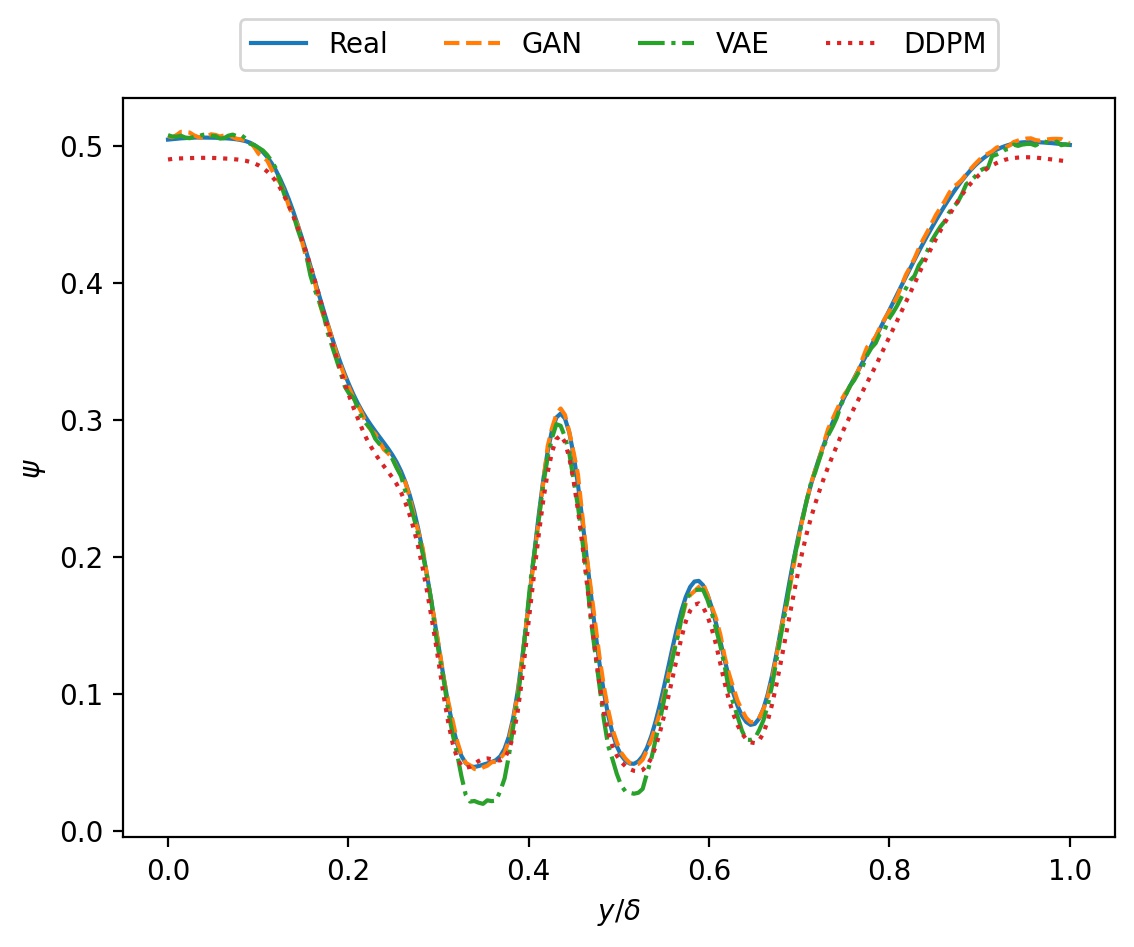}
  \caption{Mean local velocity fluctuations}
  \label{fig:mean_2V8H_small}
\end{subfigure}%
\begin{subfigure}{.5\textwidth}
  \centering
  \includegraphics[width=0.9\linewidth]{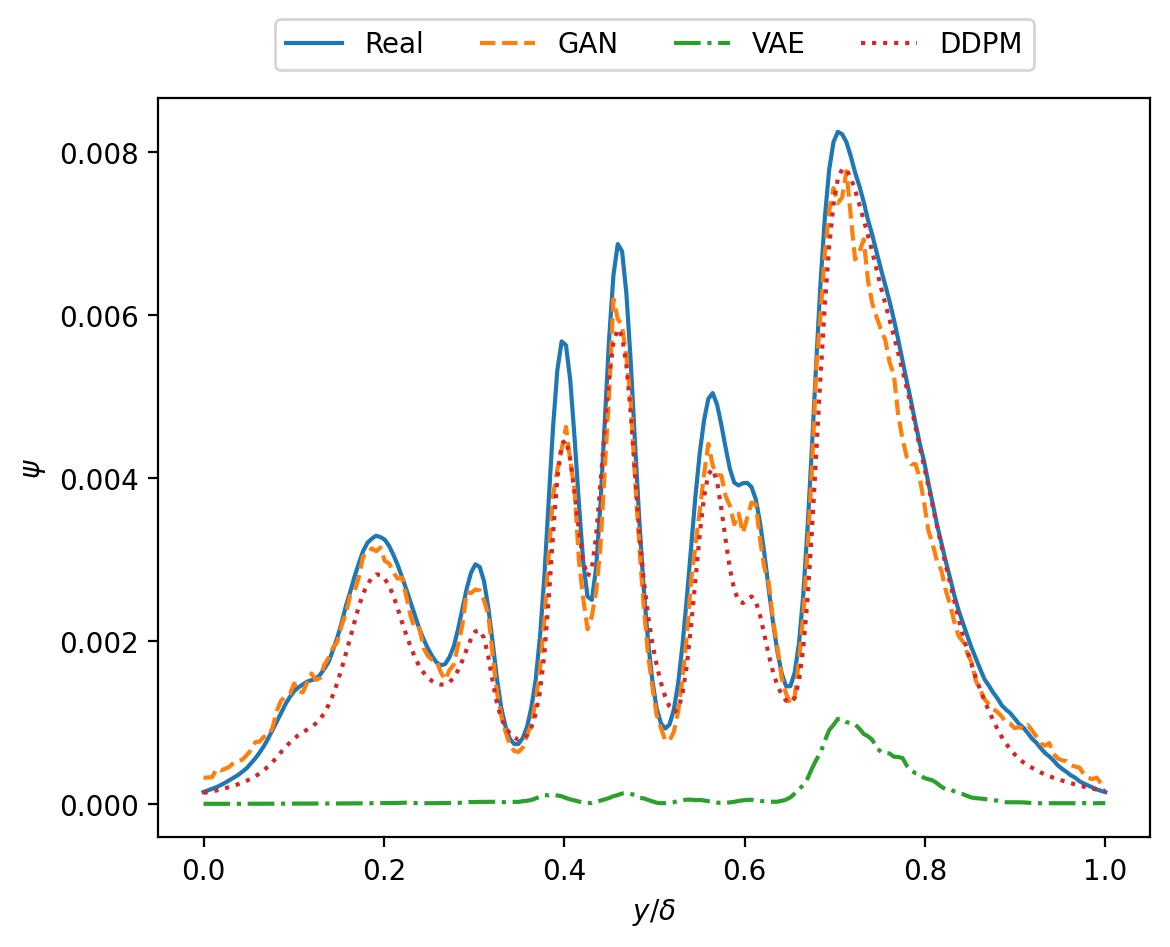}
  \caption{Variance of local velocity fluctuations}
  \label{fig:var2V8H_small}
\end{subfigure}
\caption{Comparison of (a) the mean local velocity fluctuation magnitude and (b) the variance of the local velocity fluctuations, evaluated over the wake region comprising 25\% of the wake flow behind the array of seven cylinders in the 2V8H test case, extending to \(12.6D\) within the numerical domain (see \cref{fig:35px_and_70px_lines}). Each value represents the deviation of the local flow velocity from the background (mean) flow. All datasets were normalized prior to evaluation.
}
\label{fig:eval_2V8H_small}
\end{figure}

A similar trend is observed for the second test case, 2V8H, confirming the general validity of these findings (see \cref{fig:examples_grid_2V8H}). As described in \cref{fig:eval_2V8H_small}, the VAE captures the mean local fluctuation magnitude more accurately for the first time, while the DCGAN again outperforms the DDPM in reproducing the peak fluctuation levels. Since the measured PIV data of this case exhibit weaker high-intensity turbulence in the far-wake region, the quality of the generated data remains high across all evaluated regions (see \cref{fig:eval_2V8H_large_mean} and \cref{fig:eval_2V8H_large_var}).

\subsection{Computational Cost}
We consider the training and inference time to measure the computational cost (see \cref{tab:compcost}).
Since each of the considered generative models has its own properties, summarized in \cref{tab:implementdet}, each model was trained on different amounts of A100 GPUs with 80GB each. 

While VAE and DDPM needed all $100,000$ images to train, DCGAN training can be done with $0.05\%$ of the whole dataset without losing generalizability.
In addition, the DDPM has over 135 million model parameters. To meet the challenge of a large dataset of high-resolution images and large models, VAE and DDPM were parallelized across multiple GPUs to minimize training time. Note that the speedup for training with more GPUs in parallel is almost, but slightly less than, linear. 
The DCGAN was the fastest model to train, despite having the most parameters and being trained for more than ten times as many epochs as the VAE and DDPM, but on a single GPU. 
In terms of inference time, the VAE is significantly faster than the DCGAN and the DDPM, but the generated results don't reach the quality of the LES.
Consequently, the inference performance of DCGAN and DDPM is of particular interest. 
The DCGAN generates one snapshot in $0.001\,\text{s}$, whereas the DDPM requires $36.3\,\text{s}$ per snapshot. 
Each generated sample represents a state snapshot of the flow field drawn from the distribution of the LES dataset, which was originally sampled at $68.9\,\text{Hz}$ (i.e., every $\approx 250$ LES time steps, $\Delta t = 5.8\times10^{-5}\,\text{s}$).
Thus, the reported ``seconds per frame'' values purely represent wall-clock inference times and are not related to physical time scales.

\begin{table}[!hb]
\begin{center}
\resizebox{0.75\textwidth}{!}{%
    {\tabulinesep=1.2mm
        \begin{tabu}{|l|l|l|l|}
            \hline
            Model & VAE & DCGAN & DDPM (with attention)\\
            \hline
            Training Time & $75$ hrs total across 4 $\times$ A100 
            & $62$ hrs total across 1 $\times$ A100
            & $490$ total hrs across 3 $\times$ A100\\ 
            \hline
            Inference Time & $0.000826$ sec/frame & $0.001$ sec/frame & $36.3$ sec/frame \\
            \hline
        \end{tabu}}}
        \caption{Comparison of the computational cost for the VAE, DCGAN and DDPM.}
        \label{tab:compcost}
    \end{center}
\end{table}

\begin{figure}[!h]
    \centering
    \includegraphics[width=0.7\textwidth]{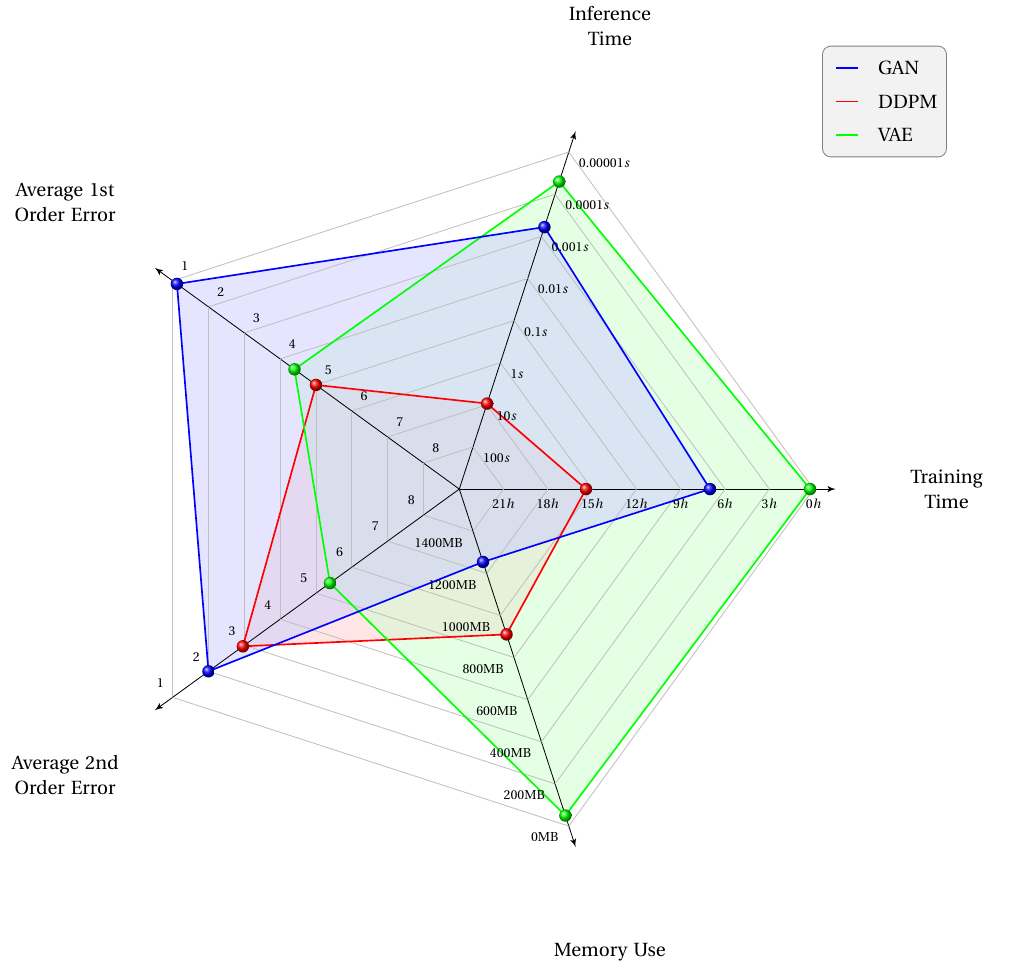}
    \caption{A Kiviat diagram comparing the three generative models for the PIV dataset 4V2H. Inference time and memory use are measured per sample, and the errors are measured by taking a mean square error between the respective error curves, and are then placed on a log-scale.}
    \label{fig:spider_ilmenau}
\end{figure}

For the experimental PIV data, computational cost plays a minor role, and we therefore focus our analysis on the LES cases. Nevertheless, the Kiviat diagram (see \cref{fig:spider_ilmenau}) confirms that the considered generative models remain computationally advantageous also for the PIV data, while the main challenge in this case lies in handling the measurement uncertainties inherent to experimental observations.


\begin{figure}[!h]
    \centering
    \includegraphics[width=0.7\textwidth]{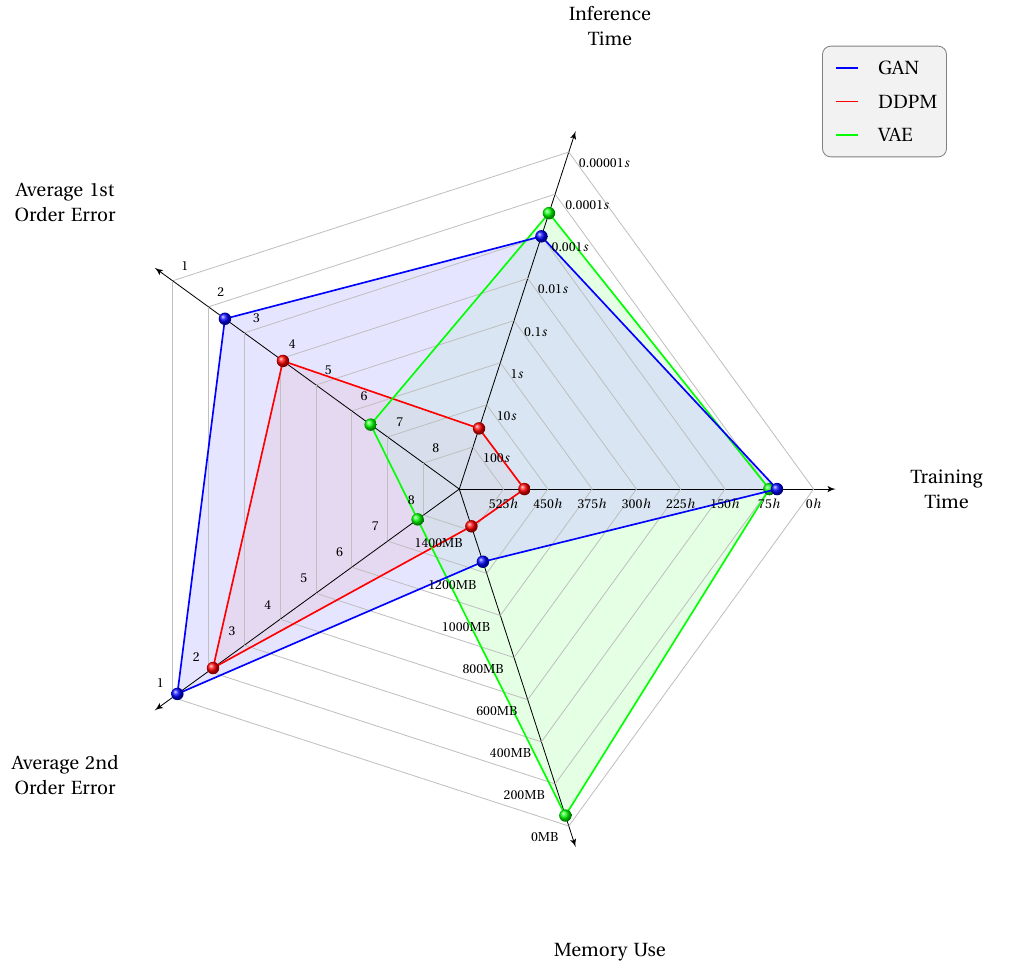}
    \caption{A Kiviat diagram comparing the three generative models for the LES flow around a cylinder dataset. Inference time and memory use are measured per sample, and the errors are measured by taking a mean square error between the respective error curves, and are then placed on a log-scale.}
    \label{fig:spider}
\end{figure}

\section{Conclusion and Outlook}
\label{sec:conclusion}
We compared three generative learning approaches, starting from a VAE, a lighter model with about 4 million parameters, to DCGAN, already widely studied in the field of neural network turbulence modeling, to DDPM, a recent generative model. 
Our experiments have shown that the VAE is not able to match the quality of the LES, which can be directly observed by just taking a look at the generated image.
In contrast, DCGAN and DDPM both synthesize turbulent flow that agrees excellently with the LES flow at the visual level and by examining physics-based metrics. Nevertheless, we found that DCGAN performs better in hitting the peaks of the distributions regarding the physics-based evaluation, but care has to be taken to reduce noise in experimental data. 
This issue is particularly relevant for the experimental PIV data. While computational cost plays a minor role, both DCGAN and DDPM accurately reproduce the measured flow statistics and small-scale structures, despite the inherent measurement uncertainties. In particular, the DCGAN captures the characteristic small-scale fluctuations of the PIV data, demonstrating that these models can robustly handle experimental noise and variability.
Returning to the overall comparison, a strong argument for DCGAN was found by analyzing the computational cost, since it requires significantly less data to train and is significantly faster in training and inference time compared to DDPM. 
In conclusion, both \cref{fig:spider} and \cref{fig:spider_ilmenau} show that DCGAN is clearly superior to DDPM, especially in terms of model efficiency.
Our experiments have shown that the DCGAN itself, as well as extended variations based on the DCGAN, remain promising for future work in turbulence modeling. 
However, with the growing amount of data and increasingly powerful computers, DDPM should not be completely discarded for this problem. They are not as efficient as DCGAN, but they are still significantly faster in generating turbulent flow than LES, and the quality is also adequate.
Overall, our experiments indicate that DCGAN and its extended variations remain highly promising for future work in turbulence modeling, including both numerical simulations and experimental datasets. The successful application to PIV data suggests that these AI-based approaches are capable not only of reproducing LES-like flows but also of accommodating the uncertainties inherent in experimental measurements, broadening their potential impact in fluid dynamics research.

\section*{Acknowledgments}
This work was funded by the Deutsche Forschungsgemeinschaft (DFG, German
Research Foundation) within the priority program "Carnot Batteries: Inverse Design from Markets to Molecules" (SPP 2403) grant no. 526152410 and 525893212 and DFG grant no. 467227170. We also thank the two anonymous referees for hints that helped to improve this paper. We also acknowledge support from the NVIDIA Academic Grant Program in the form of compute resources.

\newpage
\bibliography{bibliography}

@article{winhart2024data,
  author       = {Winhart, Benjamin and
                  di Mare, Francesca},
  title        = {Flow Around a Cylinder for Generative Learning},
  month        = {11},
  year         = {2024},
  publisher    = {Zenodo},
  version      = {1.0},
  note         = {\url{https://doi.org/10.5281/zenodo.13820259}, DOI: 10.5281/zenodo.13820259, Publisher: Zenodo}
}

@article{gopalan2013unified,
  title={A unified RANS--LES model: Computational development, accuracy and cost},
  author={Gopalan, Harish and Heinz, Stefan and St{\"o}llinger, Michael K},
  journal={Journal of Computational Physics},
  volume={249},
  pages={249--274},
  year={2013},
  publisher={Elsevier}
}

@article{wang2024patch,
  title={Patch diffusion: Faster and more data-efficient training of diffusion models},
  author={Wang, Zhendong and Jiang, Yifan and Zheng, Huangjie and Wang, Peihao and He, Pengcheng and Wang, Zhangyang and Chen, Weizhu and Zhou, Mingyuan and others},
  journal={Advances in neural information processing systems},
  volume={36},
  year={2024}
}

@misc{latentdiffusionmodel2022,
      title={High-Resolution Image Synthesis with Latent Diffusion Models}, 
      author={Robin Rombach and Andreas Blattmann and Dominik Lorenz and Patrick Esser and Björn Ommer},
      year={2022},
      eprint={2112.10752},
      archivePrefix={arXiv},
      primaryClass={cs.CV},
      url={https://arxiv.org/abs/2112.10752}, 
}

@article{menendez1997jensen,
  title={The jensen-shannon divergence},
  author={Men{\'e}ndez, Mar{\'\i}a Luisa and Pardo, JA and Pardo, L and Pardo, MC},
  journal={Journal of the Franklin Institute},
  volume={334},
  number={2},
  pages={307--318},
  year={1997},
  publisher={Elsevier}
}

@article{kullback1951information,
  title={On information and sufficiency},
  author={Kullback, Solomon and Leibler, Richard A},
  journal={The annals of mathematical statistics},
  volume={22},
  number={1},
  pages={79--86},
  year={1951},
  publisher={JSTOR}
}

@article{bayes1958essay,
  title={An essay towards solving a problem in the doctrine of chances},
  author={Bayes, Thomas},
  journal={Biometrika},
  volume={45},
  number={3-4},
  pages={296--315},
  year={1958}
}

@article{cheng2023uncertainty,
  author={Cheng, Sibo and Quilodrán-Casas, César and Ouala, Said and Farchi, Alban and Liu, Che and Tandeo, Pierre and Fablet, Ronan and Lucor, Didier and Iooss, Bertrand and Brajard, Julien and Xiao, Dunhui and Janjic, Tijana and Ding, Weiping and Guo, Yike and Carrassi, Alberto and Bocquet, Marc and Arcucci, Rossella},
  journal={IEEE/CAA Journal of Automatica Sinica}, 
  title={Machine Learning With Data Assimilation and Uncertainty Quantification for Dynamical Systems: A Review}, 
  year={2023},
  volume={10},
  number={6},
  pages={1361-1387},
  doi={10.1109/JAS.2023.123537}
}

@article{olson2001turbulence,
title = {The motion of fibres in turbulent flow, stochastic simulation of isotropic homogeneous turbulence},
journal = {International Journal of Multiphase Flow},
volume = {27},
number = {12},
pages = {2083-2103},
year = {2001},
issn = {0301-9322},
doi = {https://doi.org/10.1016/S0301-9322(01)00050-7},
url = {https://www.sciencedirect.com/science/article/pii/S0301932201000507},
author = {James A. Olson},
}

@book{lumley2007stochastic,
  title={Stochastic Tools in Turbulence},
  author={Lumley, J.L.},
  isbn={9780486462707},
  lccn={2007019198},
  series={Dover books on engineering},
  url={https://books.google.de/books?id=CsVlooboM3cC},
  year={2007},
  publisher={Dover Publications}
}

@book{box2011bayesian,
  title={Bayesian inference in statistical analysis},
  author={Box, George EP and Tiao, George C},
  year={2011},
  publisher={John Wiley \& Sons}
}

@article{kramer1991nonlinear,
  title={Nonlinear principal component analysis using autoassociative neural networks},
  author={Kramer, Mark A},
  journal={AIChE journal},
  volume={37},
  number={2},
  pages={233--243},
  year={1991},
  publisher={Wiley Online Library}
}

@inproceedings{ballard1987modular,
  title={Modular learning in neural networks},
  author={Ballard, Dana H},
  booktitle={Proceedings of the sixth National conference on Artificial intelligence-Volume 1},
  pages={279--284},
  year={1987}
}

@incollection{pinaya2020autoencoders,
  title={Autoencoders},
  author={Pinaya, Walter Hugo Lopez and Vieira, Sandra and Garcia-Dias, Rafael and Mechelli, Andrea},
  booktitle={Machine learning},
  pages={193--208},
  year={2020},
  publisher={Elsevier}
}

@inproceedings{NEURIPS2023_8df90a14,
 author = {R\"{u}hling Cachay, Salva and Zhao, Bo and Joren, Hailey and Yu, Rose},
 booktitle = {Advances in Neural Information Processing Systems},
 editor = {A. Oh and T. Naumann and A. Globerson and K. Saenko and M. Hardt and S. Levine},
 pages = {45259--45287},
 publisher = {Curran Associates, Inc.},
 title = {DYffusion: A Dynamics-informed Diffusion Model for Spatiotemporal Forecasting},
 url = {https://proceedings.neurips.cc/paper_files/paper/2023/file/8df90a1440ce782d1f5607b7a38f2531-Paper-Conference.pdf},
 volume = {36},
 year = {2023}
}

@article{yang2023diffusion,
  title={Diffusion probabilistic modeling for video generation},
  author={Yang, Ruihan and Srivastava, Prakhar and Mandt, Stephan},
  journal={Entropy},
  volume={25},
  number={10},
  pages={1469},
  year={2023},
  publisher={MDPI}
}

@inproceedings{kohl2024benchmarking,
  title={Benchmarking Autoregressive Conditional Diffusion Models for Turbulent Flow Simulation},
  author={Kohl, Georg and Chen, Liwei and Thuerey, Nils},
  booktitle={ICML 2024 AI for Science Workshop},
  year={2024}
}

@article{shu2024zero,
  title={Zero-Shot Uncertainty Quantification using Diffusion Probabilistic Models},
  author={Shu, Dule and Farimani, Amir Barati},
  journal={arXiv preprint arXiv:2408.04718},
  year={2024}
}

@inproceedings{ledig2017photo,
  title={Photo-realistic single image super-resolution using a generative adversarial network},
  author={Ledig, Christian and Theis, Lucas and Husz{\'a}r, Ferenc and Caballero, Jose and Cunningham, Andrew and Acosta, Alejandro and Aitken, Andrew and Tejani, Alykhan and Totz, Johannes and Wang, Zehan and others},
  booktitle={Proceedings of the IEEE conference on computer vision and pattern recognition},
  pages={4681--4690},
  year={2017}
}

@article{kingma2013auto,
  title={Auto-encoding variational bayes},
  author={Kingma, Diederik P},
  journal={arXiv preprint arXiv:1312.6114},
  year={2013}
}

@article{wang2021flow,
    author = {Wang, Jing and He, Cheng and Li, Runze and Chen, Haixin and Zhai, Chen and Zhang, Miao},
    title = "{Flow field prediction of supercritical airfoils via variational autoencoder based deep learning framework}",
    journal = {Physics of Fluids},
    volume = {33},
    number = {8},
    pages = {086108},
    year = {2021},
    month = {08},
    issn = {1070-6631},
    doi = {10.1063/5.0053979},
    url = {https://doi.org/10.1063/5.0053979},
    eprint = {https://pubs.aip.org/aip/pof/article-pdf/doi/10.1063/5.0053979/14151975/086108\_1\_online.pdf},
}

@article{agostini2020aelow,
    author = {Agostini, Lionel},
    title = "{Exploration and prediction of fluid dynamical systems using auto-encoder technology}",
    journal = {Physics of Fluids},
    volume = {32},
    number = {6},
    pages = {067103},
    year = {2020},
    month = {06},
    issn = {1070-6631},
    doi = {10.1063/5.0012906},
    url = {https://doi.org/10.1063/5.0012906},
    eprint = {https://pubs.aip.org/aip/pof/article-pdf/doi/10.1063/5.0012906/14754686/067103\_1\_online.pdf},
}

@article{CHENG2020113375,
title = {An advanced hybrid deep adversarial autoencoder for parameterized nonlinear fluid flow modelling},
journal = {Computer Methods in Applied Mechanics and Engineering},
volume = {372},
pages = {113375},
year = {2020},
issn = {0045-7825},
doi = {https://doi.org/10.1016/j.cma.2020.113375},
url = {https://www.sciencedirect.com/science/article/pii/S0045782520305600},
author = {M. Cheng and F. Fang and C.C. Pain and I.M. Navon}
}

@article{gundersen2021scvae,
    author = {Gundersen, Kristian and Oleynik, Anna and Blaser, Nello and Alendal, Guttorm},
    title = "{Semi-conditional variational auto-encoder for flow reconstruction and uncertainty quantification from limited observations}",
    journal = {Physics of Fluids},
    volume = {33},
    number = {1},
    pages = {017119},
    year = {2021},
    month = {01},
    issn = {1070-6631},
    doi = {10.1063/5.0025779},
    url = {https://doi.org/10.1063/5.0025779},
    eprint = {https://pubs.aip.org/aip/pof/article-pdf/doi/10.1063/5.0025779/15872101/017119\_1\_online.pdf},
}

@article{DBLP:journals/corr/abs-1903-05789,
  author       = {Bin Dai and
                  David P. Wipf},
  title        = {Diagnosing and Enhancing {VAE} Models},
  journal      = {CoRR},
  volume       = {abs/1903.05789},
  year         = {2019},
  url          = {http://arxiv.org/abs/1903.05789},
  eprinttype    = {arXiv},
  eprint       = {1903.05789},
  bibsource    = {dblp computer science bibliography, https://dblp.org}
}

@inproceedings{NIPS2016_ddeebdee,
 author = {Kingma, Durk P and Salimans, Tim and Jozefowicz, Rafal and Chen, Xi and Sutskever, Ilya and Welling, Max},
 booktitle = {Advances in Neural Information Processing Systems},
 editor = {D. Lee and M. Sugiyama and U. Luxburg and I. Guyon and R. Garnett},
 pages = {},
 publisher = {Curran Associates, Inc.},
 title = {Improved Variational Inference with Inverse Autoregressive Flow},
 url = {https://proceedings.neurips.cc/paper_files/paper/2016/file/ddeebdeefdb7e7e7a697e1c3e3d8ef54-Paper.pdf},
 volume = {29},
 year = {2016}
}

@misc{ghosh2020variationaldeterministicautoencoders,
      title={From Variational to Deterministic Autoencoders}, 
      author={Partha Ghosh and Mehdi S. M. Sajjadi and Antonio Vergari and Michael Black and Bernhard Schölkopf},
      year={2020},
      eprint={1903.12436},
      archivePrefix={arXiv},
      primaryClass={cs.LG},
      url={https://arxiv.org/abs/1903.12436}, 
}

@Article{Qi2024Combined,
AUTHOR = {Qi, Jiaheng and Ma, Hongbing},
TITLE = {A Combined Model of Diffusion Model and Enhanced Residual Network for Super-Resolution Reconstruction of Turbulent Flows},
JOURNAL = {Mathematics},
VOLUME = {12},
YEAR = {2024},
NUMBER = {7},
ARTICLE-NUMBER = {1028},
URL = {https://www.mdpi.com/2227-7390/12/7/1028},
ISSN = {2227-7390},
DOI = {10.3390/math12071028}
}

@misc{oord2018neuraldiscreterepresentationlearning,
      title={Neural Discrete Representation Learning}, 
      author={Aaron van den Oord and Oriol Vinyals and Koray Kavukcuoglu},
      year={2018},
      eprint={1711.00937},
      archivePrefix={arXiv},
      primaryClass={cs.LG},
      url={https://arxiv.org/abs/1711.00937}, 
}

@article{liu2024uncertainty,
  title={Uncertainty-Aware Surrogate Models for Airfoil Flow Simulations with Denoising Diffusion Probabilistic Models},
  author={Liu, Qiang and Thuerey, Nils},
  journal={AIAA Journal},
  pages={1--22},
  year={2024},
  publisher={American Institute of Aeronautics and Astronautics}
}

@inproceedings{higgins2017betavae,
      title={beta-{VAE}: Learning Basic Visual Concepts with a Constrained Variational Framework},
      author={Irina Higgins and Loic Matthey and Arka Pal and Christopher Burgess and Xavier Glorot and Matthew Botvinick and Shakir Mohamed and Alexander Lerchner},
      booktitle={International Conference on Learning Representations},
      year={2017},
      url={https://openreview.net/forum?id=Sy2fzU9gl}
}

@Article{atmos15010060,
AUTHOR = {Li, Tianyi and Lanotte, Alessandra S. and Buzzicotti, Michele and Bonaccorso, Fabio and Biferale, Luca},
TITLE = {Multi-Scale Reconstruction of Turbulent Rotating Flows with Generative Diffusion Models},
JOURNAL = {Atmosphere},
VOLUME = {15},
YEAR = {2024},
NUMBER = {1},
ARTICLE-NUMBER = {60},
URL = {https://www.mdpi.com/2073-4433/15/1/60},
ISSN = {2073-4433},
DOI = {10.3390/atmos15010060}
}

@article{Li_Buzzicotti_Biferale_Bonaccorso_Chen_Wan_2023, 
title={Multi-scale reconstruction of turbulent rotating flows with proper orthogonal decomposition and generative adversarial networks}, 
volume={971}, 
DOI={10.1017/jfm.2023.573}, 
journal={Journal of Fluid Mechanics}, 
author={Li, Tianyi and Buzzicotti, Michele and Biferale, Luca and Bonaccorso, Fabio and Chen, Shiyi and Wan, Minping}, 
year={2023}, 
pages={A3}
}

@article{nista2023investigation,
  title={Investigation of the generalization capability of a generative adversarial network for large eddy simulation of turbulent premixed reacting flows},
  author={Nista, Ludovico and Schumann, CDK and Grenga, Temistocle and Attili, Antonio and Pitsch, Heinz},
  journal={Proceedings of the Combustion Institute},
  volume={39},
  number={4},
  pages={5279--5288},
  year={2023},
  publisher={Elsevier}
}

@article{bode2021using,
  title={Using physics-informed enhanced super-resolution generative adversarial networks for subfilter modeling in turbulent reactive flows},
  author={Bode, Mathis and Gauding, Michael and Lian, Zeyu and Denker, Dominik and Davidovic, Marco and Kleinheinz, Konstantin and Jitsev, Jenia and Pitsch, Heinz},
  journal={Proceedings of the Combustion Institute},
  volume={38},
  number={2},
  pages={2617--2625},
  year={2021},
  publisher={Elsevier}
}

@article{sardar2023spectrally,
  title={Spectrally Decomposed Diffusion Models for Generative Turbulence Recovery},
  author={Sardar, Mohammed and Skillen, Alex and Zimo{\'n}, Ma{\l}gorzata J and Draycott, Samuel and Revell, Alistair},
  journal={arXiv preprint arXiv:2312.15029},
  year={2023}
}

@article{zheng2024high,
  title={High-fidelity reconstruction of large-area damaged turbulent fields with a physically constrained generative adversarial network},
  author={Zheng, Qinmin and Li, Tianyi and Ma, Benteng and Fu, Lin and Li, Xiaomeng},
  journal={Physical Review Fluids},
  volume={9},
  number={2},
  pages={024608},
  year={2024},
  publisher={APS}
}

@article{gao2024bayesian,
  title={Bayesian conditional diffusion models for versatile spatiotemporal turbulence generation},
  author={Gao, Han and Han, Xu and Fan, Xiantao and Sun, Luning and Liu, Li-Ping and Duan, Lian and Wang, Jian-Xun},
  journal={Computer Methods in Applied Mechanics and Engineering},
  volume={427},
  pages={117023},
  year={2024},
  publisher={Elsevier}
}

@article{yan2023local,
  title={Local turbulence generation using conditional generative adversarial networks toward Reynolds-averaged Navier--Stokes modeling},
  author={Yan, Chongyang and Zhang, Yufei},
  journal={Physics of Fluids},
  volume={35},
  number={10},
  year={2023},
  publisher={AIP Publishing}
}

@article{Kim_Kim_Lee_2024, 
title={Prediction and control of two-dimensional decaying turbulence using generative adversarial networks}, 
volume={981}, 
DOI={10.1017/jfm.2024.77}, 
journal={Journal of Fluid Mechanics}, 
author={Kim, Jiyeon and Kim, Junhyuk and Lee, Changhoon}, 
year={2024}, 
pages={A19}
}

@article{youssif2022superres,
    author = {Yousif, Mustafa Z. and Yu, Linqi and Lim, Hee-Chang},
    title = "{Super-resolution reconstruction of turbulent flow fields at various Reynolds numbers based on generative adversarial networks}",
    journal = {Physics of Fluids},
    volume = {34},
    number = {1},
    pages = {015130},
    year = {2022},
    month = {01},
    issn = {1070-6631},
    doi = {10.1063/5.0074724},
    url = {https://doi.org/10.1063/5.0074724},
    eprint = {https://pubs.aip.org/aip/pof/article-pdf/doi/10.1063/5.0074724/16625179/015130\_1\_online.pdf},
}

@article{youssif2021high,
    author = {Yousif, Mustafa Z. and Yu, Linqi and Lim, Hee-Chang},
    title = "{High-fidelity reconstruction of turbulent flow from spatially limited data using enhanced super-resolution generative adversarial network}",
    journal = {Physics of Fluids},
    volume = {33},
    number = {12},
    pages = {125119},
    year = {2021},
    month = {12},
    issn = {1070-6631},
    doi = {10.1063/5.0066077},
    url = {https://doi.org/10.1063/5.0066077},
    eprint = {https://pubs.aip.org/aip/pof/article-pdf/doi/10.1063/5.0066077/15873357/125119\_1\_online.pdf},
}

@article{subramaniam2020turbulence,
  title={Turbulence enrichment using physics-informed generative adversarial networks},
  author={Subramaniam, Akshay and Wong, Man Long and Borker, Raunak D and Nimmagadda, Sravya and Lele, Sanjiva K},
  journal={arXiv preprint arXiv:2003.01907},
  year={2020}
}

@article{deng2019super,
  title={Super-resolution reconstruction of turbulent velocity fields using a generative adversarial network-based artificial intelligence framework},
  author={Deng, Zhiwen and He, Chuangxin and Liu, Yingzheng and Kim, Kyung Chun},
  journal={Physics of Fluids},
  volume={31},
  number={12},
  year={2019},
  publisher={AIP Publishing}
}

@article{Spalart_Venkatakrishnan_2016, 
title={On the role and challenges of CFD in the aerospace industry}, 
volume={120}, 
DOI={10.1017/aer.2015.10}, 
number={1223}, 
journal={The Aeronautical Journal}, 
author={Spalart, P. R. and Venkatakrishnan, V.}, 
year={2016}, 
pages={209–232}
}

@misc{mcallester2023maths_diff,
      title={On the Mathematics of Diffusion Models}, 
      author={David McAllester},
      year={2023},
      eprint={2301.11108},
      archivePrefix={arXiv},
      primaryClass={cs.LG},
      url={https://arxiv.org/abs/2301.11108}, 
}

@misc{song2020generativemodelingestimatinggradients,
      title={Generative Modeling by Estimating Gradients of the Data Distribution}, 
      author={Yang Song and Stefano Ermon},
      year={2020},
      eprint={1907.05600},
      archivePrefix={arXiv},
      primaryClass={cs.LG},
      url={https://arxiv.org/abs/1907.05600}, 
}

@article{anderson1982reverse_sde,
title = {Reverse-time diffusion equation models},
journal = {Stochastic Processes and their Applications},
volume = {12},
number = {3},
pages = {313-326},
year = {1982},
issn = {0304-4149},
doi = {https://doi.org/10.1016/0304-4149(82)90051-5},
url = {https://www.sciencedirect.com/science/article/pii/0304414982900515},
author = {Brian D.O. Anderson}
}

@article{2023diff_survey,
author = {Yang, Ling and Zhang, Zhilong and Song, Yang and Hong, Shenda and Xu, Runsheng and Zhao, Yue and Zhang, Wentao and Cui, Bin and Yang, Ming-Hsuan},
title = {Diffusion Models: A Comprehensive Survey of Methods and Applications},
year = {2023},
issue_date = {April 2024},
publisher = {Association for Computing Machinery},
address = {New York, NY, USA},
volume = {56},
number = {4},
issn = {0360-0300},
url = {https://doi.org/10.1145/3626235},
doi = {10.1145/3626235},
journal = {ACM Comput. Surv.},
month = nov,
articleno = {105},
numpages = {39}
}

@misc{kong2021diffwaveversatile,
      title={DiffWave: A Versatile Diffusion Model for Audio Synthesis}, 
      author={Zhifeng Kong and Wei Ping and Jiaji Huang and Kexin Zhao and Bryan Catanzaro},
      year={2021},
      eprint={2009.09761},
      archivePrefix={arXiv},
      primaryClass={eess.AS},
      url={https://arxiv.org/abs/2009.09761}, 
}

@misc{kingma2013,
      title={Auto-Encoding Variational Bayes}, 
      author={Diederik P Kingma and Max Welling},
      year={2013},
      eprint={1312.6114},
      archivePrefix={arXiv},
      primaryClass={stat.ML},
      url={https://arxiv.org/abs/1312.6114}, 
}

@article{Barenghi_2014,
   title={Introduction to quantum turbulence},
   volume={111},
   ISSN={1091-6490},
   url={http://dx.doi.org/10.1073/pnas.1400033111},
   DOI={10.1073/pnas.1400033111},
   number={1},
   journal={Proceedings of the National Academy of Sciences},
   publisher={Proceedings of the National Academy of Sciences},
   author={Barenghi, Carlo F. and Skrbek, Ladislav and Sreenivasan, Katepalli R.},
   year={2014},
   month={mar}, 
   pages={4647–4652} 
}

@article{stars_turbulence,
doi = {10.1086/431734},
url = {https://dx.doi.org/10.1086/431734},
year = {2005},
month = {sep},
publisher = {},
volume = {630},
number = {1},
pages = {250},
author = {Mark R. Krumholz and Christopher F. McKee},
title = {A General Theory of Turbulence-regulated Star Formation, from Spirals to Ultraluminous Infrared Galaxies},
journal = {The Astrophysical Journal}
}

@article{cancer_turbulence,
author = {Uthamacumaran, Abicumaran},
year = {2020},
month = {10},
pages = {759-769},
title = {Cancer: A Turbulence Problem},
volume = {22},
journal = {Neoplasia (New York, N.Y.)},
doi = {10.1016/j.neo.2020.09.008}
}

@article{Maulik_2018,
   title={Subgrid modelling for two-dimensional turbulence using neural networks},
   volume={858},
   ISSN={1469-7645},
   url={http://dx.doi.org/10.1017/jfm.2018.770},
   DOI={10.1017/jfm.2018.770},
   journal={Journal of Fluid Mechanics},
   publisher={Cambridge University Press (CUP)},
   author={Maulik, R. and San, O. and Rasheed, A. and Vedula, P.},
   year={2018},
   month=nov, pages={122–144} }

@misc{shankar2023differentiableturbulenceii,
      title={Differentiable Turbulence II}, 
      author={Varun Shankar and Romit Maulik and Venkatasubramanian Viswanathan},
      year={2023},
      eprint={2307.13533},
      archivePrefix={arXiv},
      primaryClass={physics.flu-dyn},
      url={https://arxiv.org/abs/2307.13533}, 
}

@Article{zhangreview2023,
    AUTHOR = {Zhang, Yi and Zhang, Dapeng and Jiang, Haoyu},
    TITLE = {Review of Challenges and Opportunities in Turbulence Modeling: A Comparative Analysis of Data-Driven Machine Learning Approaches},
   JOURNAL = {Journal of Marine Science and Engineering},
    VOLUME = {11},
    YEAR = {2023},
    NUMBER = {7},
    ARTICLE-NUMBER = {1440},
    URL = {https://www.mdpi.com/2077-1312/11/7/1440},
    ISSN = {2077-1312},
    DOI = {10.3390/jmse11071440}
}

@misc{song2021scorebased,
      title={Score-Based Generative Modeling through Stochastic Differential Equations}, 
      author={Yang Song and Jascha Sohl-Dickstein and Diederik P. Kingma and Abhishek Kumar and Stefano Ermon and Ben Poole},
      year={2021},
      eprint={2011.13456},
      archivePrefix={arXiv},
      primaryClass={cs.LG}
}

@misc{ho2020denoising,
      title={Denoising Diffusion Probabilistic Models}, 
      author={Jonathan Ho and Ajay Jain and Pieter Abbeel},
      year={2020},
      eprint={2006.11239},
      archivePrefix={arXiv},
      primaryClass={cs.LG}
}

@misc{sohldickstein2015deep,
      title={Deep Unsupervised Learning using Nonequilibrium Thermodynamics}, 
      author={Jascha Sohl-Dickstein and Eric A. Weiss and Niru Maheswaranathan and Surya Ganguli},
      year={2015},
      eprint={1503.03585},
      archivePrefix={arXiv},
      primaryClass={cs.LG}
}

@article{drygala2022generative,
  title={Generative modeling of turbulence},
  author={Drygala, Claudia and Winhart, Benjamin and di Mare, Francesca and Gottschalk, Hanno},
  journal={Physics of Fluids},
  volume={34},
  number={3},
  year={2022},
  publisher={AIP Publishing}
}

@inproceedings{drygala2023generalization,
  title={Generalization capabilities of conditional GAN for turbulent flow under changes of geometry},
  author={Drygala, Claudia and di Mare, Francesca and Gottschalk, Hanno},
  booktitle={Proceedings of the 15th International Conference on Evolutionary and Deterministic Methods For Design, Optimization and Control (EUROGEN 2023)},
  year={2023}
}

@article{wang_2017,
  title = {Physics-informed machine learning approach for reconstructing Reynolds stress modeling discrepancies based on DNS data},
  author = {Wang, Jian-Xun and Wu, Jin-Long and Xiao, Heng},
  journal = {Phys. Rev. Fluids},
  volume = {2},
  issue = {3},
  pages = {034603},
  numpages = {22},
  year = {2017},
  month = {Mar},
  publisher = {American Physical Society},
  doi = {10.1103/PhysRevFluids.2.034603},
  url = {https://link.aps.org/doi/10.1103/PhysRevFluids.2.034603}
}

@Inbook{Buzzi2009,
author="Buzzi, J{\'e}r{\^o}me",
editor="Meyers, Robert A.",
title="Chaos and Ergodic Theory",
bookTitle="Encyclopedia of Complexity and Systems Science",
year="2009",
publisher="Springer New York",
address="New York, NY",
pages="953--978",
isbn="978-0-387-30440-3",
doi="10.1007/978-0-387-30440-3_64",
url="https://doi.org/10.1007/978-0-387-30440-3_64"
}

@article{vanilla_gan,
author = {Goodfellow, Ian and Pouget-Abadie, Jean and Mirza, Mehdi and Xu, Bing and Warde-Farley, David and Ozair, Sherjil and Courville, Aaron and Bengio, Y.},
year = {2014},
month = {06},
pages = {},
title = {Generative Adversarial Networks},
volume = {3},
journal = {Advances in Neural Information Processing Systems},
doi = {10.1145/3422622}
}

@INPROCEEDINGS{pix2pixHD,
  author={Wang, Ting-Chun and Liu, Ming-Yu and Zhu, Jun-Yan and Tao, Andrew and Kautz, Jan and Catanzaro, Bryan},
  booktitle={2018 IEEE/CVF Conference on Computer Vision and Pattern Recognition}, 
  title={High-Resolution Image Synthesis and Semantic Manipulation with Conditional GANs}, 
  year={2018},
  volume={},
  number={},
  pages={8798-8807},
  doi={10.1109/CVPR.2018.00917}
}

@inproceedings{dcgan,
  author    = {Alec Radford and Luke Metz and Soumith Chintala},
  editor    = {Yoshua Bengio and Yann LeCun},
  title     = {Unsupervised Representation Learning with Deep Convolutional Generative Adversarial Networks},
  booktitle = {4th International Conference on Learning Representations, {ICLR} 2016, San Juan, Puerto Rico, May 2-4, 2016, Conference Track Proceedings},
  year      = {2016},
  url       = {http://arxiv.org/abs/1511.06434}
}

@ARTICLE{bce,
  author={Ho, Yaoshiang and Wookey, Samuel},
  journal={IEEE Access}, 
  title={The Real-World-Weight Cross-Entropy Loss Function: Modeling the Costs of Mislabeling}, 
  year={2020},
  volume={8},
  pages={4806-4813},
  doi={10.1109/ACCESS.2019.2962617}
  }

@article{backprop,
  title={Learning representations by back-propagating errors},
  author={David E. Rumelhart and Geoffrey E. Hinton and Ronald J. Williams},
  journal={Nature},
  year={1986},
  volume={323},
  pages={533-536}
}

@INPROCEEDINGS{cnn_impact,
  author={Albawi, Saad and Mohammed, Tareq Abed and Al-Zawi, Saad},
  booktitle={2017 International Conference on Engineering and Technology (ICET)}, 
  title={Understanding of a convolutional neural network}, 
  year={2017},
  pages={1-6},
  doi={10.1109/ICEngTechnol.2017.8308186}
  }

@Inbook{cnn_explanation,
author="Kim, Phil",
title="Convolutional Neural Network",
bookTitle="MATLAB Deep Learning: With Machine Learning, Neural Networks and Artificial Intelligence",
year="2017",
publisher="Apress",
address="Berkeley, CA",
pages="121--147",
isbn="978-1-4842-2845-6",
doi="10.1007/978-1-4842-2845-6_6",
url="https://doi.org/10.1007/978-1-4842-2845-6_6"
}

@article{asatryan2020convenient,
  title={A Convenient Infinite Dimensional Framework for Generative Adversarial Learning},
  author={Asatryan, Hayk and Gottschalk, Hanno and Lippert, Marieke and Rottmann, Matthias},
  journal={arXiv preprint arXiv:2011.12087},
  year={2020}
}

@book{Hirsch:2007,
author = {Hirsch, Charles},
year = {2007},
month = {01},
pages = {},
publisher = {\href{https://doi.org/10.1016/B978-0-7506-6594-0.X5037-1}{Butterworth-Heinemann }},
title = {{"Numerical Computation of Internal and External Flows: The Fundamentals of Computational Fluid Dynamics"}},
doi = {10.1016/B978-0-7506-6594-0.X5037-1}
}

@book{Ferziger:2008,
  address = {Berlin},
  author = {Ferziger, J.H. and Peri\'{c}, M.},
  publisher = {\href{https://doi.org/10.1007/978-3-642-56026-2}{Springer}},
  title = {{"Computational Methods for Fluid Dynamics"}},
  doi = "10.1007/978-3-642-56026-2",
  year = 2008
}

@article{Kolmogorov:1991,
 ISSN = {09628444},
 author = {A. N. Kolmogorov},
 journal = {\href{https://doi.org/10.2307/51980}{Proceedings: Mathematical and Physical Sciences}},
 number = {1890},
 pages = {9--13},
 publisher = {The Royal Society},
 title = {{"The Local Structure of Turbulence in Incompressible Viscous Fluid for Very Large Reynolds Numbers"}},
 volume = {434},
 year = {1991},
 doi = {10.2307/51980}
}

@article{Parnaudeau:2008,
author = {Parnaudeau,Philippe  and Carlier,Johan  and Heitz,Dominique  and Lamballais,Eric },
title = {Experimental and numerical studies of the flow over a circular cylinder at Reynolds number 3900},
journal = {Physics of Fluids},
volume = {20},
number = {8},
pages = {085101},
year = {2008},
doi = {10.1063/1.2957018},
}

@article{Norberg:1994, 
title={An experimental investigation of the flow around a circular cylinder: influence of aspect ratio}, 
volume={258},
DOI={10.1017/S0022112094003332}, 
journal={Journal of Fluid Mechanics}, 
publisher={Cambridge University Press}, 
author={Norberg, C.}, 
year={1994}, 
pages={287–316}
}

@article{Ong:1996,
  title={The velocity field of the turbulent very near wake of a circular cylinder},
  author={Lawrence Ong and James M. Wallace},
  journal={Experiments in Fluids},
  year={1996},
  volume={20},
  pages={441-453}
}

@phdthesis{Beaudan:1995,
  title={Numerical experiments on the flow past a circular cylinder at sub-critical Reynolds number},
  author={Beaudan, Patrick Bruno},
  year={1995},
  school={Stanford University}
}

@article{Kravchenko:2000,
author = {Kravchenko,Arthur G.  and Moin,Parviz },
title = {Numerical studies of flow over a circular cylinder at ReD=3900},
journal = {Physics of Fluids},
volume = {12},
number = {2},
pages = {403-417},
year = {2000},
doi = {10.1063/1.870318},
}

@article{Cheung:2011,
title = {Bayesian uncertainty analysis with applications to turbulence modeling},
journal = {Reliability Engineering \& System Safety},
volume = {96},
number = {9},
pages = {1137-1149},
year = {2011},
note = {Quantification of Margins and Uncertainties},
issn = {0951-8320},
doi = {https://doi.org/10.1016/j.ress.2010.09.013},
url = {https://www.sciencedirect.com/science/article/pii/S0951832011000664},
author = {Sai Hung Cheung and Todd A. Oliver and Ernesto E. Prudencio and Serge Prudhomme and Robert D. Moser},
}

@article{Edeling:2014a,
title = {Predictive RANS simulations via Bayesian Model-Scenario Averaging},
journal = {Journal of Computational Physics},
volume = {275},
pages = {65-91},
year = {2014},
issn = {0021-9991},
doi = {https://doi.org/10.1016/j.jcp.2014.06.052},
url = {https://www.sciencedirect.com/science/article/pii/S0021999114004707},
author = {W.N. Edeling and P. Cinnella and R.P. Dwight},
}

@article{Edeling:2014b,
author = {Edeling, Wouter and Cinnella, Paola and Dwight, Richard and Bijl, Hester},
year = {2014},
month = {02},
pages = {73-94},
title = {Bayesian estimates of parameter variability in the $k-\varepsilon$ turbulence model},
volume = {258},
journal = {Journal of Computational Physics},
doi = {10.1016/j.jcp.2013.10.027}
}

@article{Zhang:2018a,
title = {An efficient Bayesian uncertainty quantification approach with application to $k-\omega-\gamma$ transition modeling},
journal = {Computers \& Fluids},
volume = {161},
pages = {211-224},
year = {2018},
issn = {0045-7930},
doi = {https://doi.org/10.1016/j.compfluid.2017.11.007},
url = {https://www.sciencedirect.com/science/article/pii/S0045793017304152},
author = {Jincheng Zhang and Song Fu},
}

@article{Weatheritt:2016,
title = {A novel evolutionary algorithm applied to algebraic modifications of the RANS stress–strain relationship},
journal = {Journal of Computational Physics},
volume = {325},
pages = {22-37},
year = {2016},
issn = {0021-9991},
doi = {https://doi.org/10.1016/j.jcp.2016.08.015},
url = {https://www.sciencedirect.com/science/article/pii/S0021999116303643},
author = {Jack Weatheritt and Richard Sandberg},
}

@article{Weatheritt:2017,
title = {The development of algebraic stress models using a novel evolutionary algorithm},
journal = {International Journal of Heat and Fluid Flow},
volume = {68},
pages = {298-318},
year = {2017},
issn = {0142-727X},
doi = {https://doi.org/10.1016/j.ijheatfluidflow.2017.09.017},
url = {https://www.sciencedirect.com/science/article/pii/S0142727X17303223},
author = {J. Weatheritt and R.D. Sandberg},
}

@article{Zhao:2020,
title = {RANS turbulence model development using CFD-driven machine learning},
journal = {Journal of Computational Physics},
volume = {411},
pages = {109413},
year = {2020},
issn = {0021-9991},
doi = {https://doi.org/10.1016/j.jcp.2020.109413},
url = {https://www.sciencedirect.com/science/article/pii/S002199912030187X},
author = {Yaomin Zhao and Harshal D. Akolekar and Jack Weatheritt and Vittorio Michelassi and Richard D. Sandberg},
}

@misc{Zhang:2018b,
author = {Zhang, Weiwei and Zhu, Linyang and Kou, Jiaqing and Liu, Yilang},
year = {2018},
month = {06},
pages = {},
title = {Machine learning methods for turbulence modeling in subsonic flows over airfoils}
}

@incollection{goswami2023uncertainty,
title = {5 - Artificial intelligence–based uncertainty quantification technique for external flow computational fluid dynamic (CFD) simulations},
editor = {Tilottama Goswami and G.R. Sinha},
booktitle = {Statistical Modeling in Machine Learning},
publisher = {Academic Press},
pages = {79-92},
year = {2023},
isbn = {978-0-323-91776-6},
doi = {https://doi.org/10.1016/B978-0-323-91776-6.00014-2},
url = {https://www.sciencedirect.com/science/article/pii/B9780323917766000142},
author = {Srinivas Soumitri Miriyala and Pramod D. Jadhav and Raja Banerjee and Kishalay Mitra}
}

@article{Ling:2016, 
title={Reynolds averaged turbulence modelling using deep neural networks with embedded invariance}, 
volume={807}, 
DOI={10.1017/jfm.2016.615}, 
journal={Journal of Fluid Mechanics}, 
publisher={Cambridge University Press}, 
author={Ling, Julia and Kurzawski, 
Andrew and Templeton, Jeremy}, 
year={2016}, pages={155–166}
}

@article{Jiang:2020,
author = {Jiang, Chao and Mi, Junyi and Laima, Shujin and Li, Hui},
year = {2020},
month = {01},
pages = {258},
title = {A Novel Algebraic Stress Model with Machine-Learning-Assisted Parameterization},
volume = {13},
journal = {Energies},
doi = {10.3390/en13010258}
}

@INPROCEEDINGS{King:2017,
       author = {{King}, Ryan and {Graf}, Peter and {Chertkov}, Michael},
        title = "{Creating Turbulent Flow Realizations with Generative Adversarial Networks}",
    booktitle = {APS Division of Fluid Dynamics Meeting Abstracts},
         year = 2017,
       series = {APS Meeting Abstracts},
        month = nov,
          eid = {A31.008},
        pages = {A31.008},
       adsurl = {https://ui.adsabs.harvard.edu/abs/2017APS..DFDA31008K}
}

@misc{King:2018,
      title={From Deep to Physics-Informed Learning of Turbulence: Diagnostics}, 
      author={Ryan King and Oliver Hennigh and Arvind Mohan and Michael Chertkov},
      year={2018},
      eprint={1810.07785},
      archivePrefix={arXiv},
      primaryClass={physics.flu-dyn}
}

@article{Kim:2020,
title = {Deep unsupervised learning of turbulence for inflow generation at various Reynolds numbers},
journal = {Journal of Computational Physics},
volume = {406},
pages = {109216},
year = {2020},
issn = {0021-9991},
doi = {https://doi.org/10.1016/j.jcp.2019.109216},
url = {https://www.sciencedirect.com/science/article/pii/S0021999119309210},
author = {Junhyuk Kim and Changhoon Lee},
}

@article{kim2021unsupervised,
  title={Unsupervised deep learning for super-resolution reconstruction of turbulence},
  author={Kim, Hyojin and Kim, Junhyuk and Won, Sungjin and Lee, Changhoon},
  journal={Journal of Fluid Mechanics},
  volume={910},
  year={2021},
  publisher={Cambridge University Press}
}

@article{Fukami:2019,
   title={Super-resolution reconstruction of turbulent flows with machine learning},
   volume={870},
   ISSN={1469-7645},
   url={http://dx.doi.org/10.1017/jfm.2019.238},
   DOI={10.1017/jfm.2019.238},
   journal={Journal of Fluid Mechanics},
   publisher={Cambridge University Press (CUP)},
   author={Fukami, Kai and Fukagata, Koji and Taira, Kunihiko},
   year={2019},
   month={May},
   pages={106–120}
}

@misc{Fukami:2020,
      title={Machine learning based spatio-temporal super resolution reconstruction of turbulent flows}, 
      author={Kai Fukami and Koji Fukagata and Kunihiko Taira},
      year={2020},
      eprint={2004.11566},
      archivePrefix={arXiv},
      primaryClass={physics.flu-dyn}
}

@article{Liu:2020,
author = {Liu,Bo  and Tang,Jiupeng  and Huang,Haibo  and Lu,Xi-Yun },
title = {Deep learning methods for super-resolution reconstruction of turbulent flows},
journal = {Physics of Fluids},
volume = {32},
number = {2},
pages = {025105},
year = {2020},
doi = {10.1063/1.5140772},
URL = {https://doi.org/10.1063/1.5140772},
eprint = {https://doi.org/10.1063/1.5140772}
}

@article{Xie:2017,
   title={Data-driven synthesis of smoke flows with CNN-based feature descriptors},
   volume={36},
   ISSN={1557-7368},
   url={http://dx.doi.org/10.1145/3072959.3073643},
   DOI={10.1145/3072959.3073643},
   number={4},
   journal={ACM Transactions on Graphics},
   publisher={Association for Computing Machinery (ACM)},
   author={Xie, You and Franz, Eric and Chu, Mengyu and Thuerey, Nils},
   year={2017},
   month={Jul},
   pages={1–14}
}

@article{Werhahn:2019,
   title={A Multi-Pass GAN for Fluid Flow Super-Resolution},
   volume={2},
   ISSN={2577-6193},
   url={http://dx.doi.org/10.1145/3340251},
   DOI={10.1145/3340251},
   number={2},
   journal={Proceedings of the ACM on Computer Graphics and Interactive Techniques},
   publisher={Association for Computing Machinery (ACM)},
   author={Werhahn, Maximilian and Xie, You and Chu, Mengyu and Thuerey, Nils},
   year={2019},
   month={Jul},
   pages={1–21}
}

@article{Karras2023AnalyzingAI,
    title   = {Analyzing and Improving the Training Dynamics of Diffusion Models},
    author  = {Tero Karras and Miika Aittala and Jaakko Lehtinen and Janne Hellsten and Timo Aila and Samuli Laine},
    journal = {ArXiv},
    year    = {2023},
    volume  = {abs/2312.02696},
    url     = {https://api.semanticscholar.org/CorpusID:265659032}
}

@misc{vaswani2023attentionneed,
      title={Attention Is All You Need}, 
      author={Ashish Vaswani and Noam Shazeer and Niki Parmar and Jakob Uszkoreit and Llion Jones and Aidan N. Gomez and Lukasz Kaiser and Illia Polosukhin},
      year={2023},
      eprint={1706.03762},
      archivePrefix={arXiv},
      primaryClass={cs.CL},
      url={https://arxiv.org/abs/1706.03762}, 
}

@inproceedings{CycleGAN2017,
  title={Unpaired Image-to-Image Translation using Cycle-Consistent Adversarial Networks},
  author={Zhu, Jun-Yan and Park, Taesung and Isola, Phillip and Efros, Alexei A},
  booktitle={Computer Vision (ICCV), 2017 IEEE International Conference on},
  year={2017}
}

@inproceedings{wang2018esrgan,
  title={Esrgan: Enhanced super-resolution generative adversarial networks},
  author={Wang, Xintao and Yu, Ke and Wu, Shixiang and Gu, Jinjin and Liu, Yihao and Dong, Chao and Qiao, Yu and Change Loy, Chen},
  booktitle={Proceedings of the European conference on computer vision (ECCV) workshops},
  year={2018}
}

@inproceedings{karras2019style,
  title={A style-based generator architecture for generative adversarial networks},
  author={Karras, Tero and Laine, Samuli and Aila, Timo},
  booktitle={Proceedings of the IEEE/CVF Conference on Computer Vision and Pattern Recognition},
  pages={4401--4410},
  year={2019}
}

@incollection{pytorch2019,
title = {PyTorch: An Imperative Style, High-Performance Deep Learning Library},
author = {Paszke, Adam and Gross, Sam and Massa, Francisco and Lerer, Adam and Bradbury, James and Chanan, Gregory and Killeen, Trevor and Lin, Zeming and Gimelshein, Natalia and Antiga, Luca and Desmaison, Alban and Kopf, Andreas and Yang, Edward and DeVito, Zachary and Raison, Martin and Tejani, Alykhan and Chilamkurthy, Sasank and Steiner, Benoit and Fang, Lu and Bai, Junjie and Chintala, Soumith},
booktitle = {Advances in Neural Information Processing Systems 32},
pages = {8024--8035},
year = {2019},
publisher = {Curran Associates, Inc.},
url = {http://papers.neurips.cc/paper/9015-pytorch-an-imperative-style-high-performance-deep-learning-library.pdf}
}

@book{frisch1995turbulence,
  title={Turbulence: the legacy of AN Kolmogorov},
  author={Frisch, Uriel and Kolmogorov, Andre{\u{\i}} Nikolaevich},
  year={1995},
  publisher={Cambridge university press}
}

@misc{winhart2021large,
  title={Large-eddy simulation of a modified T106 low-pressure turbine stator under periodic wake impact},
  author={Winhart, Benjamin},
  year={2021},
  note={Dissertation, Bochum, Ruhr-Universit{\"a}t Bochum, pages=32--34, 39, 40, 50--54}
}

@article{duraisamy2019turbulence,
  title={Turbulence modeling in the age of data},
  author={Duraisamy, Karthik and Iaccarino, Gianluca and Xiao, Heng},
  journal={Annual review of fluid mechanics},
  volume={51},
  number={1},
  pages={357--377},
  year={2019},
  publisher={Annual Reviews}
}

@article{fang2025breakthroughs,
  title={Breakthroughs and Perspectives of Artificial Intelligence in Turbulence Research: From Data Parsing to Physical Insights},
  author={Fang, Junjie and Qin, Xujiang and Zuo, Yanqiu and Wang, Hongqiang},
  journal={Archives of Computational Methods in Engineering},
  pages={1--36},
  year={2025},
  publisher={Springer}
}

@inbook{sagaut2005large,
  title={Large eddy simulation for incompressible flows: an introduction},
  author={Sagaut, Pierre},
  year={2001},
  publisher={Springer Science \& Business Media},
  pages={33-48}
}

@article{von1911mechanismus,
  title={{{\"U}ber den Mechanismus des Widerstandes, den ein bewegter K{\"o}rper in einer Fl{\"u}ssigkeit erf{\"a}hrt}},
  author={von Kármán, Theodore},
  journal={{N}achrichten von der {G}esellschaft der {W}issenschaften zu {G}{\"o}ttingen, {M}athematisch-{P}hysikalische {K}lasse},
  volume={1911},
  pages={509--517},
  year={1911}
}

@Article{Sharifi_Ghazijahani_2025_eaai,
  author    = {Sharifi Ghazijahani, Mohammad and Cierpka, Christian},
  journal   = {Engineering Applications of Artificial Intelligence},
  title     = {On the spatial prediction of the turbulent flow behind an array of cylinders via echo state networks},
  year      = {2025},
  pages     = {110079},
  volume    = {144},
  doi       = {10.1016/j.engappai.2025.110079},
  publisher = {Elsevier BV},
}

@Article{Sharifi_Ghazijahani_2024_scirep,
  author    = {Sharifi Ghazijahani, Mohammad and Cierpka, Christian},
  journal   = {Scientific Reports},
  title     = {Echo state networks for modeling turbulent convection},
  year      = {2024},
  volume    = {14},
pages ={1637},
  doi       = {10.1038/s41598-024-79756-7},
  publisher = {Springer Science and Business Media LLC},
}

@Article{Ghazijahani_2024_MLST,
  author    = {Ghazijahani, M Sharifi and Cierpka, C},
  journal   = {Machine Learning: Science and Technology},
  title     = {On the prediction of the turbulent flow behind cylinder arrays via echo state networks},
  year      = {2024},
  pages     = {035005},
  volume    = {5},
  doi       = {10.1088/2632-2153/ad5414},
  publisher = {IOP Publishing},
}

@Article{SharifiGhazijahani2023,
  author    = {Sharifi Ghazijahani, Mohammad and Cierpka, Christian},
  journal   = {Physics of Fluids},
  title     = {Flow structure and dynamics behind cylinder arrays at Reynolds number $\sim$ 100},
  year      = {2023},
  volume    = {35},
  doi       = {10.1063/5.0155102},
  publisher = {AIP Publishing},
}

@Article{Teutsch_2025,
  author    = {Teutsch, Philipp and Pfeffer, Philipp and Sharifi Ghazijahani, Mohammad and Cierpka, Christian and Schumacher, Jörg and Mäder, Patrick},
  journal   = {APL Machine Learning},
  title     = {Slim multi-scale convolutional autoencoder-based reduced-order models for interpretable features of a complex dynamical system},
  year      = {2025},
  volume    = {3},
  doi       = {10.1063/5.0244416},
  publisher = {AIP Publishing},
}

@Article{Teutsch2025_aplml,
  author    = {Teutsch, Philipp and Ghazijahani, Mohammad Sharifi and Heyder, Florian and Cierpka, Christian and Schumacher, Jörg and Mäder, Patrick},
  journal   = {APL Machine Learning},
  title     = {Large-scale-aware data augmentation for reduced-order models of high-dimensional flows},
  year      = {2025},
  volume    = {3},
  doi       = {10.1063/5.0213700},
  publisher = {AIP Publishing},
}

@Article{sharifi_ghazijahani_2025_16794036,
  author       = {Sharifi Ghazijahani, Mohammad and
                  Cierpka, Christian},
  title        = {Turbulent Flow Behind Cylinder Arrays},
  month        = {8},
  year         = {2025},
  note = {\url{https://doi.org/10.5281/zenodo.16794036}, DOI: 10.5281/zenodo.16794036, Publisher: Zenodo}
}

@article{neumann1932proof,
  title={Proof of the quasi-ergodic hypothesis},
  author={Neumann, J v},
  journal={Proceedings of the National Academy of Sciences},
  volume={18},
  number={1},
  pages={70--82},
  year={1932},
  publisher={National Academy of Sciences}
}

@article{birkhoff1931proof,
  title={Proof of the ergodic theorem},
  author={Birkhoff, George D},
  journal={Proceedings of the National Academy of Sciences},
  volume={17},
  number={12},
  pages={656--660},
  year={1931},
  publisher={National Academy of Sciences}
}

@inbook{eisner2015operator,
  title={Operator theoretic aspects of ergodic theory},
  author={Eisner, Tanja and Farkas, B{\'a}lint and Haase, Markus and Nagel, Rainer},
  volume={272},
  year={2015},
  publisher={Springer}, 
  pages={1--4, 104--106}
}

\newpage
\appendix

\section{Evaluation results for larger grid of pixels}
\label{sec:appendix_eval}

\subsection{Flow around a cylinder}

\begin{figure}[!h]
\centering
\begin{subfigure}{.33\textwidth}
  \centering
  \includegraphics[width=\linewidth]{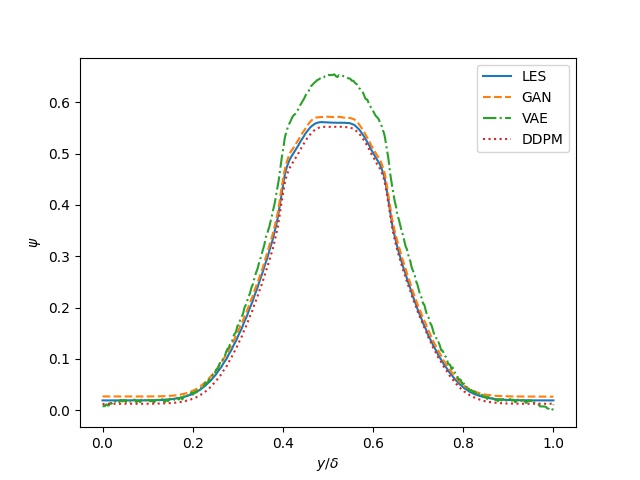}
  \caption*{13D}
\end{subfigure}%
\begin{subfigure}{.33\textwidth}
  \centering
  \includegraphics[width=\linewidth]{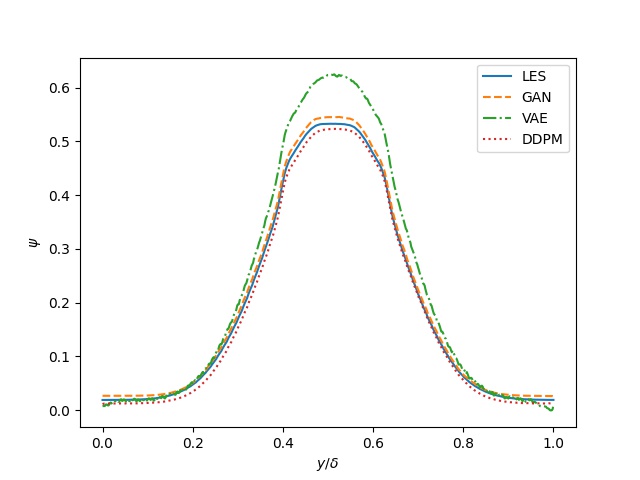}
  \caption*{16D}
\end{subfigure}
\begin{subfigure}{.33\textwidth}
  \centering
  \includegraphics[width=\linewidth]{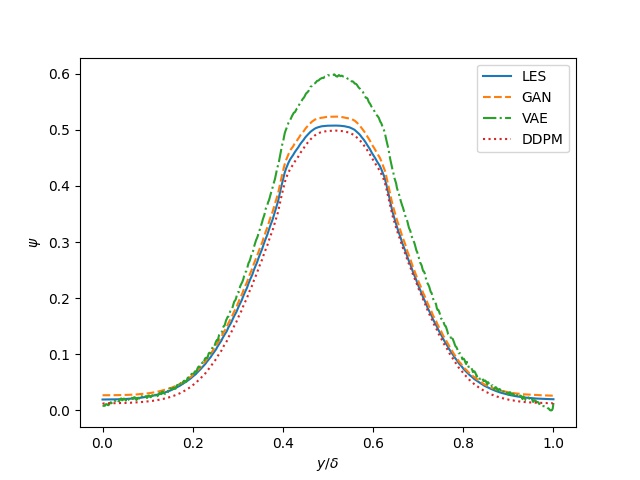}
  \caption*{20D}
\end{subfigure}
\caption{Comparison of the mean local velocity fluctuation magnitude was performed over increasingly large portions of the downstream wake region, extending to \(13D\), \(16D\), and \(20D\) within the numerical domain behind the cylinder (see \cref{fig:300px_and_500px_lines}), corresponding to \(50\%\), \(75\%\), and \(100\%\) of the investigated wake. Each value represents the deviation of the local flow velocity from the background (mean) flow. All datasets were normalized prior to evaluation.
}
\label{fig:eval_kvs_large_mean}
\end{figure}
\begin{figure}[!h]
\centering
\begin{subfigure}{.33\textwidth}
  \centering
  \includegraphics[width=\linewidth]{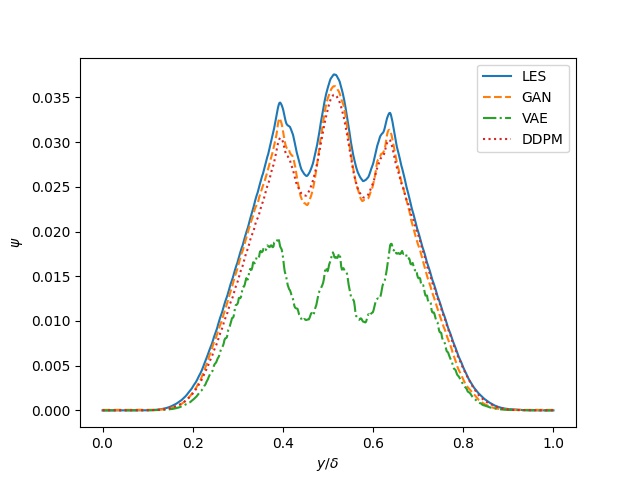}
  \caption*{13D}
\end{subfigure}%
\begin{subfigure}{.33\textwidth}
  \centering
  \includegraphics[width=\linewidth]{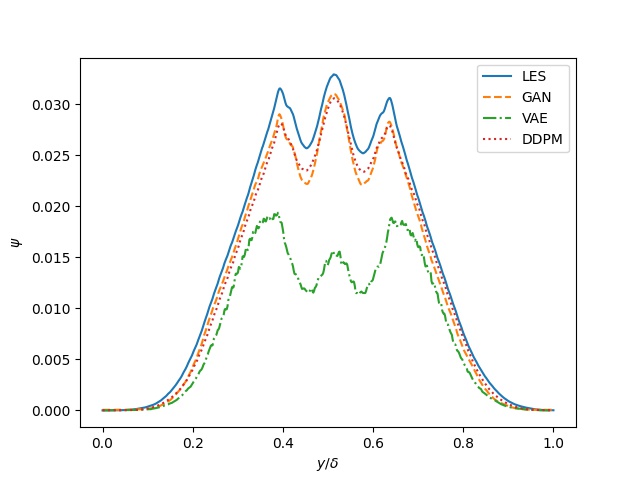}
  \caption*{16D}
\end{subfigure}
\begin{subfigure}{.33\textwidth}
  \centering
  \includegraphics[width=\linewidth]{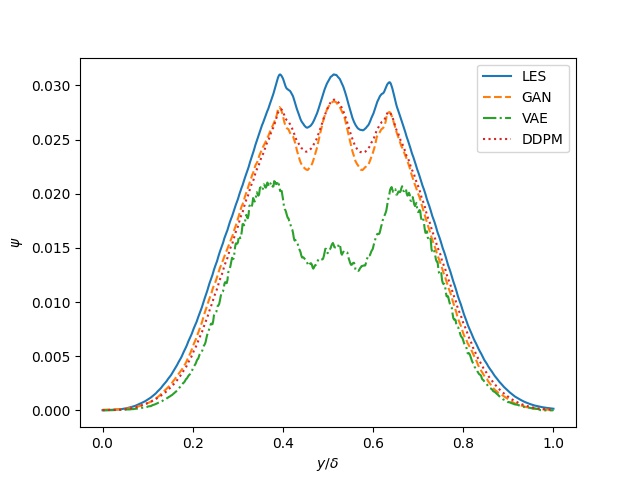}
  \caption*{20D}
\end{subfigure}
\caption{Comparison of the variance of the local velocity fluctuation was performed over increasingly large portions of the downstream wake region, extending to \(13D\), \(16D\), and \(20D\) within the numerical domain behind the cylinder (see \cref{fig:300px_and_500px_lines}), corresponding to \(50\%\), \(75\%\), and \(100\%\) of the investigated wake. Each value represents the deviation of the local flow velocity from the background (mean) flow. All datasets were normalized prior to evaluation.}
\label{fig:eval_kvs_large_var}
\end{figure}

\subsection{Flow behind array of seven cylinders}

\begin{figure}[!h]
\centering
\begin{subfigure}{.33\textwidth}
  \centering
  \includegraphics[width=\linewidth]{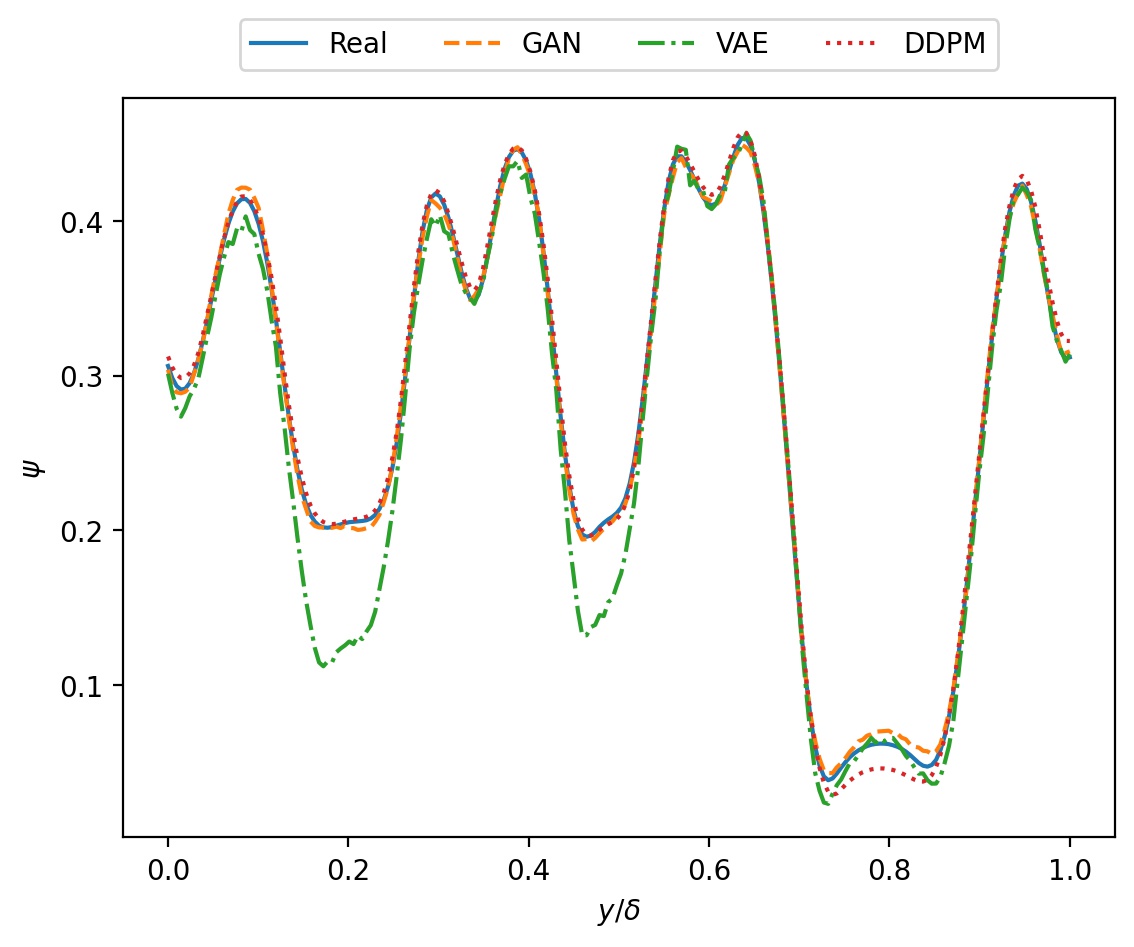}
  \caption*{25.2D}
\end{subfigure}%
\begin{subfigure}{.33\textwidth}
  \centering
  \includegraphics[width=\linewidth]{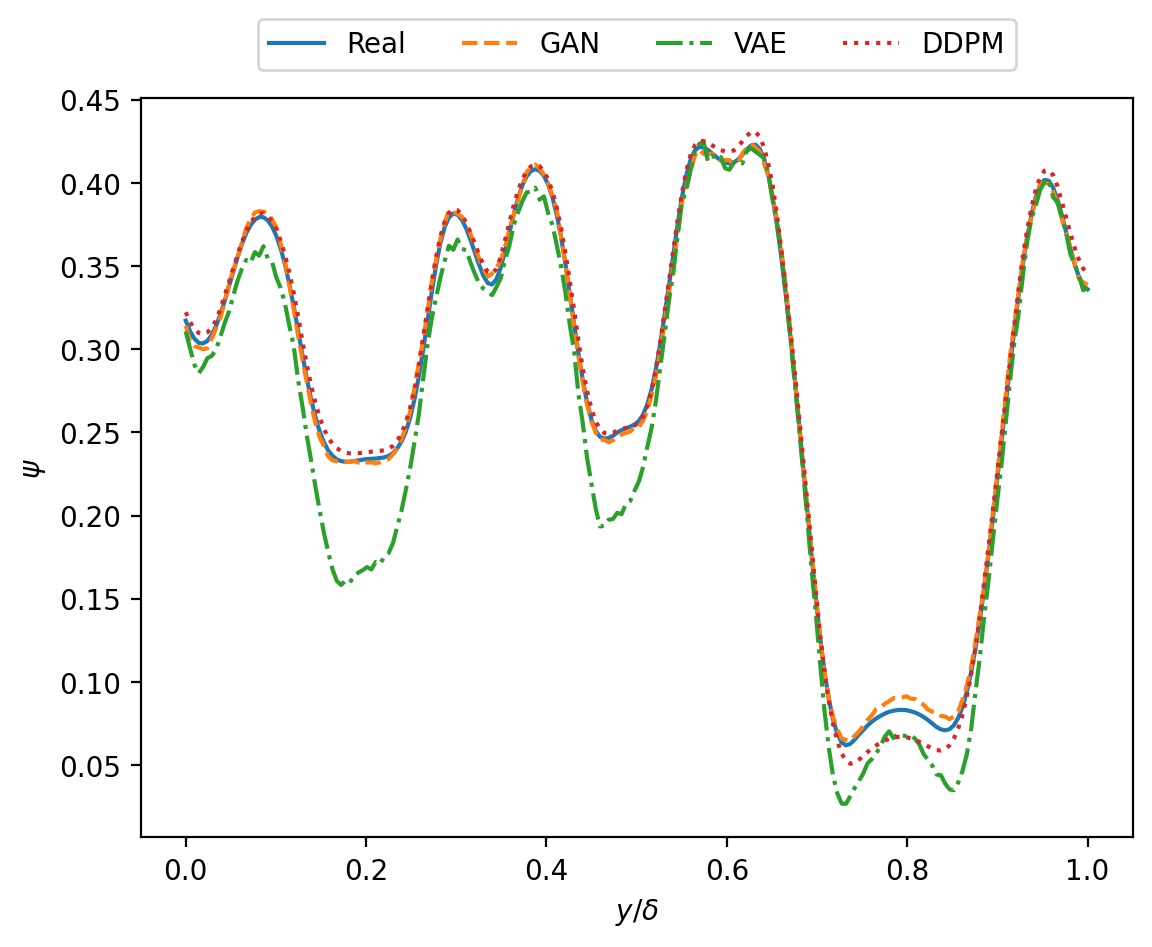}
  \caption*{37.5D}
\end{subfigure}
\begin{subfigure}{.33\textwidth}
  \centering
  \includegraphics[width=\linewidth]{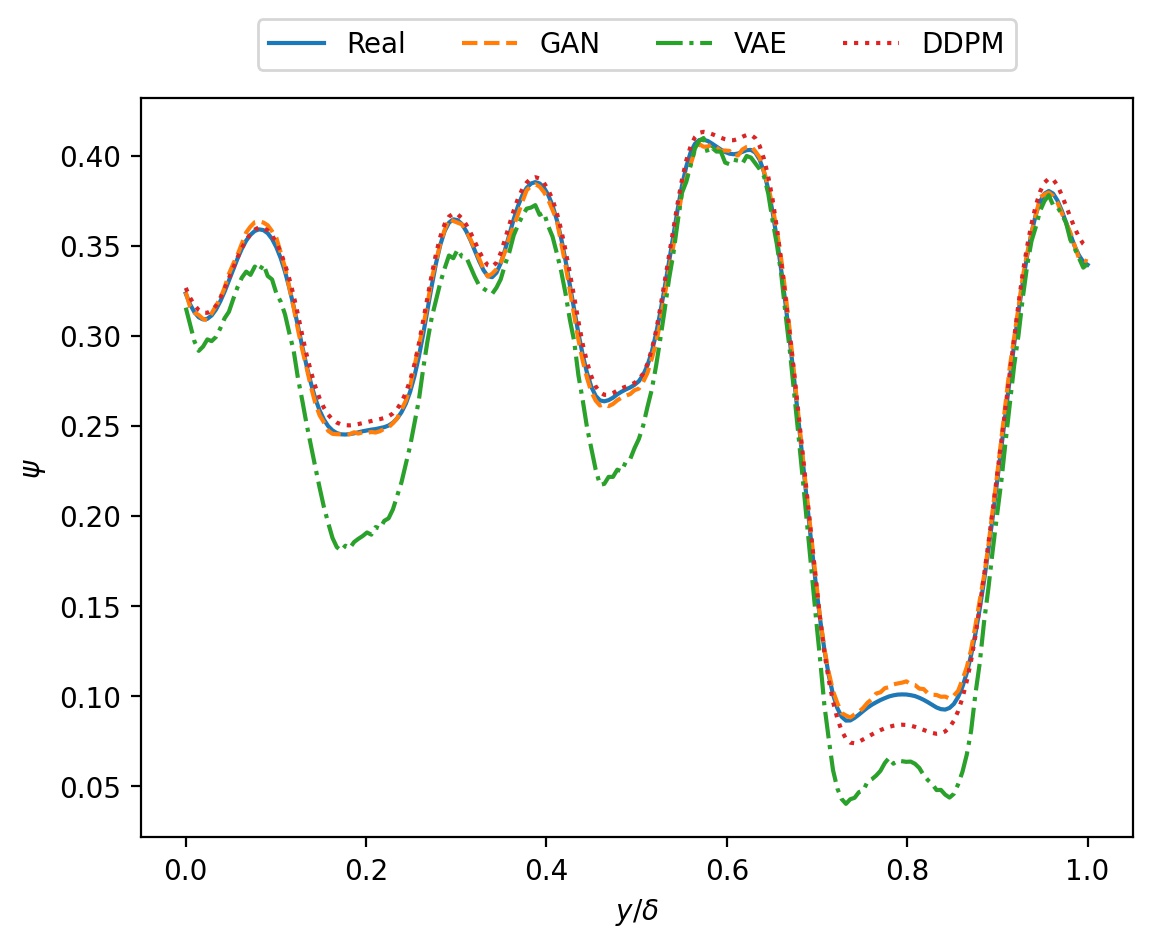}
  \caption*{50D}
\end{subfigure}
\caption{Comparison of the mean local velocity fluctuation magnitude was performed over increasingly large portions of the downstream wake region, extending to \(25.2D\), \(37.5D\), and \(50D\) within the numerical domain behind the array of seven  cylinders in the 4V2H test case (see \cref{fig:35px_and_70px_lines}), corresponding to \(50\%\), \(75\%\), and \(100\%\) of the investigated wake. Each value represents the deviation of the local flow velocity from the background (mean) flow. All datasets were normalized prior to evaluation.}
\label{fig:eval_4V2H_large_mean}
\end{figure}

\begin{figure}[!h]
\centering
\begin{subfigure}{.33\textwidth}
  \centering
  \includegraphics[width=\linewidth]{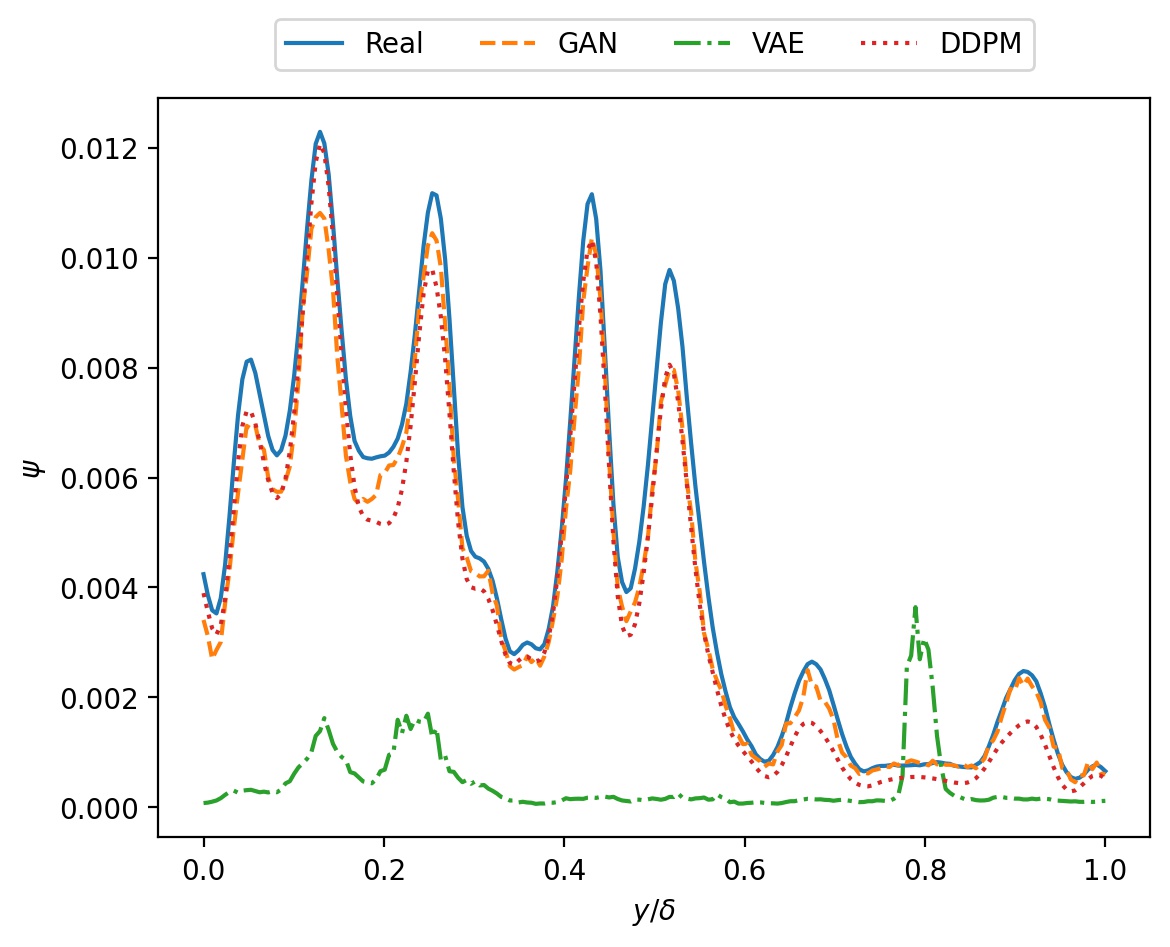}
  \caption*{25.2D}
\end{subfigure}%
\begin{subfigure}{.33\textwidth}
  \centering
  \includegraphics[width=\linewidth]{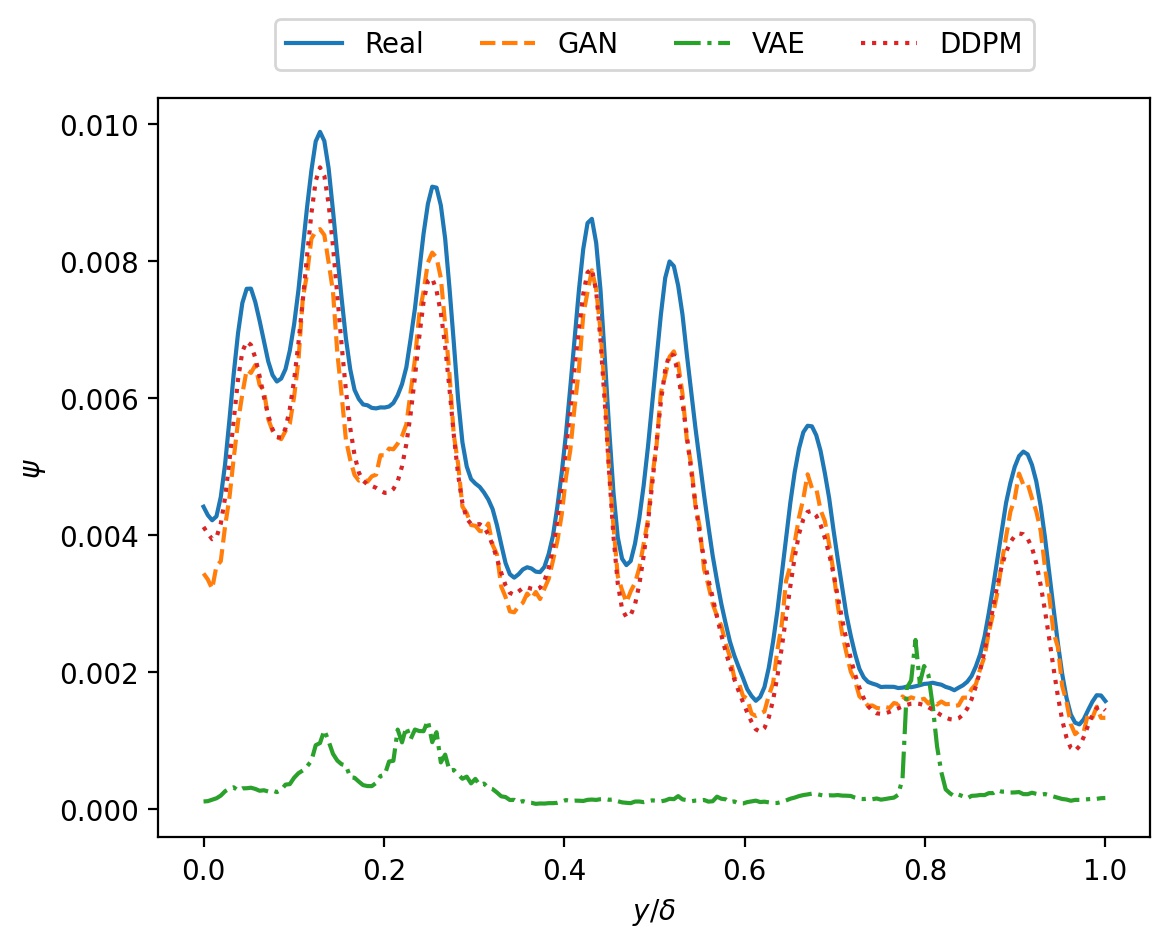}
  \caption*{37.5D}
\end{subfigure}
\begin{subfigure}{.33\textwidth}
  \centering
  \includegraphics[width=\linewidth]{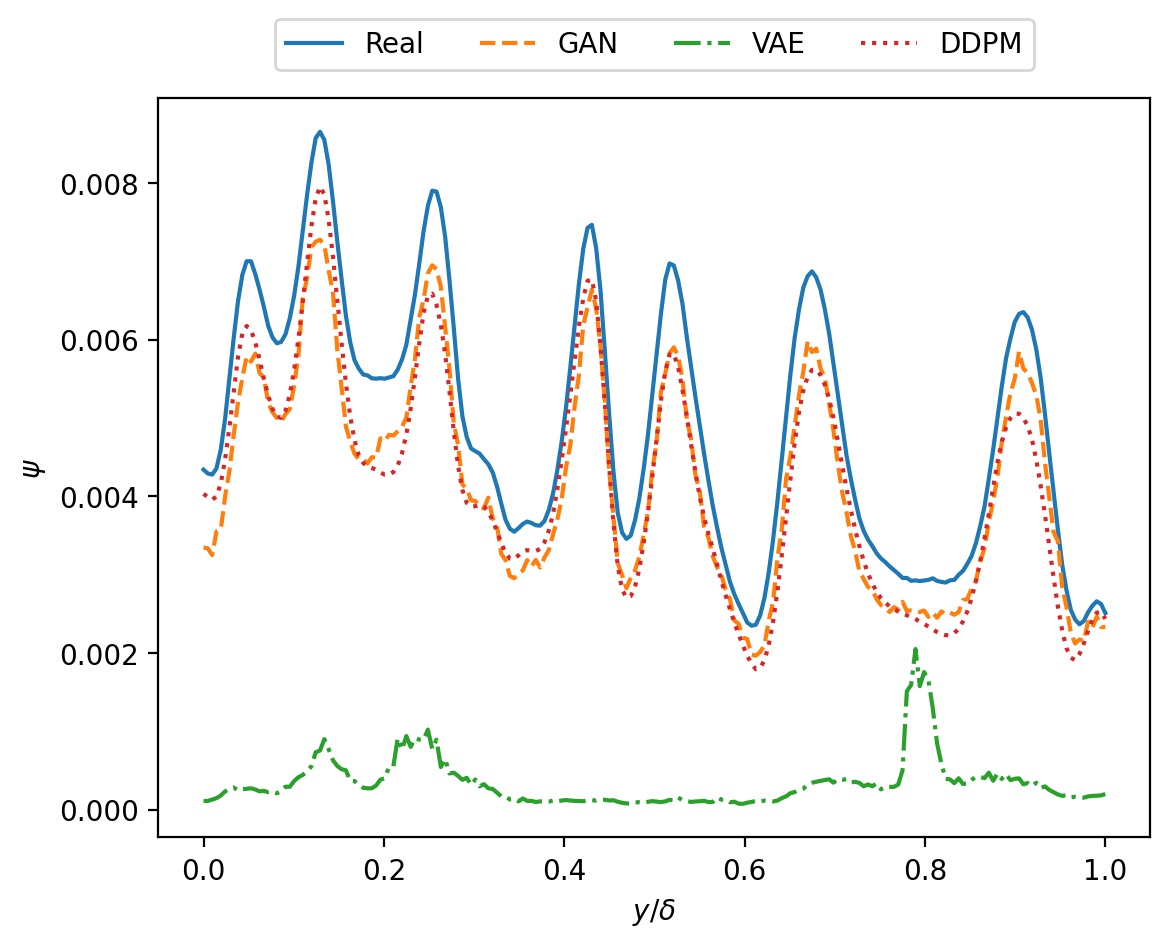}
  \caption*{50D}
\end{subfigure}
\caption{Comparison of the variance of the local velocity fluctuation was performed over increasingly large portions of the downstream wake region, extending to \(25.2D\), \(37.5D\), and \(50D\) within the numerical domain behind the array of seven  cylinders in the 4V2H test case  (see \cref{fig:35px_and_70px_lines}), corresponding to \(50\%\), \(75\%\), and \(100\%\) of the investigated wake. Each value represents the deviation of the local flow velocity from the background (mean) flow. All datasets were normalized prior to evaluation.}
\label{fig:eval_4V2H_large_var}
\end{figure}

\begin{figure}[!h]
\centering
\begin{subfigure}{.33\textwidth}
  \centering
  \includegraphics[width=\linewidth]{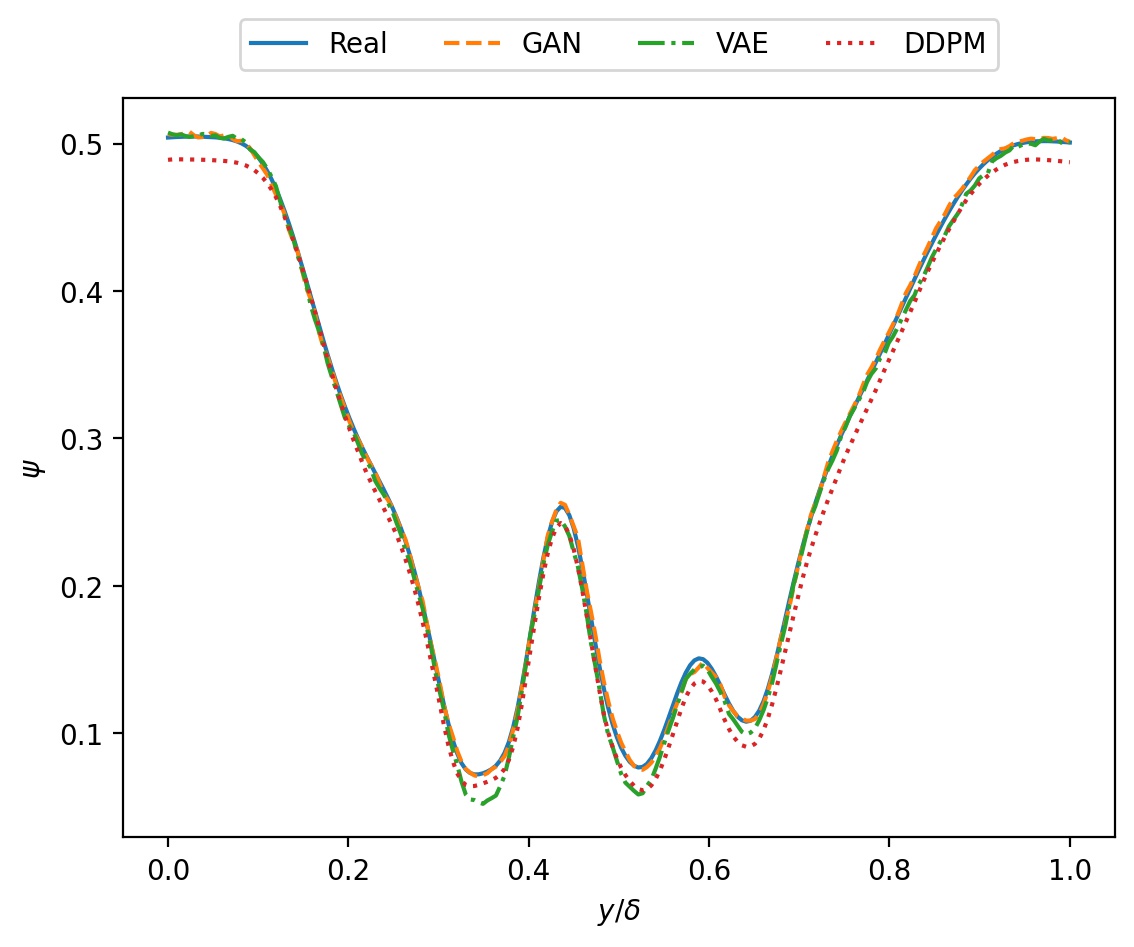}
  \caption*{25.2D}
\end{subfigure}%
\begin{subfigure}{.33\textwidth}
  \centering
  \includegraphics[width=\linewidth]{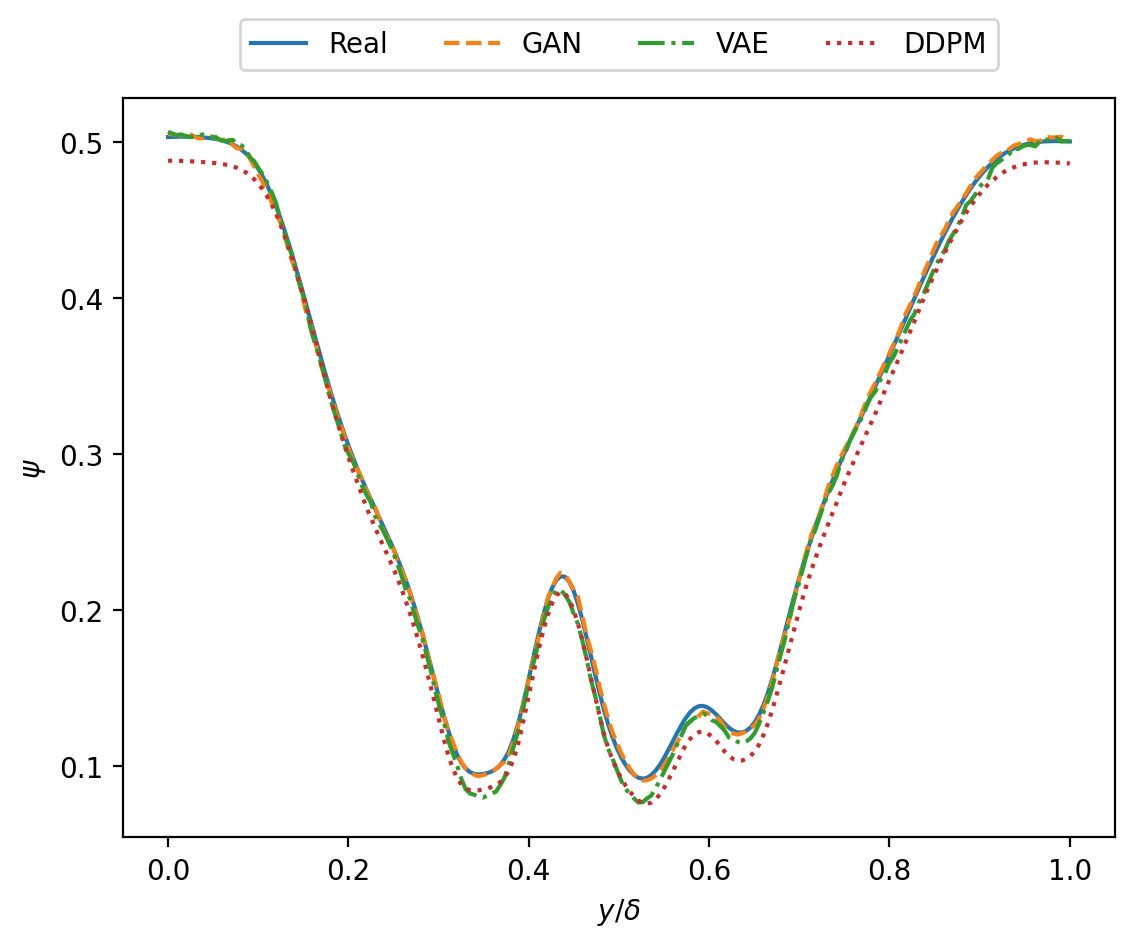}
  \caption*{37.5D}
\end{subfigure}
\begin{subfigure}{.33\textwidth}
  \centering
  \includegraphics[width=\linewidth]{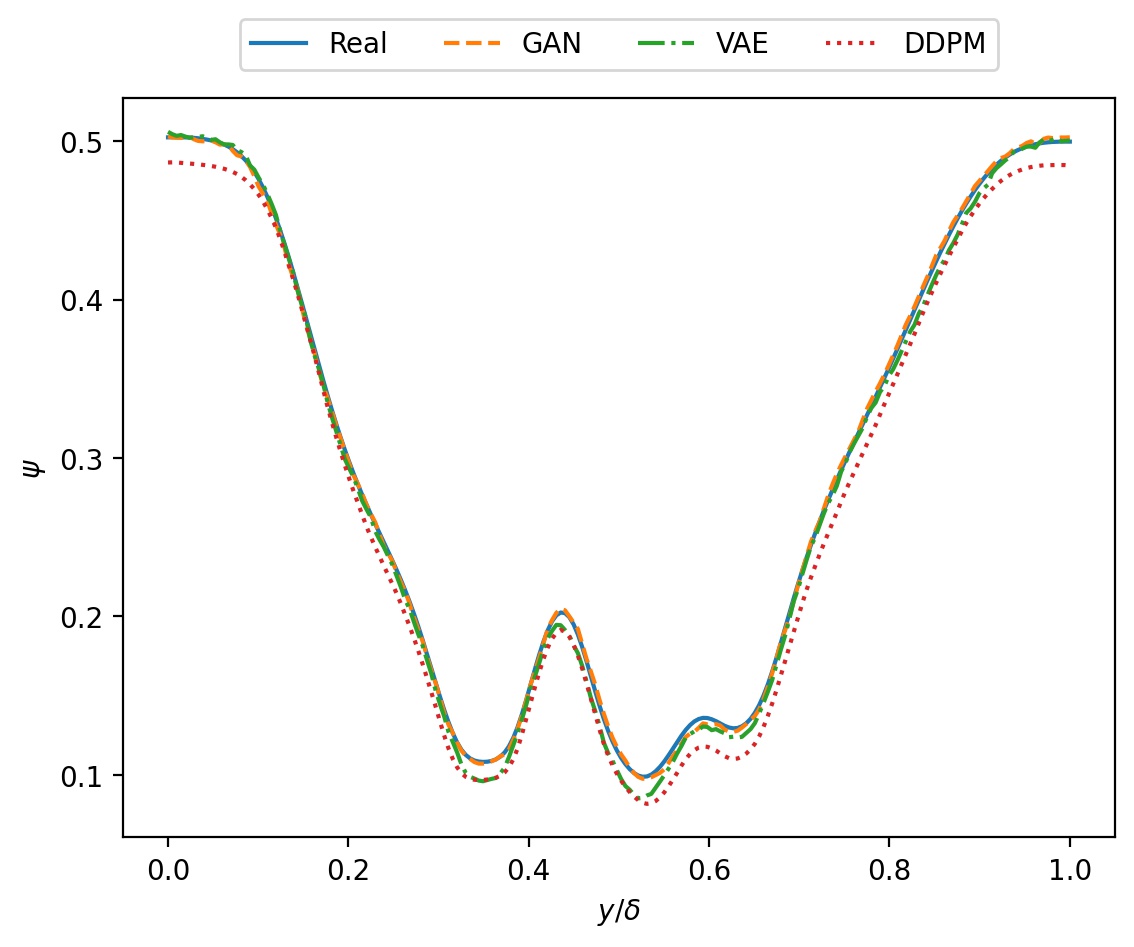}
  \caption*{50D}
\end{subfigure}
\caption{Comparison of the mean local velocity fluctuation magnitude was performed over increasingly large portions of the downstream wake region, extending to \(25.2D\), \(37.5D\), and \(50D\) within the numerical domain behind the array of seven  cylinders in the 2V8H test case (see \cref{fig:35px_and_70px_lines}), corresponding to \(50\%\), \(75\%\), and \(100\%\) of the investigated wake. Each value represents the deviation of the local flow velocity from the background (mean) flow. All datasets were normalized prior to evaluation.}
\label{fig:eval_2V8H_large_mean}
\end{figure}

\begin{figure}[!h]
\centering
\begin{subfigure}{.33\textwidth}
  \centering
  \includegraphics[width=\linewidth]{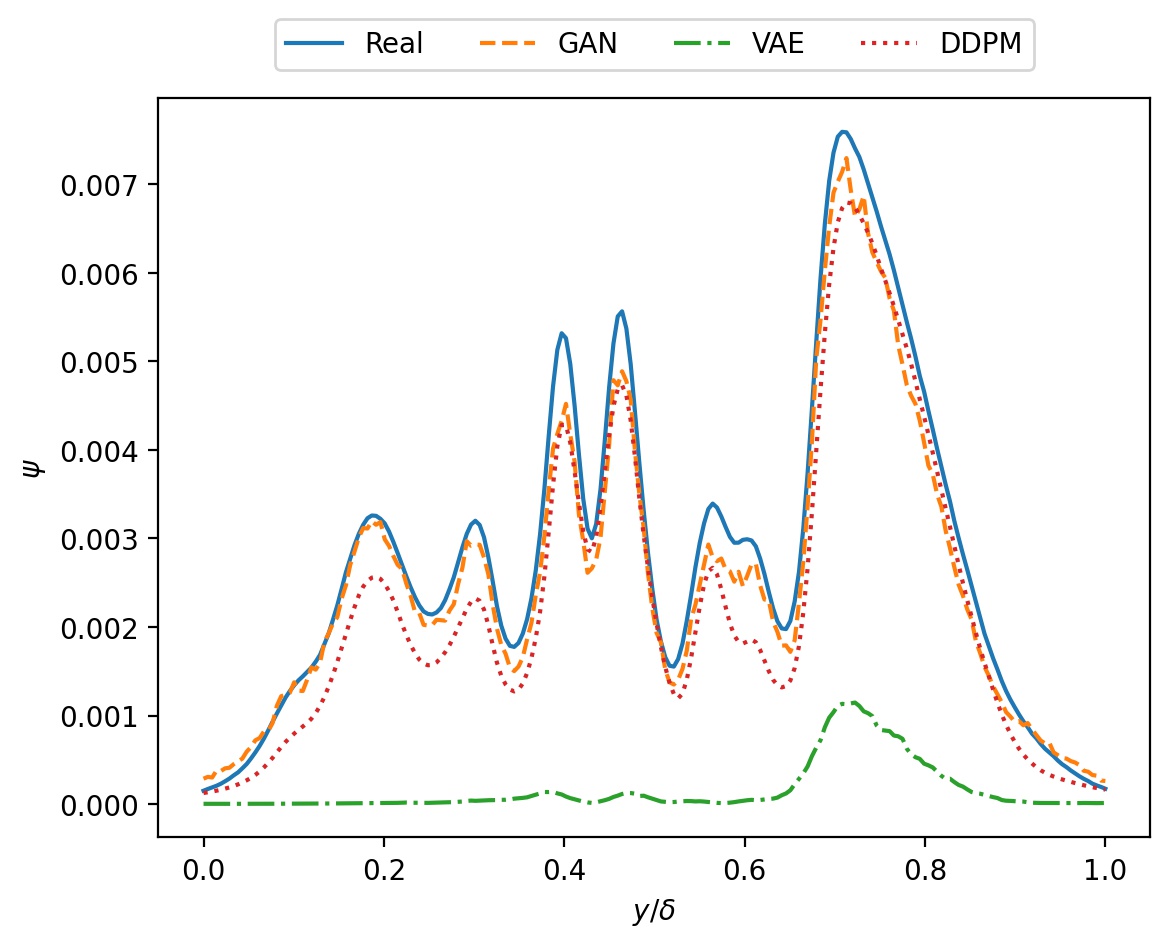}
  \caption*{25.2D}
\end{subfigure}%
\begin{subfigure}{.33\textwidth}
  \centering
  \includegraphics[width=\linewidth]{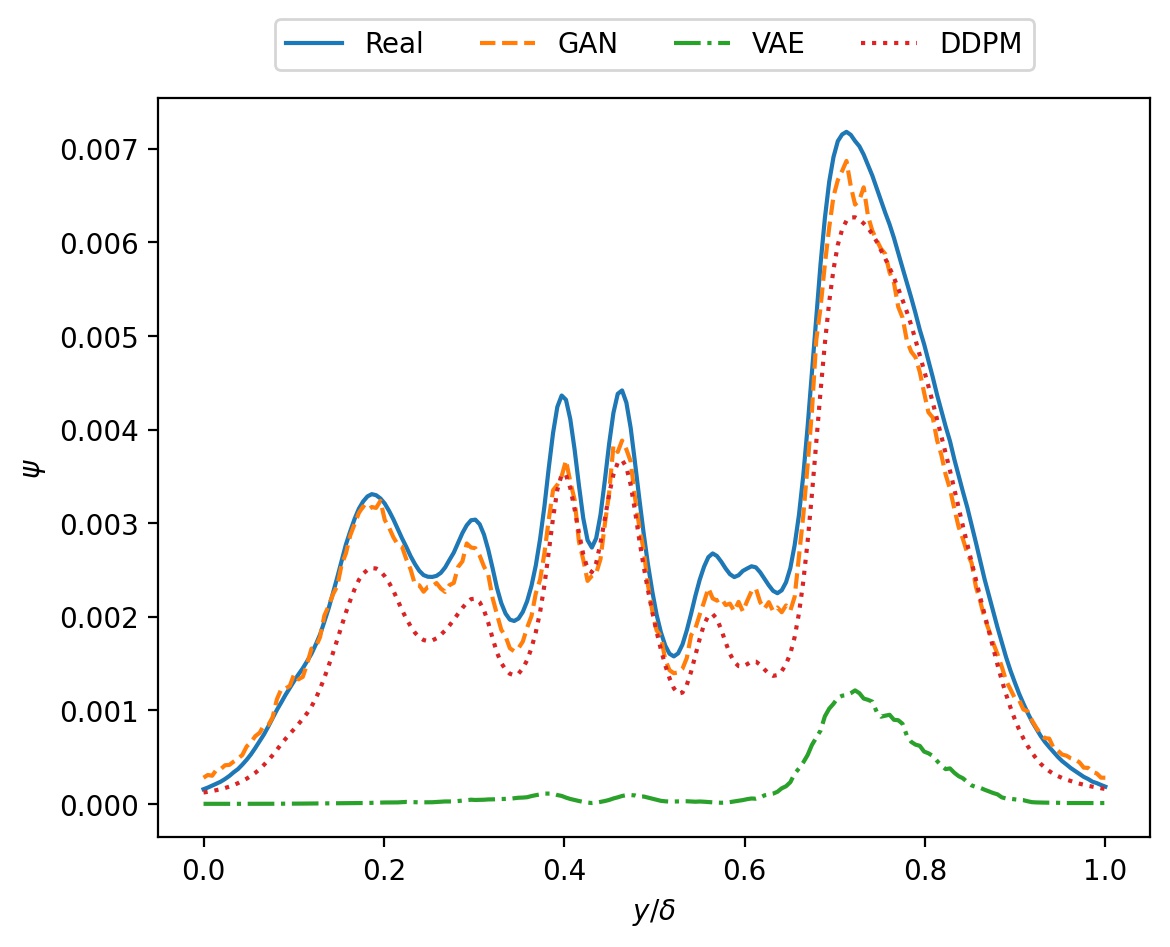}
  \caption*{37.5D}
\end{subfigure}
\begin{subfigure}{.33\textwidth}
  \centering
  \includegraphics[width=\linewidth]{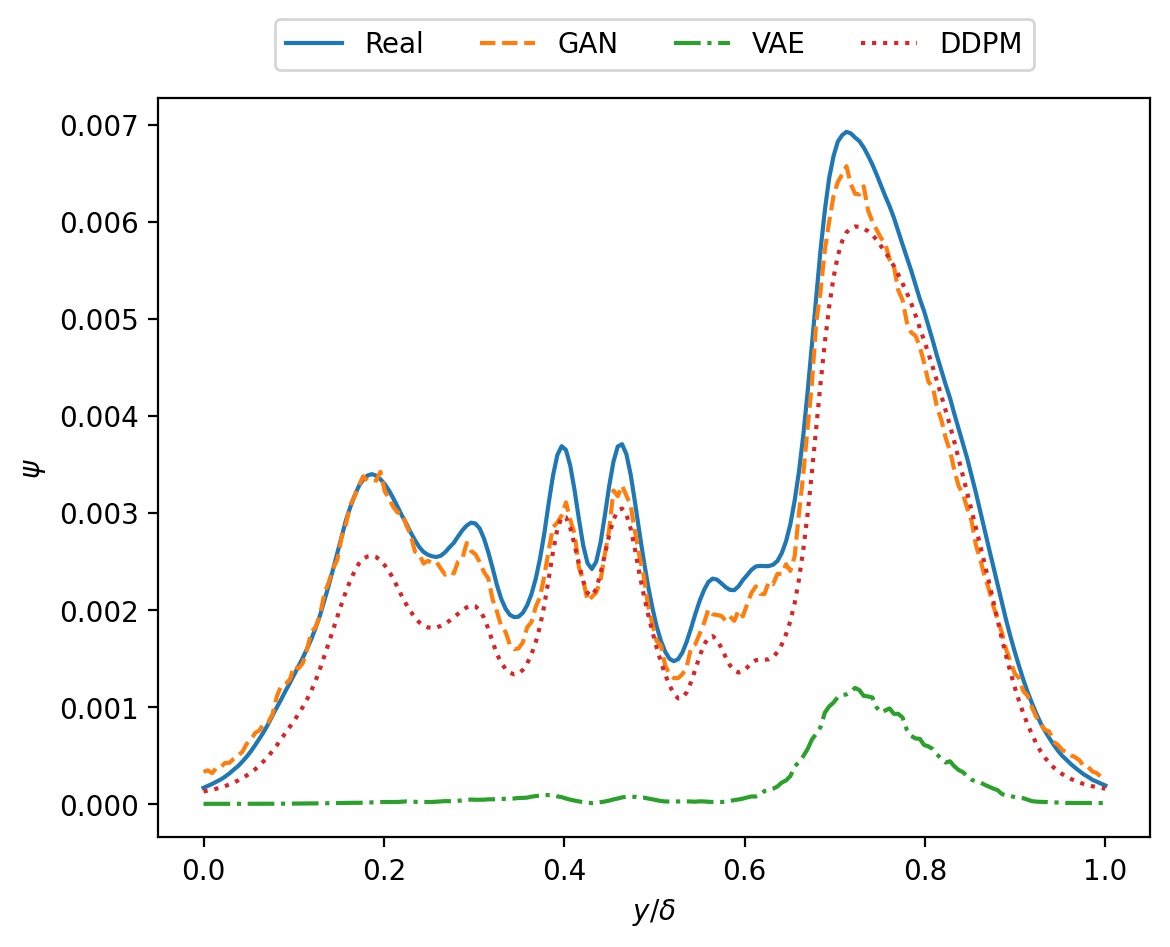}
  \caption*{50D}
\end{subfigure}
\caption{Comparison of the variance of the local velocity fluctuation was performed over increasingly large portions of the downstream wake region, extending to \(25.2D\), \(37.5D\), and \(50D\) within the numerical domain behind the array of seven  cylinders in the 2V8H test case  (see \cref{fig:35px_and_70px_lines}), corresponding to \(50\%\), \(75\%\), and \(100\%\) of the investigated wake. Each value represents the deviation of the local flow velocity from the background (mean) flow. All datasets were normalized prior to evaluation.}
\label{fig:eval_2V8H_large_var}
\end{figure}

\end{document}